\documentclass[12pt,a4paper,titlepage]{article}
\usepackage{graphics,color}
\newlength{\extraspace}
\newlength{\extraspaces}
\newlength{\interfigspace}
\newlength{\figwidth}
\newlength{\halffigwidth}
\newcommand{\be}{\begin{equation}
\addtolength{\abovedisplayskip}{\extraspaces}
\addtolength{\belowdisplayskip}{\extraspaces}
\addtolength{\abovedisplayshortskip}{\extraspace}
\addtolength{\belowdisplayshortskip}{\extraspace}}
\newcommand{\ee}{\end{equation}}

\newcommand{\bq}{\begin{eqnarray}
\addtolength{\abovedisplayskip}{\extraspaces}
\addtolength{\belowdisplayskip}{\extraspaces}
\addtolength{\abovedisplayshortskip}{\extraspace}
\addtolength{\belowdisplayshortskip}{\extraspace}}
\newcommand{\eq}{\end{eqnarray}}

\newcommand{\nsection}[1]{
\section{#1}
\setcounter{equation}{0} \setcounter{figure}{0} }

\newlength{\figsize}
\newcommand{\bra}[1]{\ensuremath{\langle{#1}|}}
\newcommand{\ket}[1]{\ensuremath{|{#1}\rangle}}
\newcommand{\set}[1]{\ensuremath{\{{#1}\}}}
\newcommand{\commut}[2]{\ensuremath{[{#1},{#2}]}}
\newcommand{\expect}[1]{\ensuremath{\langle{#1}\rangle}}

\newcommand{\var}{\ensuremath{\mathrm{var}}}

\newcommand{\vb}{\verb}

\begin{document}
\addtolength{\baselineskip}{.8mm}

\thispagestyle{empty}
\begin{flushright}
{\sc OUTP}-00-54P\\
cond-matt/0101474
\end{flushright}
\vspace{.3cm}

\begin{center}
{\large\sc{Numerical analysis of the Minimal and Two-Liquid models of the Market Microstructure  }}
\\ [12mm]

{\sc  David L.C. Chan\footnote{david.chan@sjc.ox.ac.uk}}\\
\vspace{.3cm}
{St.John's College, Oxford \\
 OX1 3JP, UK}\\
\vspace{.3cm}
{\sc  David Eliezer\footnote{deliezer@numerix.com }}\\
\vspace{.3cm}
{Research Group, Numerix, Inc.\\546 5th Ave.\\ New York \\ NY,  USA}\\
\vspace{.3cm}
 and\\
\vspace{.3cm}
 {\sc Ian I. Kogan\footnote{i.kogan1@physics.ox.ac.uk}}\\
\vspace{.3cm}
{Theoretical Physics, Department of Physics \\[2mm]
 University  of Oxford, 1 Keble Road \\[2mm]
 Oxford, OX1 3NP, UK}
\\[12mm]

{\sc Abstract}
\end{center}
\noindent
We present results of numerical analysis  of several  simple 
models for the microstructure of a
double auction market without intermediaries which were introduced in 
\cite{kogan}. These markets can be represented 
as a set of buyers and a set of sellers, whose
numbers vary in time, and who diffuse in price space and interact
through an annihilation interaction. In this paper two models
suggested in \cite{kogan} are studied -  the minimal model and the
two-liquid model.
 We perform computer simulations of the
minimal model in order to verify three of the liquidity scaling laws
postulated in \cite{kogan}. It is found that midmarket variance, bid-offer
spread, and fluctuation of the bid-offer spread scale according to $D/J$ where
$D$ is the diffusion coefficient (trader volatility) and $J$ is the deal rate. A logarithmic
correction to the scaling law for midmarket variance is observed, but not for
bid-offer spread or its fluctuation, because they are fundamentally different
quantities. Scaling parameters are obtained. We show both analytically 
and numerically that the total number
of traders in the market scales as $JL^2/4D$, where $L$ is the width
of a price space.
 Time to midmarket sale ($\tau_S$) is found to
scale as $1/J$ while its fluctuation goes as $0.73/J$. A ``reduced'' time
($\tau_{reduced}$) is also studied, and found to scale in a non-trivial way.
Asymmetric fluxes are introduced to the minimal model and analytical
result  derived in
\cite{kogan}, for the speed of the moving midmarket agrees with
numerical results.
Simulation of the two-liquid model which describes a market with both 
market order and limit order traders,  reveals widening of the bid-offer spread
when the flux of market order traders exceeds that of limit order traders. The
variation of the spread with the fraction of market-order traders is
investigated. The formula for asymmetric fluxes is applied to the two-liquid
model and its predictions are found to agree with experiment. The critical
point is approximately determined, and the ratio of the midmarkets for $f =
0.0$ and $f = 0.5$ (where $f$ is the fraction of market-order traders) is
calculated.


\nocite{*}
\newpage
\nsection{Introduction} \label{intro}

 Financial markets are the object of considerable interest to
an increasing number of physicists, because of their complex nature and the
applicability of stochastic techniques to such dynamic systems. This is a
relatively new area of interdisciplinary physics research, and is catching the
attention of quite a few theoretical and condensed matter physicists,
especially those working in the fields of complexity and chaos. 
The relationship between physics and finance is not
something that started in recent years. In fact, as far back as 1900, Louis
Bachelier proposed the random walk model of the stock market. Historically,
much of the initial work on Brownian motion and fractals originated from
studies of stock market behaviour, before being absorbed by mainstream physics.
In the 1960s and 1970s these ideas became very popular and eventually led to
the famous Black-Scholes option pricing formula. 

Recently, many papers have appeared in which markets are treated as
far-from-equilibrium dynamical systems. Econophysics is a quickly growing
field of research, and work is being done on the scaling behaviour of exchange
rates, ``log-periodic'' oscillations as crash precursors, dynamics of the
interest rate curve, and market fluctuations, to name but a few examples.
 Some  list of references (very incomplete) can be found in 
\cite{kogan}. For   more  detailed discussion on  physical approaches to
financial markets and economy   we refer
 the reader to  recent publications  \cite{bookMS}, \cite{bookBP}, 
 \cite{farmer}  which provide a lot of references.

 In this paper we present results of numerical simulations of two
models - Minimal and Two-liquid, which had been formulated by two of
us (D.E. and I.I.K.) in the paper on market microstructure
\cite{kogan}. In our approach the market microstructure is described
by a diffusion-annihilation model. Different states of market are
described by different possible state of a system in which diffusion
controlled annihilation reaction takes place and the simplest one is
the steady state.

 The first application of   diffusion-controlled annihilation to
financial markets was suggested by  Bak, Paczuski and Shubik\cite{bps}
 to describe {\it market evolution}. These authors
introduced a series of models based on  diffusion-controlled
annihilation as a
route towards recovering the observed Levy-Pareto ``fat-tail'' distributions
which are said to describe the medium term evolution of the stock market.
 In BPS model {\it market microstructure}, i.e. the structure of
reaction zone  was not considred at
all. For discussions about similarities and differences  between BPS
and our approaches see \cite{kogan}. Let us also note that in this
paper we do not give references on economical and financial papers on
market microstructure. The list of more than half a hundred important 
 publications is given in \cite{kogan}.

In our model, we shall be studying market microstructure measured over short
time scales, so that sociological interactions between traders may be ignored.
In the next few sections, we will describe the details of the models (\S
\ref{fin:minimal} and \S\ref{fin:two-liquid}).

Much research has been done in steady-state diffusion-driven annihilation
reactions, especially in physics and chemistry. In applications in physics and
chemistry, such reactions are usually in three dimensions (e.g.\ in a solid or
a liquid), but in our model of financial markets, we deal with one dimension
only. In the initial paper by G\'{a}lfi and R\'{a}cz\cite{galfi&racz}, the
properties of the reaction front in a system with segregated initial
conditions were studied in the mean field approximation. The mean field
approximation, which ignores higher, non-Gaussian corrections, known as
fluctuations, works well for dynamics in three dimension, but breaks down in
lower dimensions due to the important role of microscopic density fluctuations
in one and two dimensions. Numerical simulations were performed in
\cite{corn&droz, araujo, araujo:let, corn:let, corn:sims}. Analytical
calculations by Cardy et.\ al.\ \cite{lee&cardy, howard&cardy, bhc} confirmed
these numerical results. 

This paper is organised as follows. In the next section
brief introduction into diffusion driven annihilation reactions
(sometimes called DCR, i.e. diffusion controlled reactions) is
given. Then applications of DCRs to market microstructure are
discussed (mostly based on \cite{kogan}) in the section $3$. 
In the last two sections the 
 results of numerical simulations for Minimal and Two-Liquid models 
are  presented. Some of this results confirm earlier known analytical 
or numerical results, but most of them are new - for example results
about bid-offer spread, time to midmarket sale in Minimal model  and  
critical ratio of market order traders in the Two-Liquid model. 
 In the conclusion we discuss obtained results and  new interetsing problems
 worth studying in a future.

\nsection{Diffusion driven annihilation reactions} \label{dcr}
\subsection{Basic principles of diffusion controlled reactions}
\label{dcr:principles} Diffusion controlled reactions, or DCRs, are reactions
which include as one of the stages the transport of components which can be
described by diffusion equations (see for example \cite{dcr}). In DCRs, the rate of transport of
reacting particles to within reacting distance of one another is the
rate-determining step, and the rate of actual chemical combination is assumed
to be much faster than that of transport by diffusion. This is a common
situation that occurs not only in chemical reactions between atoms, molecules
and ions, but also in cases involving electrons, lattice defects,
quasi-particles, dislocations, etc.

In classical chemical kinetics, the rate of chemical combination is assumed to
be sufficiently low compared with the rate of transport, that the uniform
particle distribution is not disturbed and the system of two species may be
considered to be in quasi-equilibrium. The rate of reaction is therefore a
function of the state of the system, determined by the temperature, pressure
and particle concentration. Diffusion space relaxation is fast enough so that
diffusion is not the rate-determining step. In the classical approximation,
therefore, we may characterize the dynamics by simple functions of state.

This is not the case in the diffusion-controlled regime. Inhomogeneities arise
because of local depletion of reactants in regions of high reaction rate. The
overall rate of reaction is determined not only by the mean reactant
concentration and other functions of state but also by the relative positions
and velocities of the particles. The inhomogeneities have linear dimensions of
the order of the inter-particle separation. In such cases, the notion of a
local concentration is meaningless and the system must be described by a
multi-particle distribution function. The search for the probability of
particle collisions is much complicated by this. The mutual correlation in the
positions of the particles is determined by the chemical reaction, making the
probability of collision between two particles a complex function of the
coordinator of all the particles in the system. In short, we have come up
against the many body problem. Methods of quantum mechanics and statistical
mechanics are widely applied in the search for a solution.

DCR theory, which seeks a quantitative understanding of DCRs, consists of two
stages. The first stage involves simple models based on single-particle
distribution functions. The second stage incorporates the role of correlation
effects. There are four basic assumptions:
\begin{enumerate}
  \item A System containing chemically active `primary' reagents is in a state of
  thermal equilibrium. In a process consisting of a fast and a slow stage, the
  fast stage produces species which first achieve a thermal equilibrium, before
  these unstable species react further by the slow process to reach a chemical,
  and therefore thermodynamic, equilibrium. This is well illustrated by
  high energy radiation chemical processes, whereby radiation incident on a sample causes
  excitations to be produced in the molecules, forming ions, radicals and other
  reactive species. These species are non-uniformly distributed, usually in the
  form of tracks or cords. The system first comes to thermal equilibrium
  through exchange of heat energy between parts at different local temperatures.
  There then follows a progression towards chemical equilibrium, and thus
  thermodynamic equilibrium, through the ensuing chemical transformations of
  the unstable species.

  \item Initially, chemically active particles are assumed to be distributed inhomogeneously
  throughout the volume in the general case.

  \item The main postulate of DCR theory is that active particles diffuse
  according to Fick's law, $J = -D\mathbf{\nabla} n(\mathbf{r})$ where $J$ is the diffusion current,
  $D$ is the diffusion coefficient and $n(\mathbf{r})$ is the local active particle
  concentration. This is applicable in the presence of strong macroscopic
  concentration gradients. However, in a spatially inhomogeneous distribution,
  the notion of concentration is meaningless, as the characteristic length
  scales involved are of the order of the intermolecular separation. Here, we
  have to resort to analyzing diffusive mobility on a microscopic molecular
  level. We can imagine the particles situated on an imaginary lattice and
  hopping from site to site within a short time interval. After each jump, the
  particle comes into thermal equilibrium with the medium, from assumption 1.
  The jumps are statistically independent, represented by a random walk. This
  leads to the use of the theory of stochastic processes.

  \item Chemical reactions determine boundary conditions of the differential
  equations. The system is initially in a state of quasi-equilibrium. The
  chemical reactions do not change the inhomogeneous state of the system. The
  problem of how to take into account the limiting transport (diffusive) stages
  of the chemical transformation was solved by Smoluchowski by considering the probability for a certain
  particle to remain unreacted by a certain time. The dynamics of unreacted
  particles are still determined by the diffusion equations. Thus, the
  probability of reaction can be determined by the flux of mobile particles
  through the boundary assigned by the selected particle surface. Taking into
  account chemical transformations in DCR kinetics is reduced to the problem of
  applying boundary conditions to the diffusion equations.
\end{enumerate}



\subsection{Applications of diffusion driven annihilation reactions}
\label{dcr:applics} Diffusion driven annihilation reactions have been well
studied in physics and chemistry, partly because of their wide applications
across many fields of these subjects. The applications in chemistry are
obvious; after all, chemistry is about the study of chemical reactions, a
large number of which are diffusion-limited. There are also applications in
physics, especially in the field of condensed matter where large numbers of
particles are being considered. A newer, and perhaps less obvious, application
is in economics, where the market microstructure determined by the short term
behaviour of traders may be viewed as a form of diffusion through price space.
We shall not go into it in any more detail here as the entire \S\ref{fin} is
devoted to this application of DCRs. Here are three examples of applications
in chemistry and physics:
\begin{enumerate}
  \item Atomic species $A$ and $B$ diffuse through a given volume with
  diffusion coefficients $D_A$ and $D_B$. When $A$ and $B$ are a distance $R$
  from each other, they recombine in time $dt$ with \emph{a priori} probability
  $\frac{1}{\tau(R)}dt$. Thus
  \begin{eqnarray*}
    A + B   & \rightarrow & C \\
            &             & \mathrm{(inert)}
  \end{eqnarray*}
  in a unidirectional reaction. If we have a bidirectional reaction where the
  product is not inert,
  \[
    A + B \rightleftharpoons C
  \]
  the reverse process has a decomposition rate of $\frac{1}{\tau(R)}e^{-\beta
  U}$ where $\beta = 1/kT$. Normally, $U \geq 0$ but if the reaction is mainly that
  of decomposition, then $U < 0$.

  \item Fluorescence of certain liquid or amorphous semiconductors. The
  electrons and holes within the semiconductor diffuse separately and
  independently. When they are in close proximity to one another, they
  have a finite probability to recombine and emit gamma rays. Here, the
  unidirectional model applies. It also applies to the diffusion and decay of
  excitons at recombination (`scavenger') sites, an alternative and important
  mechanism for delayed illumination which involves only a single species.

  \item The structure of imperfect solids contains an excess of vacancies
  (missing atoms) such as those created by radiation damage. Vacancies diffuse
  throughout the solid until they recombine with interstitial atoms, or diffuse
  to the surface and effectively disappear. Alternatively, vacancies or
  interstitials created at the surface can diffuse to the interior.
  Diffusion-limited aggregation, whereby certain solids grow from vapour or
  liquid, has proved a popular subject of research.
\end{enumerate}



\subsection{Uses of quantum field theory in diffusion-limited reactions}
\label{dcr:qft} Over the past two decades or so, it has been found possible,
by many researchers, to use quantum field theory\cite{qft}, 
\cite{cardylectures} to study the motion,
diffusion, recombination, and other dynamics of many body systems of
\emph{non} quantum mechanical objects, i.e. objects for which $ \Delta p\Delta
x \approx \hbar $ plays an insignificant role. It is used as a counting
device, and is familiar to many physicists, especially particle and
theoretical physicists. In quantum field theory, particles are constrained to
execute random walks on the vertices of a space lattice in $d$ dimensions.
Quantum field theory allows a formal solution of the `master equation' that
governs many-body probabilities. To illustrate the method, we describe its
application with an example.

The gist of the method may be demonstrated by considering the simplest example
of all, radioactive decay. This is a zero dimensional problem, so we can
proceed without worrying about diffusion operators. In a group of $N$
radioactive atoms, we expect the number remaining undecayed at time $t$ to be
$n(t) = N\exp(-t/\tau)$ where $\tau$ is the mean lifetime. $n(t)$ is the
expected number remaining undecayed, averaged over a large ensemble of such
$N$ atoms. It is known that initial conditions\cite[p981]{qft} affect the
short and medium term behaviour of the system, but the asymptotic behaviour
for $t \gg \tau$ is independent of initial conditions (as borne out by
simulations in \cite{corn:sims}).

If $1/\tau$ is the rate at which a single nucleus decays, then, for the set of
$n$ undecayed nuclei, the rate at which a single decay occurs is $n/\tau$. The
master equation for the probabilities is a linear differential-difference
equation,
\begin{equation}\label{eq:master}
  dP(n|t) = \{(n+1) P(n+1|t) - n P(n|t) \} \frac{dt}{\tau},
\end{equation}
which incorporates the two unidirectional processes: decay \emph{into} the
state of occupancy $n$ from $n+1$, and decay \emph{out of} it into $n-1$. This
master equation may be solved by indirection, by associating the state of $n$
particles with the $n$th excited state of a harmonic oscillator\cite{sak}.
Consider the harmonic-oscillator raising operator $a^*$ and its conjugate
operator $a$. The commutation relation $\commut{a}{a^*} = 1$ holds. If we
define \ket{0} as the ground state with zero occupation number, such that
$a\ket{0} \equiv 0$, the following relations may be obtained. We define $|n)$
by
\begin{eqnarray*}
    a^*|n) = |n+1), & a|n) = n |n-1).
\end{eqnarray*}
This implies
\begin{eqnarray*}
    |n) & = & (a^*)^n \ket{0},  \\
    \Rightarrow a(a^*)^n\ket{0} & = & n(a^*)^{n-1}\ket{0},  \\
    a^*a(a^*)^n\ket{0} & = & n(a^*)^n\ket{0}.
\end{eqnarray*}
It follows that $a^*a$ is the number operator. We now introduce the crucial
concept of a \emph{probability state vector} in which to embed the $P(n|t)$:
\begin{equation}
\label{eq:statevec}
    |\Psi (t)) \equiv \sum_{n=0}^{\infty} P(n|t)|n) = \sum_{n=0}^{\infty}
    P(n|t)(a^*)^n \ket{0}.
\end{equation}
The initial condition is that there are $N$ atoms at $t = 0$, so $P(n|0) =
\delta_{n,N}$, which is equivalent to the probability state $|\Psi (0)) =
(a^*)^N\ket{0}$. The right-hand basis states in this vector space (Fock space)
are the $(a^*)^n\ket{0}$. Given
\[
    1/n! \bra{0}(a)^n (a^*)^{n'} \ket{0} = \delta_{n,n'}
\]
they form a complete orthogonal set. However, the states are not normalized in
the conventional manner. Instead, they are normalized with respect to a
``reference'' state $(\mathcal{S}| \equiv \bra{0}e^a$ which is merely a
special case of a \emph{Glauber state} $\bra{\alpha \mathcal{S}} \equiv
\bra{0} e^{\alpha a}$ (itself an eigenvector of the operator $a^*$ with
eigenvalue $\alpha$).

The norm of any right-hand state $|\Phi)$ is defined in terms of
$(\mathcal{S}|$ by the inner product $(\mathcal{S}|\Phi)$, so that a state is
normalized if it satisfies $(\mathcal{S}|\Phi) = 1$. Given that
$(\mathcal{S}|{a^*}^n = (\mathcal{S}|$ for all $n \geq 0$, and
$(\mathcal{S}\ket{0} = 1$, each of the right-hand basis states $(a^*)^n
\ket{0}$ is normalized in this fashion, i.e.\ $(\mathcal{S}|(a^*)^n \ket{0} =
1$ for any $n \geq 0$. Note that the probabilities remain normalized at all
times $t$ if they were normalized initially, provided the equations of motion
conserve probability (which they always do):
\[
    (\mathcal{S}|\Psi(t)) \equiv \sum_n P(n|t) (\mathcal{S}|(a^*)^n\ket{0} =
    \sum_n P(n|t) = 1.
\]
Expectation values may be computed as follows. To evaluate the average number
remaining at any time $t > 0$, we make use of the number operator $a^*a$,
\[
    (\mathcal{S}|a^*a|\Psi(t)) \equiv \sum_n n P(n|t) \equiv \expect{n}.
\]
In general, the expectation value of a function of the number operator
$F(a^*a)$ may be computed thus:
\[
    (\mathcal{S}|F(a^*a)|\Psi(t)) \equiv \sum_n F(n)P(n|t) \equiv \expect{F}(t).
\]

The state vector $|\Psi(t))$ has all the properties of a generating function.
First, it satisfies an elementary differential equation equivalent to
Eq.~(\ref{eq:master}). Second, it yields individual $P(n|t)$ by projection onto
$1/n!\bra{0}a^n$. Third, it can be used to obtain the various moments, through
contractions with $(\mathcal{S}|$. The differential equation satisfied by
$|\Psi(t))$ is
\begin{equation}
\label{eq:decayde}
    \partial_t |\Psi(t)) = \frac{-1}{\tau} \, \Omega |\Psi(t)),
\end{equation}
in which $\Omega = a^* a - a$, the dimensionless rate operator for this
process, is sometimes referred to as the ``quantum Hamiltonian'' of the model.
We illustrate how Eq.~(\ref{eq:decayde}) may be derived from the state vector
equation~(\ref{eq:statevec}). The technique is to differentiate
Eq.~(\ref{eq:statevec}) with respect to time, substitute in the master
equation~(\ref{eq:master}) and then try to arrange it into the form of a
Schr\"{o}dinger equation, as in (\ref{eq:decayde}):
\begin{eqnarray}
    \partial_t |\Psi(t))    & = & \sum_{n=0}^{\infty} \frac{dP(n|t)}{dt} |n)   \nonumber \\
                            & = & \sum_{n=0}^{\infty} \{ (n+1)P(n+1|t) -
                            nP(n|t) \} \frac{1}{\tau} |n)
                            \label{eq:psimaster}
\end{eqnarray}
Now, we observe that
\begin{eqnarray}
    (a^*a-a)|\Psi)  & = & (a^*a-a) \sum_{n=0}^{\infty} P(n|t) |n) \nonumber \\
                    & = & \sum_{n=0}^{\infty} n P(n|t) |n) - \sum_{n=1}^{\infty} n P(n|t)
                    |n-1) \nonumber \\
                    & = & \sum_{n=0}^{\infty} n P(n|t) |n) -
                    \sum_{n=0}^{\infty} (n+1) P(n+1|t) |n)
\end{eqnarray}
which  we can identify with the right hand side of Eq.~(\ref{eq:psimaster}).
Therefore, we can transform the master equation involving probabilities to a
Schr\"{o}dinger-like equation involving $|\Psi)$:
\begin{eqnarray*}
    \partial_t |\Psi(t)) & = & - \frac{(a^*a-a)}{\tau} |\Psi(t)) \\
                         & = & - \frac{1}{\tau} \Omega |\Psi(t))
\end{eqnarray*}
giving us Eq.~(\ref{eq:decayde}). The solution is simply
\[
    |\Psi(t)) = e^{-t/\tau \Omega} |\Psi(0)) = e^{-t/\tau \Omega} (a^*)^N
    \ket{0},
\]
where the right hand side represents the initial condition of having precisely
$N$ undecayed nuclei. The exponential in the solution may be expanded in series
form and the whole expression substituted back into Eq.~(\ref{eq:decayde}) to
check that it satisfies the differential equation.

The above illustration of the method for 0 dimensions can be extended to 1, 2
and 3 dimensions, with the introduction of diffusion and annihilation
operators. Although the example shown above is trivial, quantum field theory is
a powerful method for solving for the dynamics of particles diffusing in a
lattice, and allows analytic solutions to be obtained. It is beyond the scope
of this report to discuss further examples of such applications, though the
interested reader is referred to \cite{qft}, \cite{cardylectures}
  for a detailed review.



\nsection{The market microstructure of the interdealer broker
markets as a reaction zone of 
 one-dimensional diffusion driven annihilation reaction} \label{fin}

\subsection{The minimal model}
\label{fin:minimal} 

Because we are only interested in scaling properties, and not in
modelling very detailed behaviour, the model is limited to only a few necessary
parameters.  This allows us to focus attention on the main model feature under examination.  
In \cite{kogan}, it was proposed that the feature dominating double auction 
market microstructure was the great press of numbers of traders, so that the behaviour 
of, e.g. moments of measures of liquidity, were largely determined by the statistics
of the traders.  The required model is one in which a large number of individual traders
interact according to any plausible dynamcics for the traders, and it is natural to 
choose the simplest one.  Trader dynamics \emph{require} that when a buyer and seller 
arrive at the same point in price space, they have agreed on a price, and must do
a deal and vanish from the price space.  And the simplest plausible model of trader motion is
a random walk.  Since we only wish to model markets in a steady state, we give these
traders no net drift.  

\begin{figure}
\centering \resizebox{\figwidth}{!}{\includegraphics{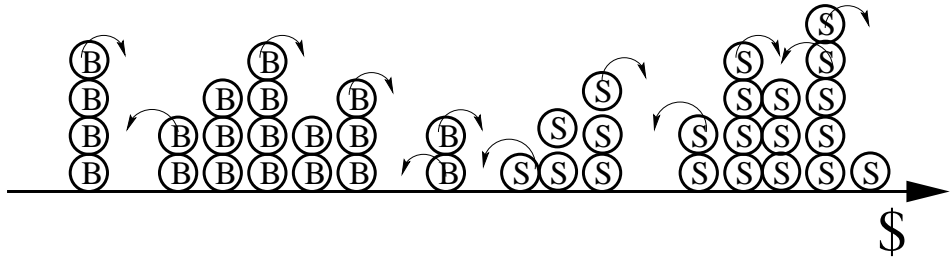}} \caption{A
typical distribution of buyers (balls with letter B) and sellers (balls with
letter S). Arrows describe ``hops''.} \label{fig:eli1}
\end{figure}

\begin{figure}
\centering
\resizebox{\figwidth}{!}{\includegraphics{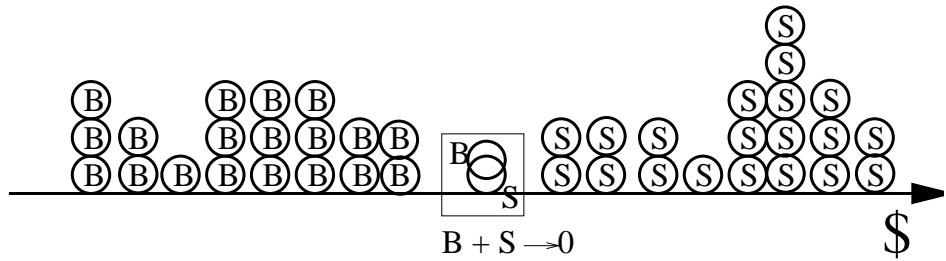}}\caption{Annihilation $B +
S \rightarrow 0$} \label{fig:eli2}
\end{figure}

\begin{figure}
\centering \resizebox{\figwidth}{!}{\includegraphics{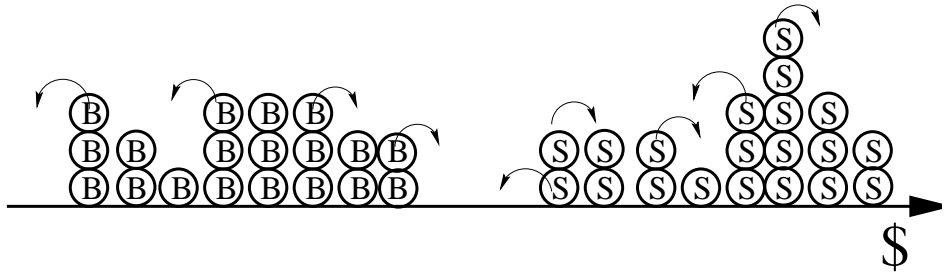}} \caption{A new
distribution of buyers (balls with letter B) and sellers (balls with letter S)
after an annihilation (trade) has occurred. Note that the bid-offer spread
fluctuates. Here it is bigger than in Fig.~\ref{fig:eli1}} \label{fig:eli3}
\end{figure}

This model of microstructure of the interdealer broker market can be
mapped onto a statistical field theory, in which particles diffuse and
annihilate.  Thus, we associate to each buyer his bid
price, and each seller his offer price, and imagine the two types of trade
prices moving around in one-dimensional price space (Fig.~\ref{fig:eli1}).
This price space is a discrete lattice, with a finite size. The traders, which
may be thought of as particles, hop from one lattice site to the next with a
certain probability (or they may choose to stay where they are). In this
model, each trader trades in a standard size. When a buyer and a seller meet
up at the same point in price space, they annihilate, leaving the market
(Figs.~\ref{fig:eli2} and \ref{fig:eli3}).
The reaction may therefore be represented by $ B + S \rightarrow 0 $, which
is a \emph{diffusion controlled (or diffusion driven) annihilation reaction}.
This type of reaction has been studied extensively in the physics and chemistry
literature\cite{dcr}.

We treat the case of a quiescent market, one which is an \emph{approximately
steady state} market, and we define our evolving state measure, at any time
$t$, to be that probability measure which is the steady state solution of our
diffusion-annihilation dynamics. As a first approximation, we allow traders to
enter only at the ends, as though the buyers start by bidding very low and the
sellers offering very high. One can solve this model by numerical simulation.
However, because of its similarity to many well-known analytically soluble
models (for example, in \cite{qft}), we can go much further. Approximate
analytical methods are available from Cardy et
al\cite{lee&cardy,howard&cardy,bhc}.

\subsection{The two-liquid model}
\label{fin:two-liquid} The minimal model is a suitable approximation of the
quiescent periods of a market's behaviour, but when it is in the process of a
long crash, it does not look quiescent and the model breaks down. It is not
sufficient merely to cause an imbalance in the rates of insertion of buyers
and sellers, because such a method would not reproduce the well-known
phenomenon of the widening of the bid-offer spread. It is believed that the
missing element is market order traders. During a crash, most traders,
desperate to sell, attempt to hit the best bid directly, as a market order. In
such a scenario, although limit order trading continues as before, the large
proportion of market order trades cannot be neglected and indeed they have a
significant effect on market dynamics. In the two-liquid model, we introduce a
second type of trader, the market order trader, who tries to hit the best
bid/offer directly, but without displaying his/her bid/offer price on the
trading screen. These market order traders trade with limit order traders, who
display their prices on the trading screen, but not with one another. If limit
order traders are represented by $B$ and $S$ and market order traders by $B'$
and $S'$, then the reactions $B + S \rightarrow 0$, $B' + S \rightarrow 0$,
and $B + S' \rightarrow 0$ can occur, but \emph{not} $B' + S' \rightarrow 0$,
since market order traders are invisible to one another. This model is able to
reproduce the familiar widening of the bid-offer spread during a crash.



\subsection{Dimensional analysis}
\label{fin:dimanal} Because we have relatively few degrees of freedom
(parameters) in our model, we can use dimensional analysis to obtain the
scaling forms of many market quantities, as a check on intuition. Starting
from the operator evolution equation, we may differentiate it to yield an
operator differential equation. This partial differential equation consists of
one coefficient only, namely the diffusion coefficient $D$, with dimensions
$x^2/t$, i.e. $D \sim \mathrm{(dollar )^2/sec}$ (see \S\ref{app:master} for
derivation of the diffusion coefficient). In addition, there is another
dimensionful quantity in the boundary conditions, $J$, the rate of insertion
of traders at the boundaries, which is equal to the deal rate in a steady
state market, with dimensions $J \sim 1/\mathrm{(sec)}$. From $D$ and $J$ we
can construct quantities with any dimensions containing (dollar) and (sec),
since $\mathrm{(dollar)} \sim \sqrt{D/J}$ and $\mathrm{(sec)} \sim 1/J$.
Therefore, the expectation value \expect{X} of a general quantity $X$ with
dimensions $[X] = \mathrm{(dollar)^m / sec^n}$ must be proportional to
$(D/J)^{m/2} J^n \sim D^{m/2} J^{n-m/2}$. 
%

So far, we have ignored, in our dimensional analysis, of the lengths scales
$L$, the size of price space, and $\delta S$, the lattice spacing. These two
lengths represent the upper and lower length scales in our model. We are
justified in doing so by the fact that if we imagine a Fourier analysis of the
dynamics, most of the `action' would be taking place away from these extreme
length scales, in the intermediate region between the two. Ultimately, the
justification comes from numerical simulations. However, these length scales
do add logarithmic corrections to the scaling laws, as shown by Cardy et
al\cite{lee&cardy,howard&cardy,bhc,kogan}. This is described in \S
\ref{fin:analytic}.

\subsection{Market parameters}
\label{fin:parameters} The model allows us to calculate the relationships
described in the introduction, but also many more. In general, whatever
initial conditions we start with, in the asymptotic limit, we will reach a
steady-state, with a reaction front, where the buyers meet the sellers. At the
centre of the reaction front is the best bid, best offer and midmarket. Beyond
the best bid/best offer, we expect to see an increasing density of
buyers/sellers. Traders are inserted at the edges at a rate equal to the deal
rate.

The statistics describing the state of the system can be read off a trading
screen. The best bid locates the top of the lower edge and the best offer the
bottom of the upper edge (see Figs.~\ref{fig:eli1} and \ref{fig:eli3}). 
%
%
Although there is a large number of interesting parameters, in this project we
shall be restricting our attention to a small subset of them: best bid $B(t)$,
best offer $O(t)$, bid-offer spread $Spr(t) \equiv O(t) - B(t)$, and midmarket
$M(t) \equiv (1/2)(B(t) + O(t))$. Note that $B(t)$, $O(t)$, $Spr(t)$ and
$M(t)$ are not Markov random variables, and do not satisfy a stochastic
differential equation, because they are subject to jumps. In addition to the
above parameters, we shall also be interested in recording the instantaneous
deal rate $DR$ and the total number of traders in the market $NUM$.

\subsection{The scaling laws}
\label{fin:scalinglaws} Financial markets, like many other systems, exhibit
scaling laws which are clearly observed by practitioners in the field. These
scaling laws are related to the concept of liquidity, which is measured in
several different ways. It may be measured by the deal rate $J$, the time to
midmarket sale, the bid-offer spread, and the trader densities near the best
bid/offer. Here we shall concentrate on several  scaling laws
which were introduced in \cite{kogan}.
\begin{itemize}
  \item Bid Offer Spread
    \begin{equation}
        Spr = \expect{O(t) - B(t)}
    \end{equation}
  \item Trade price variance (or midmarket variance)
    \begin{equation}
        w^2 = \expect{M^2(t)} - \expect{M(t)}^2
    \end{equation}
  \item Fluctuations in the Bid-Offer Spread
    \begin{equation}
        (\Delta Spr)^2 = \var(Spr) = \expect{(O(t) - B(t))^2} - \expect{O(t) -
        B(t)}^2
    \end{equation}
  \item Instantaneous Dealing Time, and its Fluctuations

    We can define $\tau_S$ as the time between successive trades. If after a
    time, more than one trade occurs in the same time step, then, in order to
    have a valid continuum limit for the quantity, we must consider the time
    between the multiple trades themselves to be zero. We may visualize it
    thus, assuming time is continuous: one trade occurs at $t = 0$, and then no trades occur for some time.
    The next trade occurs at $t = 9$, and the next at $t = 9 + \delta$. These
    are distinct times, if we use continuous time, and we would say that the
    two values of $\tau_S$ are 9 and $\delta$. If we now allow $\delta$ to tend
    to zero, we have 9 and 0 as the values of $\tau_S$, but we see that such a
    configuration is equivalent to the discrete model of time, where we have no
    trade for 9 time steps followed by 2 at the same time. Thus, the correct
    way to deal with ``simultaneous'' annihilations is to treat them as
    separate ones, separated by zero time.

    We can define its fluctuation as
    \begin{equation}
        {\Delta \tau_S}^2 = \var(\tau_S) = \expect{{\tau_S}^2} - \expect{\tau_S}^2
    \end{equation}
  \item Asymmetric fluxes

    It is possible to consider the minimal model with different fluxes for the
    buyers and the sellers. In \cite{kogan} the following
    theoretical result, valid for ``small'' times, was found:
    \begin{equation} \label{eq:asym}
        x_0(t) \sim \frac{-2D\Delta J}{\bar{J} L} t,
    \end{equation}
    where $x_0(t)$ is the position of the midmarket, which varies with time
    $t$. The flux of sellers from the upper boundary is $\bar{J} + \Delta
    J$ whilst the corresponding flux of buyers from the lower boundary is
    $\bar{J} - \Delta J$.
\end{itemize}
We expect $Spr(J)$ and $w^2$ to tend to 0 as $J \rightarrow \infty$. From
dimensional analysis, $Spr$ and $w$ have dimensions of \emph{dollars}, so
$Spr(J) \sim \sqrt(D/J)$, $w(J) \sim \sqrt(D/J)$ consistent with intuition.
Using the results of \cite{lee&cardy, howard&cardy, bhc}, we see that there is
a logarithmic correction for $w$. It is also possible to measure fluctuations
in the spread, $\Delta Spr$ or $\var(Spr)$. This quantity should have the same
scaling law as $Spr$, since it has the same dimensions. Time to midmarket sale
$\tau_S$ is expected to go as $\sim 1/J$, from dimensional analysis; in fact,
its average should be equal to $1/J$. Similarly, its fluctuation, $\Delta
\tau_S$ should have the same scaling law. There are many other scaling laws
which we will not go into here as they will not be tested later; for more
information, see \cite{kogan}.

\subsection{Analytical results}
\label{fin:analytic} Although numerical results are more precise, it is
desirable to work out some scaling laws for the model using analytical means,
as these methods supplement numerical simulations by providing intuition.
Cardy et.\ al.\cite{lee&cardy,howard&cardy,bhc} has developed an approximation
scheme known as mean field theory, replacing the evolution equations for the
operators with a partial differential equation for the density configuration
with the greatest probability mass, ignoring the ``fluctuations''. There is no
system of higher order corrections or error estimate with this method. In
spite of its limitations, its predictions have been found to coincide with
numerical simulations \cite{lee&cardy,
howard&cardy,bhc,araujo,araujo:let,corn:let,corn:sims}. It is beyond the scope
of this paper to describe in detail the derivation of the analytical results,
so we shall simply state the most important result for our simulation---the
logarithmic correction to the scaling law for $w$:
\[
    w = \left[ \frac{\ln(cL/w)}{\pi(J/D)} \right]^{1/2}
\]
This was derived by the addition of an effective noise term to the
differential equation satisfied by the density difference of buyers and
sellers.  However the numerical value of constant $c$ is unknown from
analytical calculations and we shall find it later from  numerical data.

The second analytic result we shall be using relates to the speed of
the moving midmarket in the case of asymmetric fluxes of buyers and sellers.
The result here is  (for more details see  page 34 in \cite{kogan})
\[
    x_0(t) \sim \frac{-2D\Delta J}{\bar{J} L} t
\]
and was derived using a time-dependent non-stochastic part in the density
difference equation. It is only valid for small times, such that $T < L^2/D$.


\nsection{Computer simulations of the minimal model} \label{cs:min}
\subsection{Description of simulation}
\label{cs:min:sim} 
We shall start from  a Monte-Carlo
simulation of the minimal model.

The simulation starts by randomly inserting traders into either half of price
space to set up a random initial configuration. The main loop is entered.
Traders are inserted at the edges, and then each type of trader (buyer and
seller) is allowed to diffuse separately and annihilate. Each trader hops to
the left or the right at every time step with probability 0.5 ($p = 1$). This
leads to a diffusion coefficient of 1/2, because $D = a^2/2\tau$ (see
\S\ref{app:master}), and $a = \tau = 1$ in our units. The annihilation routine
is called after each diffusion routine to ensure that pairs of mutually
annihilating traders do not cross over each other. This is in accordance with
the procedure outlined in the appendix of \cite{kogan}, where the
time-evolution operator is constructed from diffusion and annihilation
operators, with annihilation occurring after each diffusion of buyers and
sellers (cf.\ \cite{corn:sims} which has a slightly different way of doing it).
The tasks performed within each time quantum $\tau$ may be summarized as:
\begin{enumerate}
  \item Diffuse buyers
  \item Annihilate overlapping traders
  \item Diffuse sellers
  \item Annihilate overlapping traders
  \item Call \vb"getstats()" routine to store market parameters
\end{enumerate}
This method we shall call MINIMAL. At the end of each time step, the
\vb"getstats()" routine is called, which obtains interesting statistics from
the state variable, such as best bid, best offer, bid-offer spread, midmarket,
instantaneous deal rate, etc. Thus the diffusion-annihilation process is
repeated over and over again. Now, not all the data generated at each time
step is recorded, simply because it would be impractical and unwieldy to have
a set of two million results to deal with. Instead, results such as bid-offer
spreads and midmarkets are averaged over a large number \vb"DISP_INT" of time
steps. In the end, we would like to take the average\footnote{Note that the
average taken of the results generated by computer simulation is a \emph{time}
average, whilst the ``average'' used in the scaling laws and the analytical
results (\expect{\ldots}) is an \emph{ensemble} average. These two are
generally not the same. However, for a stationary ensemble (i.e.\ one that has
no preferred origin in time), the ergodic assumption, that the system will in
the course of a sufficiently long time pass through all the states accessible
to it, leads to the fact that the time average is equal to the ensemble
average. The financial market, when in equilibrium, is well-approximated by a
stationary ensemble, and so the equality of the ensemble and time averages
applies. That is why it is acceptable to compare theoretical ensemble averages
with the time averages obtained from numerical simulations, which are the same
as ensemble averages. See \cite{reif} for proof.} and variance of all the
results, not just the results averaged over \vb"DISP_INT" steps. At first, one
might be inclined to think that the process of averaging over some steps
causes one to lose information. In fact, it is possible, by simultaneously
keeping track of the mean of the squares of the quantities, to allow the mean
and variance of the entire ensemble to be recovered, as though we had kept
track of every single result (for proof  see Appendix \S\ref{app:meanvar}).

The data produced from the program were recorded in a data file. Note that in
the simulation, the quantities considered were dimensionless, and may be
defined as follows:
\begin{eqnarray*}
  T = N_1 \tau  \\
  L = N_2 a   \\
  J\tau = N_3
\end{eqnarray*}
where $\tau$ is the time between each step in the simulation (the time
quantum), $T$ is the total time of the simulation, $L$ is the width of price
space, $a$ is the lattice spacing of price space, and $J$ is the rate of
insertion of traders (no.\ of traders per second). $N_1$, $N_2$, and $N_3$ may
be thought of as the dimensionless counterparts of $T$, $L$ and $J$,
respectively. 
We may simply choose
our length and time units such that $a = \tau = 1$,
 so that lengths are measured in units of the
length quantum $a$ and time intervals in units of the time quantum $\tau$. In
these units, $N_1 = T$, $N_2 = L$ and $N_3 = J$, and so we may revert to the
natural units, which have been made dimensionless by a judicious choice of
units.

\subsection{Preliminary results for $L = 200$ using MINIMAL} The simulation
of the minimal model was initially performed for 100,000 time steps, with
averaged results recorded every 100 or 1000 steps. The number of steps in the
simulation was limited by the length of time taken to complete it: for each
run of 100,000 steps and one particular value of $J$, the run-time would be
between 1 and 2 hours. The time taken to obtain an entire set of results was a
few days. There were 200 points in price space ($L = 200$), and the initial
condition consisted of 100 buyers and 100 sellers distributed randomly on
either side of price space, with a gap of 20 empty price points in the middle.
>From this condition, the system was allowed to evolve according to the
diffusion and annihilation laws outlined  earlier. Every
\vb"DISP_INT" time steps, the following averaged quantities were recorded:
best bid ($B$), best offer ($O$), best bid size ($BS$), best offer size
($OS$), midmarket ($M$), bid-offer spread ($Spr$), deal rate ($DR$), and total
number of traders in the market-place ($NUM$).

\begin{figure}
\centering
\resizebox{\figwidth}{!}{\includegraphics{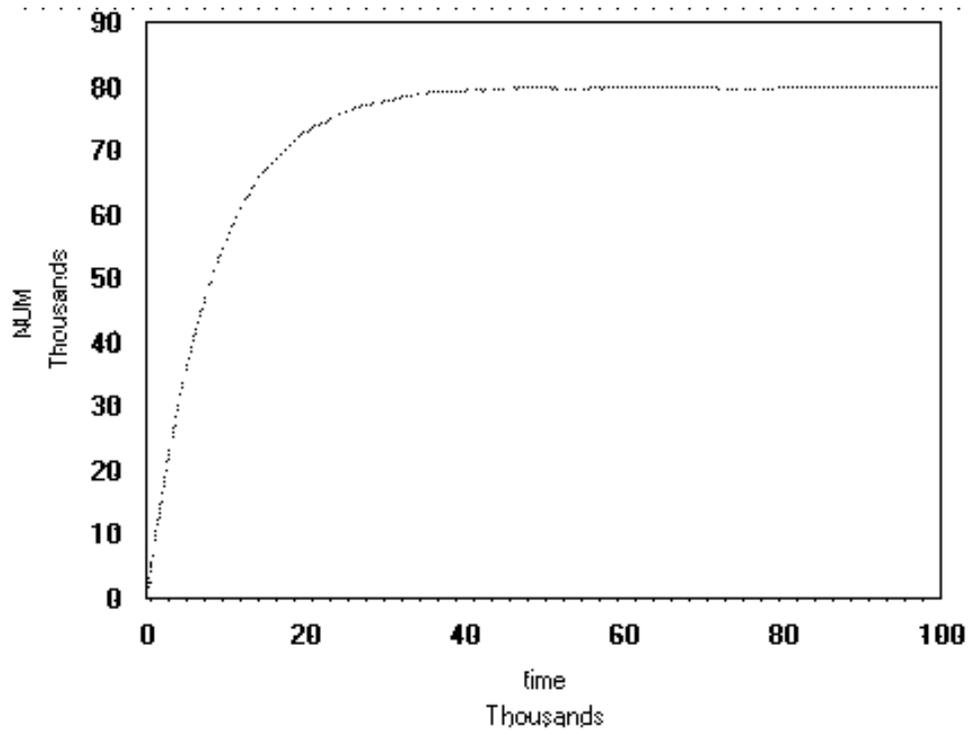}} \caption{Graph of $NUM$
against time for $J = 4$} \label{fig:num1}
\end{figure}

\begin{figure}
\centering
\resizebox{\figwidth}{!}{\includegraphics{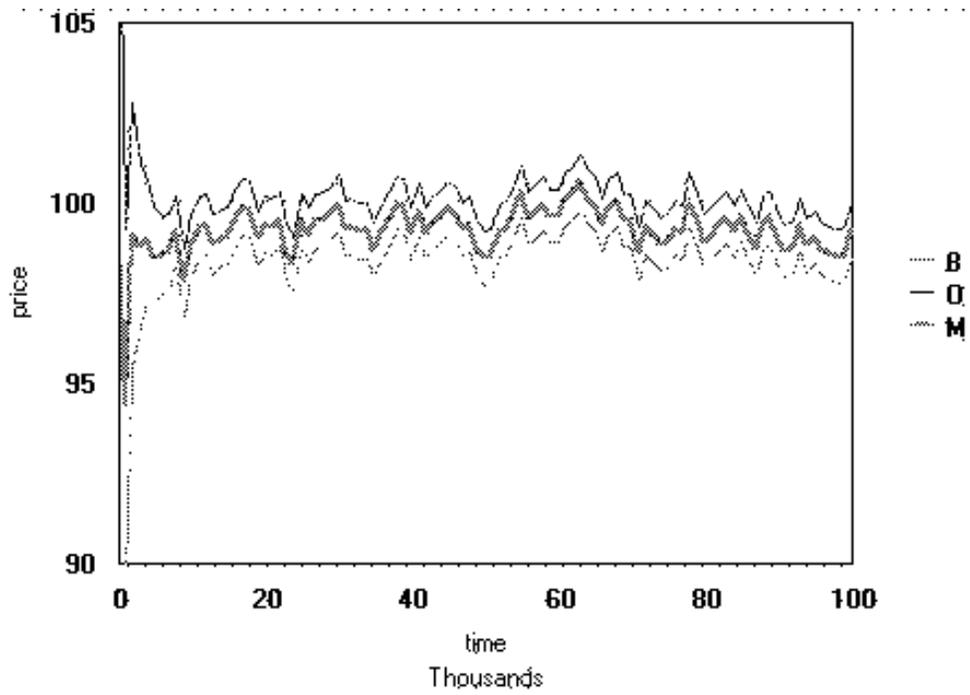}}
\caption{Variation of bid, offer and midmarket with time for $J = 1$}
\label{fig:bom1}
\end{figure}

It was found that the system came to equilibrium after about 50,000 time
steps. The total number of traders in the market ($NUM$) was used as a suitable
measure of the state of equilibrium of the system, because it was discovered
that $NUM$ grew with time and tended asymptotically to an equilibrium value
(Fig.~\ref{fig:num1}).  We 
 have also included a picture of how the bid, offer and
midmarket vary with time in the simulation (Fig.~\ref{fig:bom1}).

Thus, averages of quantities were only taken over the range of time steps for
which the system was in equilibrium, in this case, the last 50,000.

The scaling laws we are interested in  are the following:
\begin{eqnarray}
  w^2 & = & \frac{D}{\pi J} \ln\left(\frac{cL}{w}\right),   \label{eq:w2} \\
  Spr & \sim & \sqrt{D/J},  \label{eq:Spr}  \\
  \var(Spr) & \sim & \sqrt{D/J}. \label{sq:varSpr}
\end{eqnarray}
To this end, we plot graphs of $\ln w^2$ against $\ln J$, and $\ln Spr$ against
$\ln J$. We can see that each run of the simulation produces \emph{one} point
on a graph, since we average over a long time to obtain mean values for our
market parameters. To obtain any reasonable graph requires at least 10 points
or so, hence the long times taken to obtain results.

\begin{figure}
    \centering
    \begin{minipage}[t]{\halffigwidth}
    \resizebox{\halffigwidth}{!}{\includegraphics{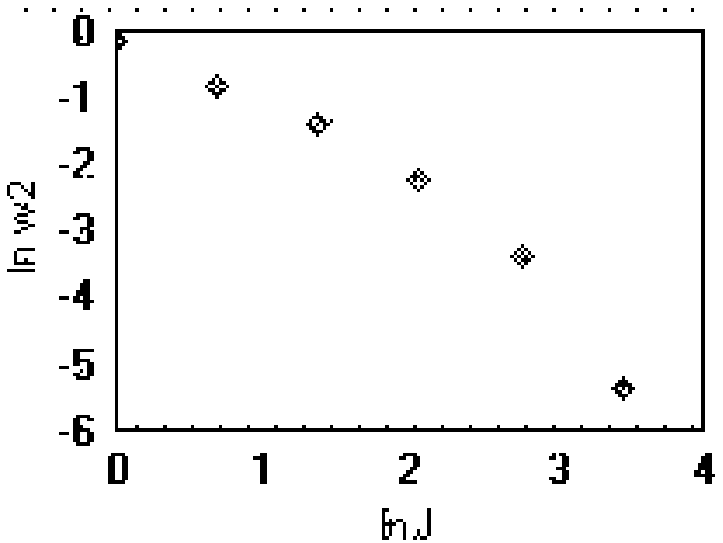}}
    \caption{Graph of $\ln w^2$ against $\ln J$ for $J \geq 1$ and $L = 200$}
    \label{fig:min200-100Kw2}
    \vspace{\interfigspace}
    \end{minipage}
    \hfill
    \begin{minipage}[t]{\halffigwidth}
    \resizebox{\halffigwidth}{!}{\includegraphics{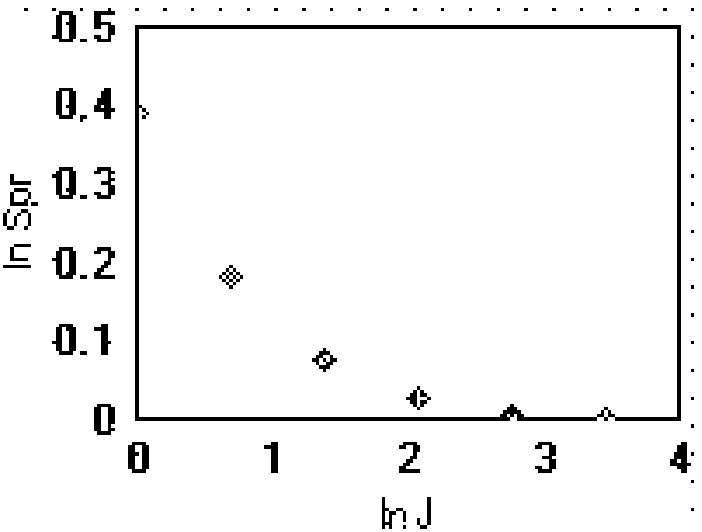}}
    \caption{Graph of $\ln Spr$ against $\ln J$ for $J \geq 1$ and $L = 200$}
    \label{fig:min200-100KSpr}
    \end{minipage}
    \begin{minipage}[t]{\halffigwidth}
    \resizebox{\halffigwidth}{!}{\includegraphics{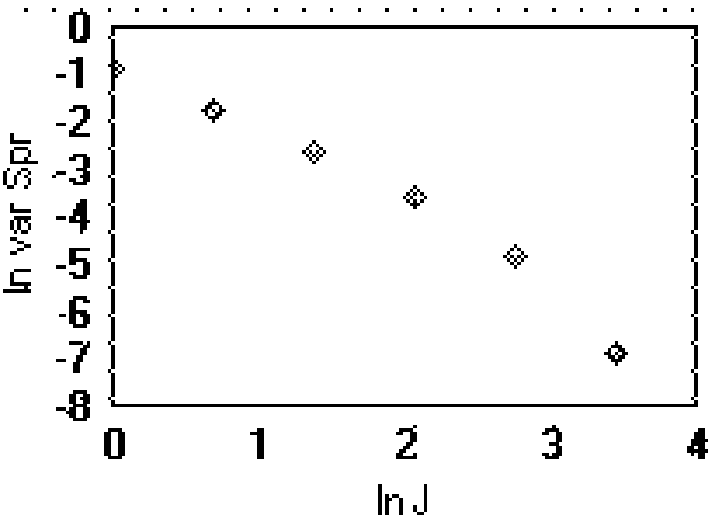}}
    \caption{Graph of $\ln \var(Spr)$ against $\ln J$ for $J \geq 1$ and $L = 200$}
    \label{fig:min200-100KvarSpr}
    \end{minipage}
    \hfill
    \begin{minipage}[t]{\halffigwidth}
    \resizebox{\halffigwidth}{!}{\includegraphics{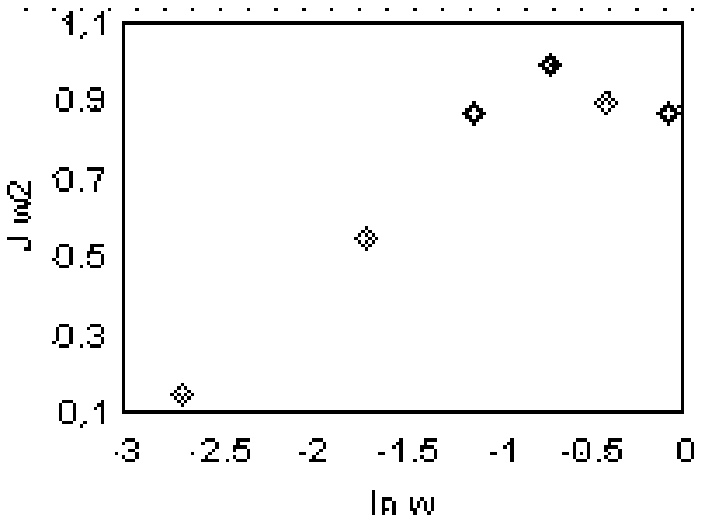}}
    \caption{Graph of $J w^2$ against $\ln w$ for $J = 1$ and $L = 200$}
    \label{fig:min200-100KJw2}
    \end{minipage}
\end{figure}

The results obtained (Figures \ref{fig:min200-100Kw2},
\ref{fig:min200-100KSpr}, \ref{fig:min200-100KvarSpr} and
\ref{fig:min200-100KJw2}) were of a poor quality and did not yield the
expected straight lines. The values of $J$ used were 1, 2, 4, 8, 16, 32 and 64.
 One can see that these values of $J$ are 
too high and that in this regime, the market is  very dense and effectively
frozen. This is not the regime described by the minimal model, and it is not
surprising that the results do not agree with the predictions. From the data,
one can see that $w < a$, $\var(Spr) < a$, and $Spr \sim a$. The reason for
the poor results is the following: the scaling laws, derived using dimensional
analysis, relied on the fact that the length scales of meaningful quantities
were far from $L$ and $a$. Thus, for dimensional analysis to hold, and the
scaling laws obtained thereby to be valid, we require that
\begin{equation}
    a \ll w, Spr, \var(Spr) \ll L.
\end{equation}
This means that $J$ must be much less than 1, since all the length quantities
are of order $\sqrt(D/J)$.

\subsection{Successive improvements to preliminary results}
We outline a sequence of improvements to the simulations as follows:
\begin{center}
\begin{tabular}{cccc}
  Method & $L$ & $T$ & Range of $J$\\
  \hline
  MINIMAL & 200 & $1 \times 10^5$ & $J > 1$ \\
  MINIMAL & 200 & $1 \times 10^5$ & $J < 1$ \\
  MINIMAL & 200 & $1 \times 10^6$ & $J < 1$ \\
  MINIMAL & 1000 & $2 \times 10^6$ & $J < 1$ 
\end{tabular}
\end{center}
The average time taken to obtain each set of results was about 3 or 4
days. The simulations were repeated for fractional $J$, from 0.001 to 0.1, for
$L = 200$. The results obtained (MINIMAL, $L = 200$, $T = 1 \times 10^5$) were
better than before, and the points lay more closely on a straight line. As the
simulation time is proportional to $J$ (see \S \ref{cs:min:num}), it 
becomes possible, with $J$ reduced by several 
orders of magnitude, to increase the number of time steps per run to 1 million
whilst keeping the total duration reasonable, with the statistics being taken
over the last 950,000 steps. Thus, it was decided to do the simulations again,
but with 1 million steps. The results obtained (MINIMAL, $L = 200$, $T = 1
\times 10^6$) were even better, though the graphs were still not very
convincing, especially the one of $J w^2$ against $\ln w$, which should have
been a straight line, according to Eq.~(\ref{eq:w2}), but in fact the points
were so scattered that it was difficult to draw any definite conclusion.

The smallness of $L$ (200) restricted the choice of $J$ too much: it was
desired to keep $w$ at least a factor of 5 away from either $a$ or $L$, so the
range of allowed $w$ was from 5 to 40, which was just one order of magnitude.
Thus, it was resolved to increase $L$ to 1000, which expanded the range of
permitted $w$ to $5 < w < 200$. However, the increase in the size of the
system caused a corresponding increase in the time taken for the system
(especially trader number $NUM$) to reach equilibrium. It was noticed that
quasi-equilibrium was not attained until after about 1 million steps. The
duration of the simulation was therefore increased to 2 million steps, with
only the last 1 million steps used for obtaining statistics.

\subsection{Encouraging results using MINIMAL for $L = 1000$ and $T = 2 \times 10^6$}
Many different values of $J$ were used, all of which produced values of $w$ and
$Spr$ roughly within the range specified above, with most interesting lengths
being at least a factor of 5 away from either $a$ or $L$. Here, the initial
condition consisted of 50 buyers and 50 sellers, separated by a gap of 20. One
of the things learned from these simulations (obtainable by intuition, and
supported by \cite{corn:sims}) is that the asymptotic behaviour of the market
does not depend on the initial conditions. 

\begin{figure}
\centering \resizebox{\figwidth}{!}{\includegraphics{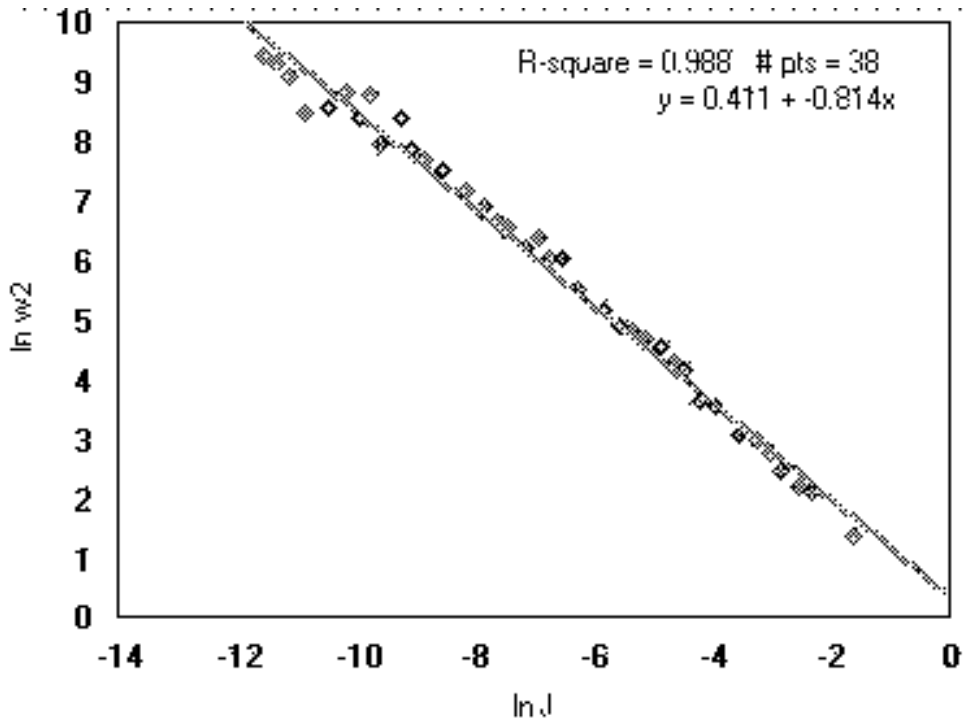}}
\caption{Graph of $\ln w^2$ against $\ln J$ for $L = 1000$}
\label{fig:min1000-2Mw2}
\end{figure}

\begin{figure}
\centering \resizebox{\figwidth}{!}{\includegraphics{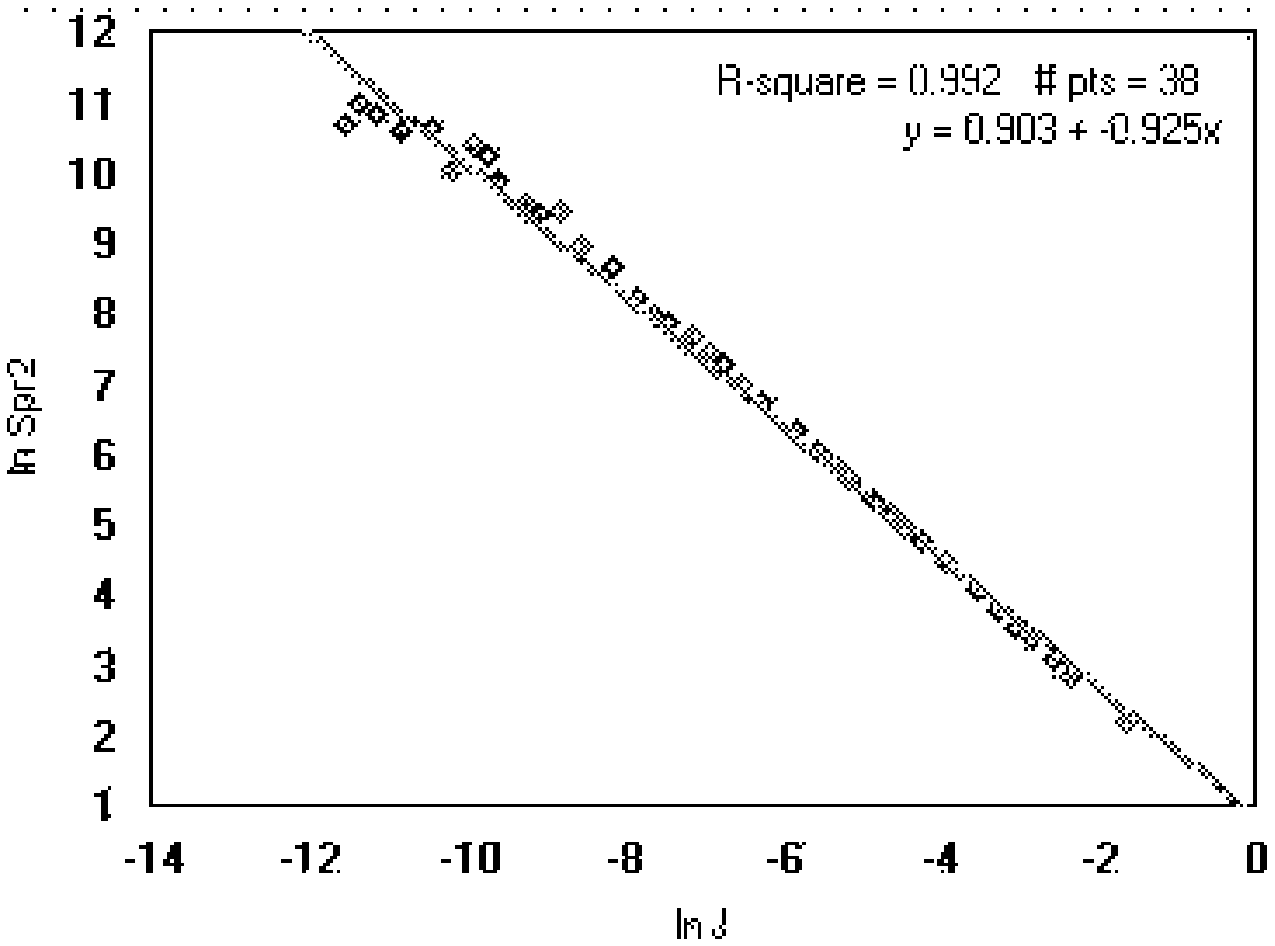}}
\caption{Graph of $\ln Spr^2$ against $\ln J$ for $L = 1000$}
\label{fig:min1000-2MSpr}
\end{figure}

\begin{figure}
\centering \resizebox{\figwidth}{!}{\includegraphics{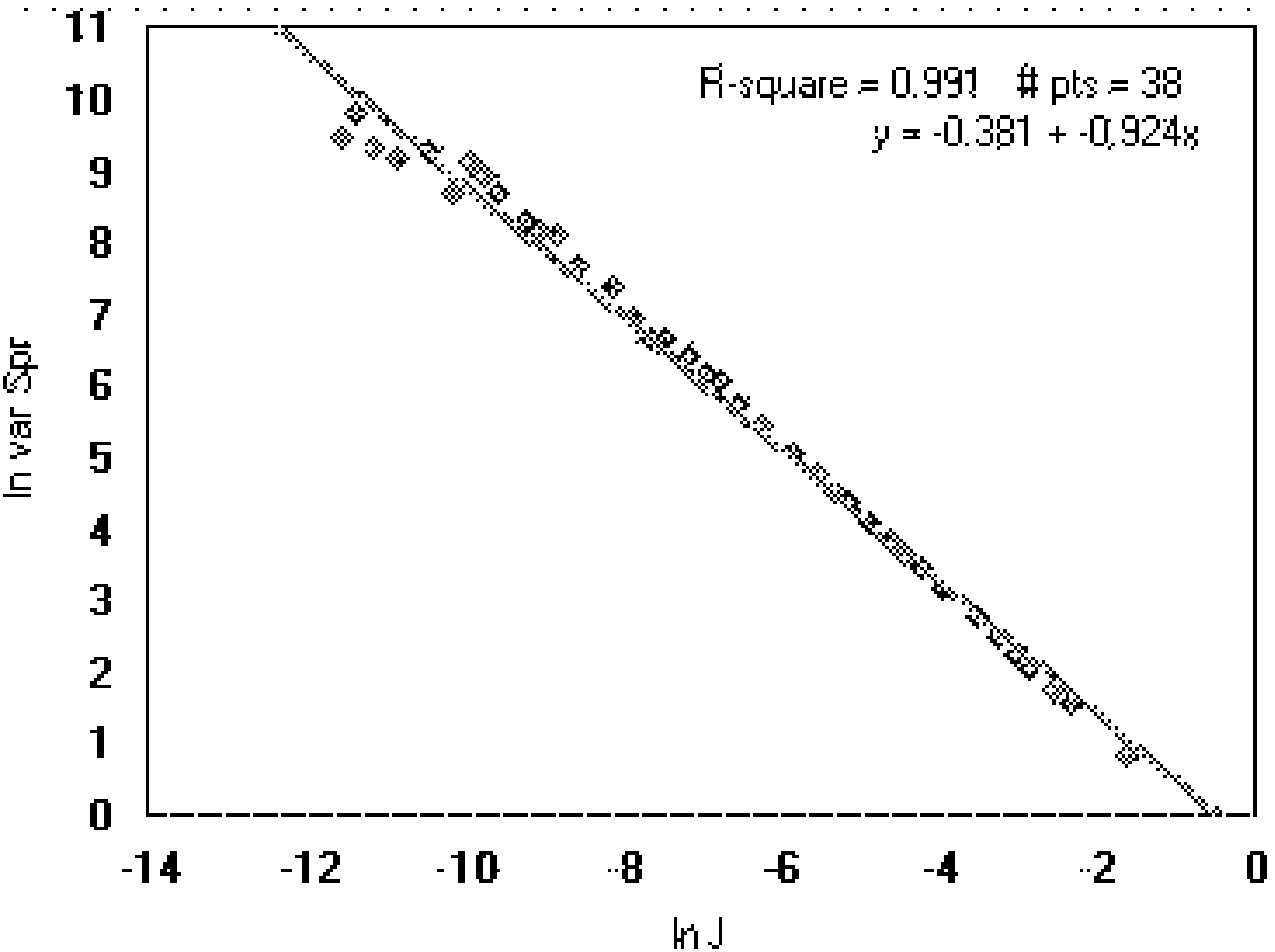}}
\caption{Graph of $\ln \var(Spr)$ against $\ln J$ for $L = 1000$}
\label{fig:min1000-2MvarSpr}
\end{figure}

\begin{figure}
\centering \resizebox{\figwidth}{!}{\includegraphics{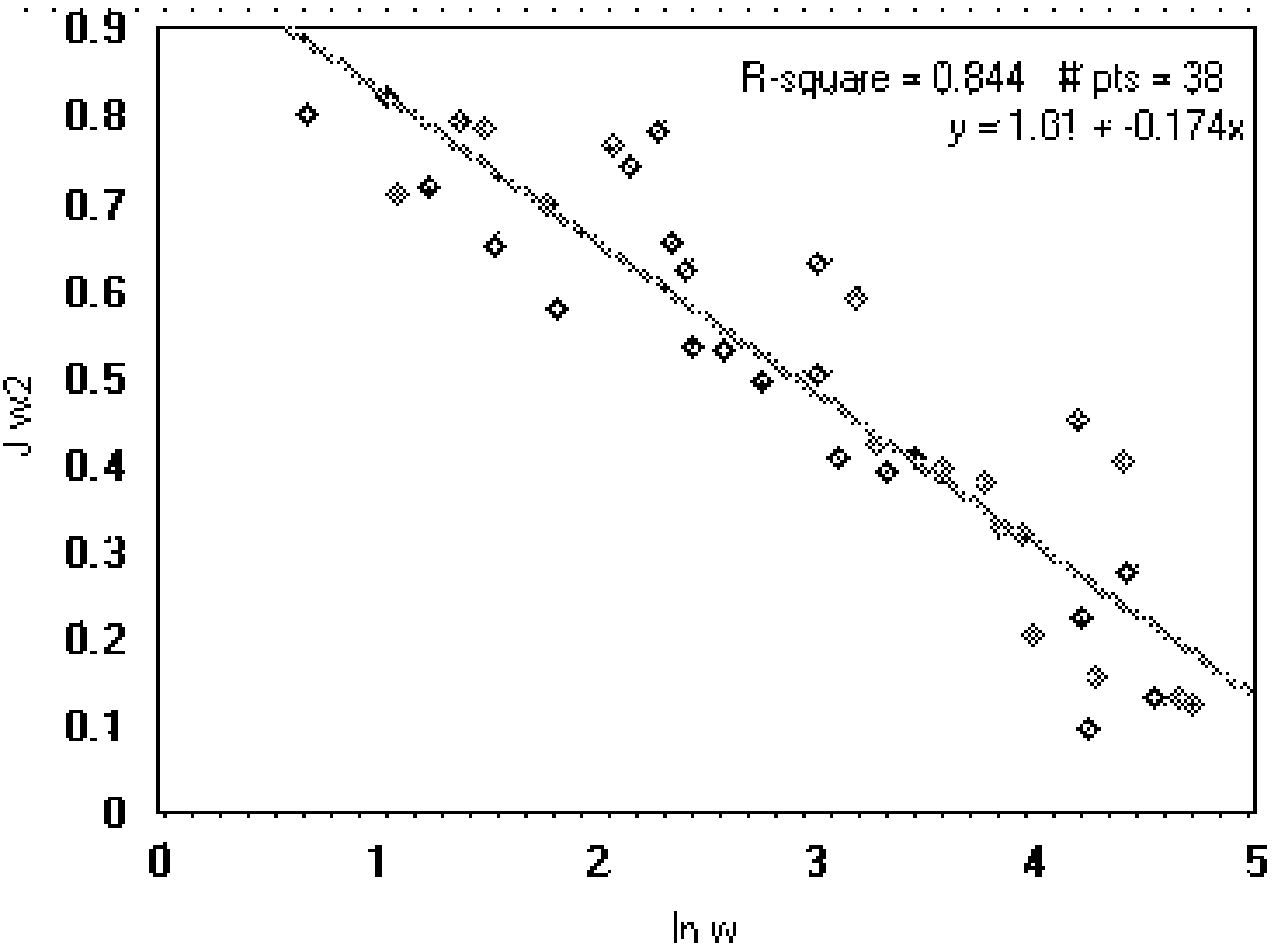}}
\caption{Graph of $J w^2$ against $\ln w$ for $L = 1000$}
\label{fig:min1000-2MJw2}
\end{figure}

38 different values of $J$ were used, producing a set of 38 points on the
graphs (Figs.~\ref{fig:min1000-2Mw2}, \ref{fig:min1000-2MSpr},
\ref{fig:min1000-2MvarSpr} and \ref{fig:min1000-2MJw2}).
It was noticed that for the first three graphs at least, the points fit a line
better towards the bottom right of the graph, while the points towards the top
left tended to be more scattered. It was likely that the points towards the top
left (small $J$ and large $w^2$) were less accurate because of the possibility
of the reaction front crashing into either edge of price space. In addition,
for graphs of $\ln w^2$, $\ln Spr$ and $\ln \var(Spr)$ against $\ln J$, the
analysis in \S\ref{cs:min:w2} requires that $\ln w \ll \ln cL$ be satisfied,
and this is not satisfied for the points towards the top left, for which $\ln
w \sim 5$. Therefore, we must choose a small section of the bottom right of
the graph and obtain the gradient and intercept from that.

\begin{figure}
  \centering
  \resizebox{\figwidth}{!}{\includegraphics{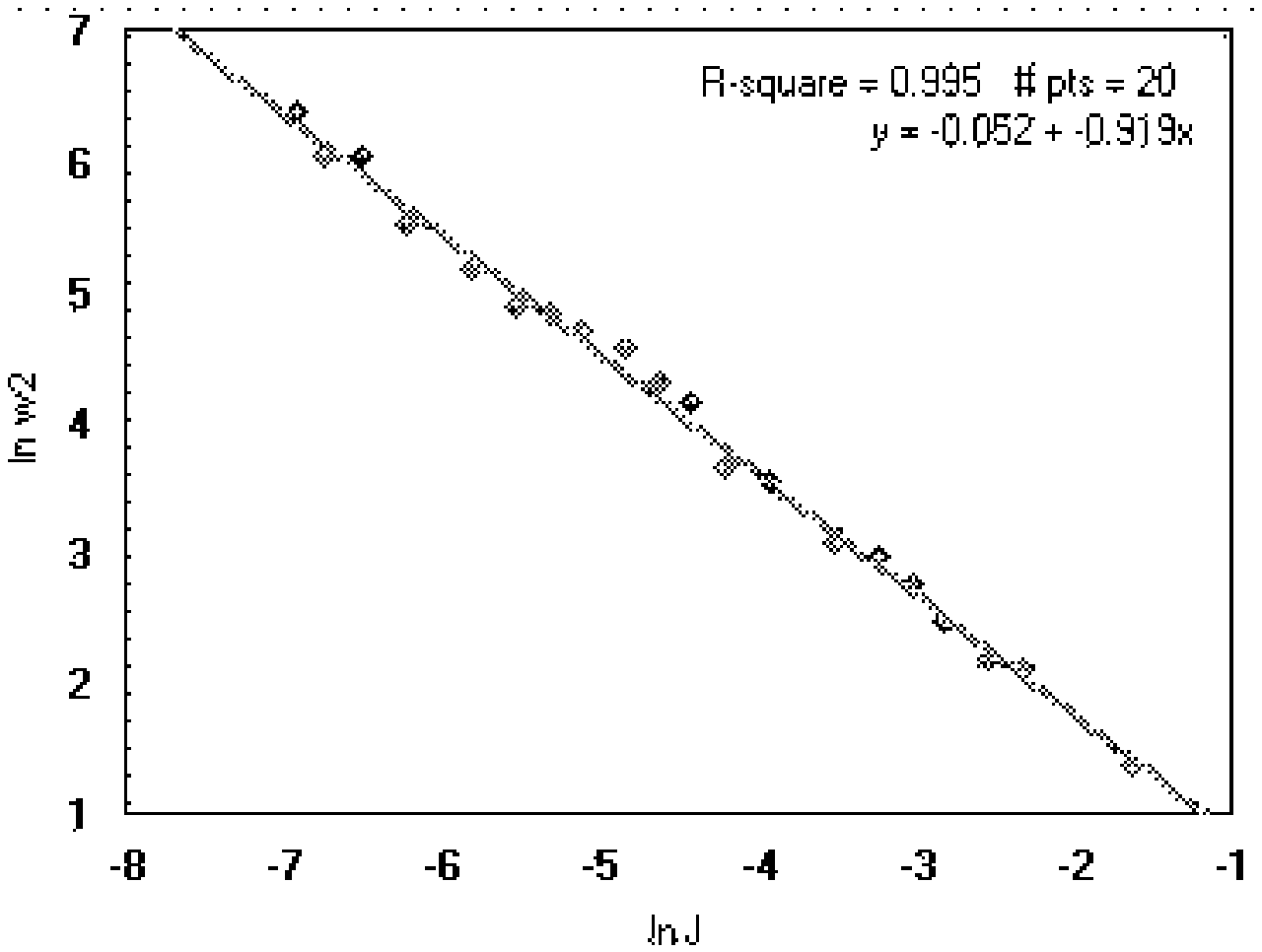}}
  \caption{Graph of $\ln w^2$ against $\ln J$ for $L = 1000$ (last 20 points)}
  \label{fig:min1000-2Mw2a}
\end{figure}

\begin{figure}
  \centering
  \resizebox{\figwidth}{!}{\includegraphics{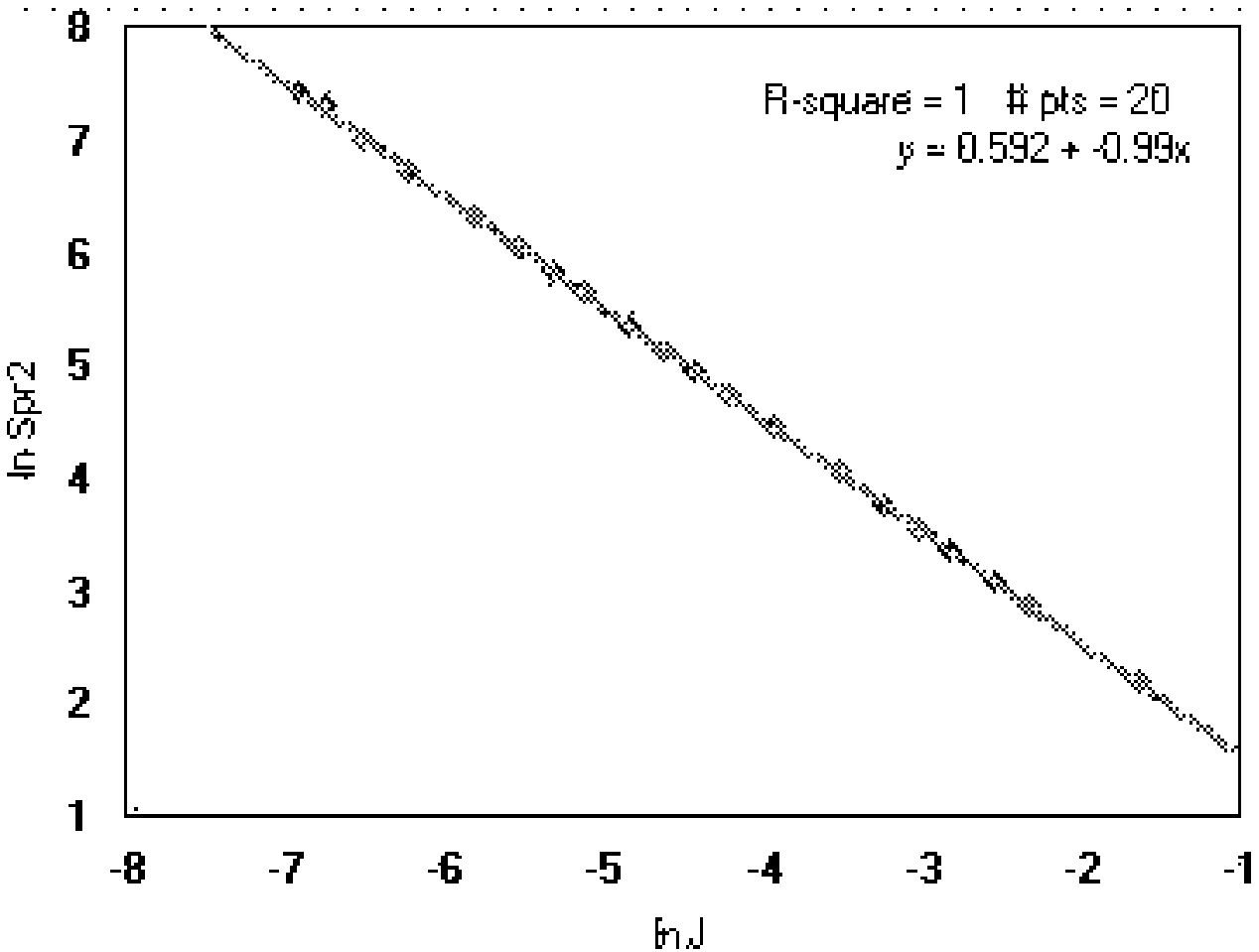}}
  \caption{Graph of $\ln Spr$ against $\ln J$ for $L = 1000$ (last 20 points)}
  \label{fig:min1000-2MSpr2}
\end{figure}

\begin{figure}
  \centering
  \resizebox{\figwidth}{!}{\includegraphics{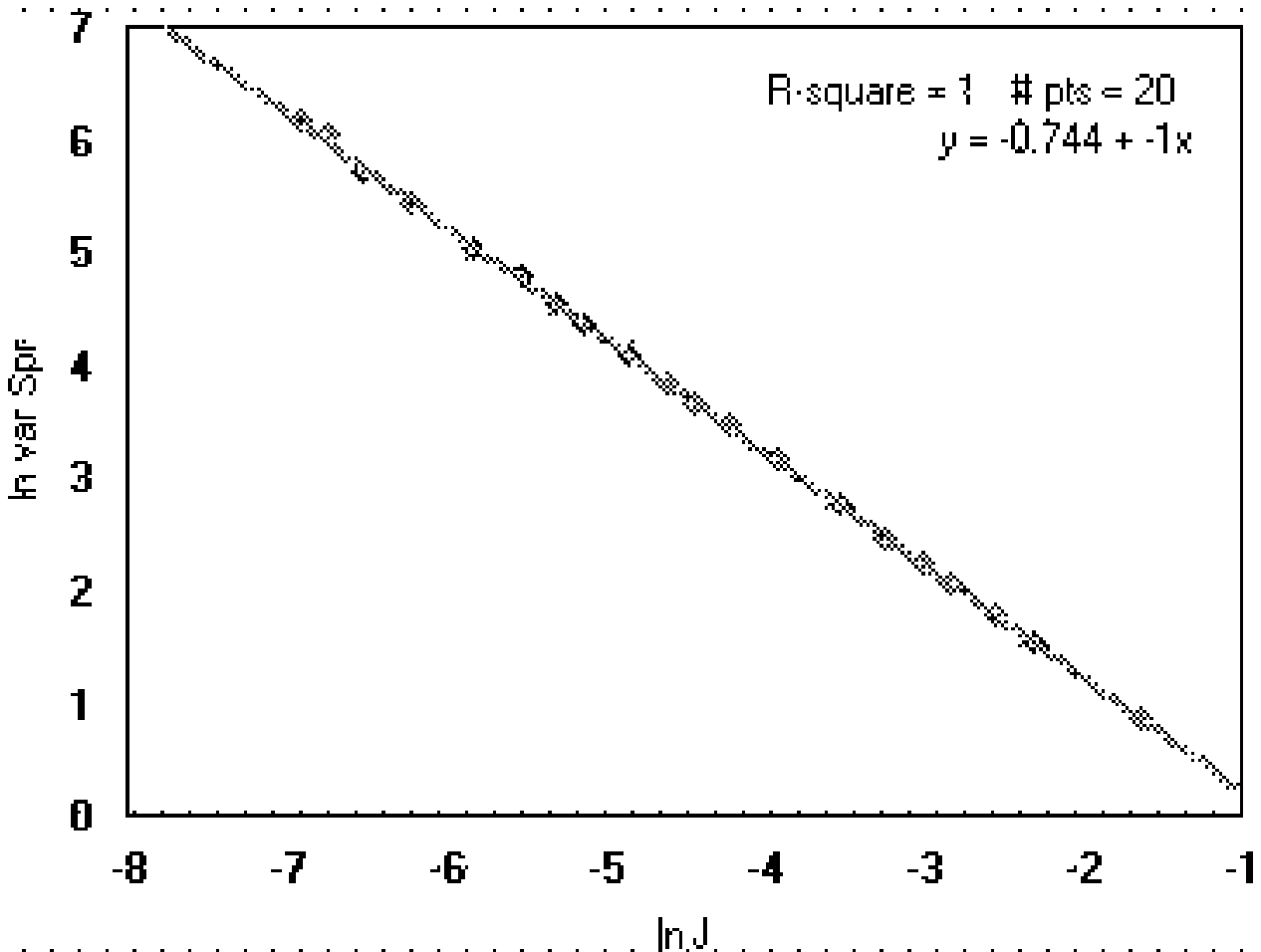}}
  \caption{Graph of $\ln \var(Spr)$ against $\ln J$ for $L = 1000$ (last 20 points)}
  \label{fig:min1000-2MvarSpr2}
\end{figure}

\begin{figure}
  \centering
  \resizebox{\figwidth}{!}{\includegraphics{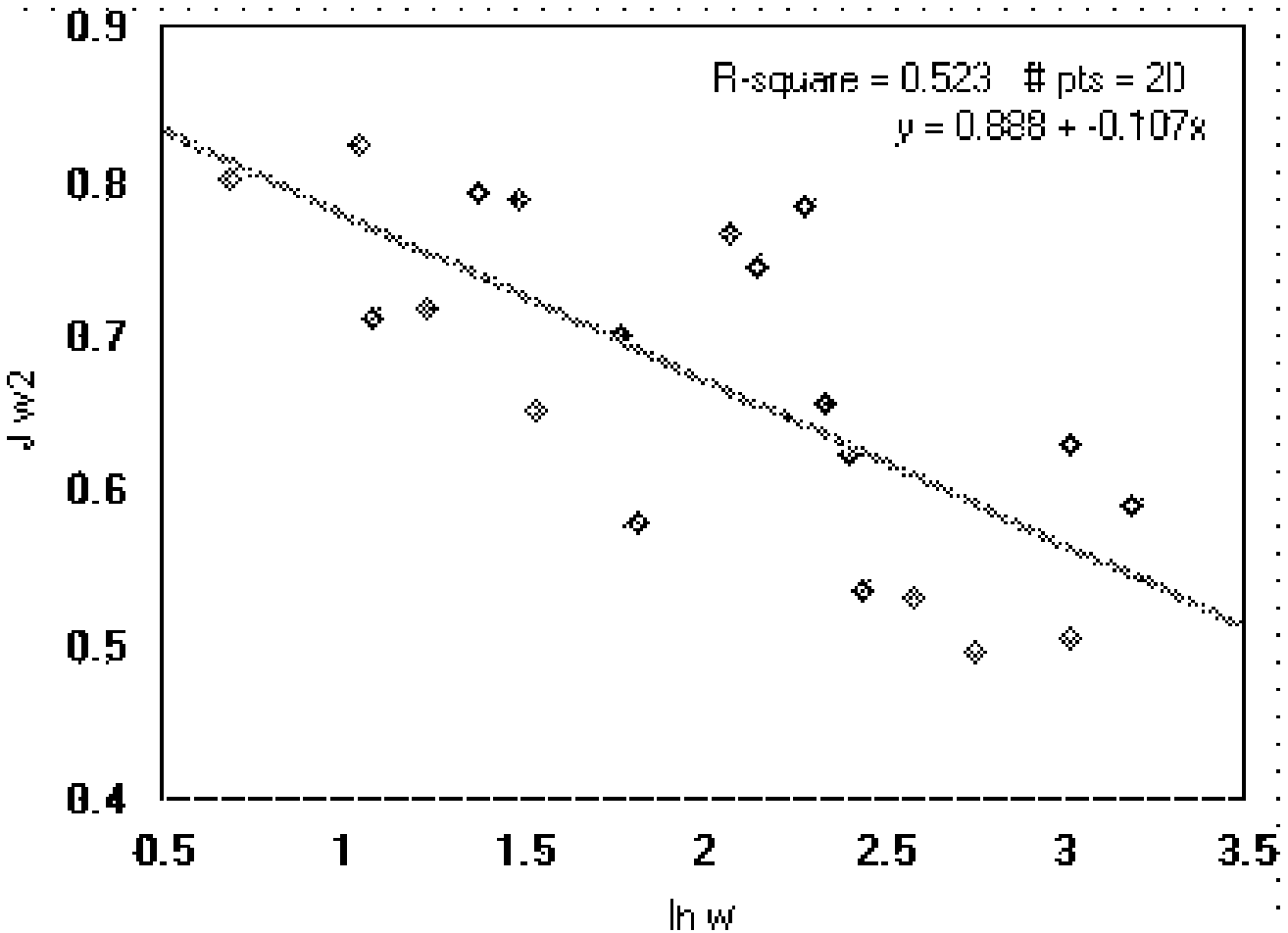}}
  \caption{Graph of $J w^2$ against $\ln w$ for $L = 1000$ (last 20 points)}
  \label{fig:min1000-2MJw2a}
\end{figure}

Thus, it was decided to take only the last 20 points and fit a line through
them, in the hope of acquiring more accurate statistics
(Figs.~\ref{fig:min1000-2Mw2a}, \ref{fig:min1000-2MSpr2},
\ref{fig:min1000-2MvarSpr2} and \ref{fig:min1000-2MJw2a}).
Here is a summary of the results, for the first set of data (all 38 points):
\begin{center}
\begin{tabular}{cccccc}
Figure & Abscissa & Ordinate & Gradient & Intercept & $R^2$ \\
\hline
\ref{fig:min1000-2Mw2} & $\ln J$ & $\ln w^2$ & $-0.814 \pm 0.015$ & $0.41 \pm 0.26$ & 0.988\\
\ref{fig:min1000-2MSpr} & $\ln J$ & $\ln Spr^2$ & $-0.925 \pm 0.014$ & $0.903 \pm 0.24$ & 0.992\\
\ref{fig:min1000-2MvarSpr} & $\ln J$ & $\ln \var(Spr)$ & $-0.924 \pm 0.014$ & $-0.38 \pm 0.25$ & 0.991\\
\ref{fig:min1000-2MJw2} & $\ln w$ & $J w^2$ & $-0.174 \pm 0.013$ & $1.006 \pm
0.089$ & 0.844
\end{tabular}
\end{center}
and here is the same for the second set of data (last 20 points):
\begin{center}
\begin{tabular}{cccccc}
Figure & Abscissa & Ordinate & Gradient & Intercept & $R^2$\\
\hline
\ref{fig:min1000-2Mw2a} & $\ln J$ & $\ln w^2$ & $-0.919 \pm 0.016$ & $0.052 \pm 0.1$ & 0.995\\
\ref{fig:min1000-2MSpr2} & $\ln J$ & $\ln Spr^2$ & $-0.9904 \pm 0.0024$ & $0.592 \pm 0.017$ & 0.99989\\
\ref{fig:min1000-2MvarSpr2} & $\ln J$ & $\ln \var(Spr)$ & $-1.001 \pm 0.003$ & $-0.744 \pm 0.023$ & 0.99980\\
\ref{fig:min1000-2MJw2a} & $\ln w$ & $J w^2$ & $-0.107 \pm 0.024$ & $0.888 \pm
0.076$ & 0.523
\end{tabular}
\end{center}
It must be noted that for all the graphs, with the exception of $Jw^2$ against
$\ln w$, the correlation coefficients are all 0.98 or above, which means that
the results strongly support a scaling relationship of some sort, the exact
details of which we shall now discuss.

\subsection{The $w^2$ scaling law}
\label{cs:min:w2} Now that we have some reasonably detailed results, we may
begin our analysis of them in earnest. The fact that the plot of $\ln w^2$
against $\ln J$ is a straight line with a gradient close to --1 suggests that
$w^2$ is inversely proportional to $J$, which is correct according to
Eq.~(\ref{eq:w2}) if we ignore the logarithmic correction. This is simply a
consequence of the dimensional analysis in \S\ref{fin:dimanal}.

Next, we seek to observe the logarithmic correction. Consider the plot of $\ln
w^2$ against $\ln J$. From Eq.~(\ref{eq:w2}), and from $D = a^2/2\tau = 1/2$,
we can write
\begin{eqnarray*}
    \ln w^2 & = & -\ln J + \ln \left[ \frac{1}{2\pi} \ln \left( \frac{cL}{w}
    \right) \right] \\
    & = & -\ln J + \ln \left\{ \frac{1}{2 \pi} \left[ \ln (cL) - \ln w \right] \right\} \\
    & = & -\ln J + \ln \left\{ \frac{1}{2\pi} \ln(cL) \left[ 1 - \frac{\ln w}{\ln cL} \right] \right\} \\
    & = & -\ln J + \ln \left[ \frac{1}{2\pi} \ln(cL) \right] + \ln \left[ 1 -
    \frac{\ln w}{\ln cL} \right].
\end{eqnarray*}
Here, we make the approximation $\ln(1-x) \approx -x$, ignoring the higher
terms $-x^2/2 - x^3/3 - \ldots$, which is valid for $x \ll 1$, on the third
term on the right hand side. In this case, the approximation is valid if $\ln w
\ll \ln cL$. Letting $\ln w = 1/2 \, \ln w^2$, and grouping terms of $\ln w^2$
on the left hand side, we obtain,
\begin{eqnarray*}
    \left( 1 + \frac{1}{2\ln cL} \right)\ln w^2 & = & -\ln J + \ln \left(
    \frac{\ln cL}{2\pi} \right).
\end{eqnarray*}
For $2 \ln cL \gg 1$, we can make the further approximation of $\left( 1 +
\frac{1}{2\ln cL} \right)^{-1} \approx 1 - \frac{1}{2\ln cL}$, yielding
\begin{eqnarray}
    \ln w^2 & = & -\left( 1 - \frac{1}{2\ln cL} \right) \ln J + \left(1 -
    \frac{1}{2\ln cL} \right) \ln \left( \frac{\ln cL}{2\pi} \right) \\
    & = & -\alpha \ln J + \alpha \ln \left( \frac{\ln cL}{2\pi} \right),
    \label{eq:alpha}
\end{eqnarray}
which defines $\alpha$. If we assume that $c \sim O(1)$ and take $L = 1000$,
then the logarithm in $\alpha$ is positive, making $\alpha$ slightly smaller
than unity. From measuring $\alpha$, we may therefore deduce the value of $c$.
In this way, it is possible to see the subtle effect of the logarithmic
correction. Note that the second approximation is not strictly necessary and
one can use the full expression for $\alpha$ if one wishes, and this is indeed
what we shall do.

We make use of the graph of the last 20 points because we want the
approximation $\ln w \ll \ln cL$ to hold, as otherwise the foregoing analysis
would be invalid. From Fig.~\ref{fig:min1000-2Mw2a}, and using the full
expression for $\alpha$ (i.e.\ without making the binomial expansion for the
reciprocal) we find that the gradient of the graph is $-0.919 \pm 0.016$,
making $\alpha = 0.919$ and $c = 0.29$, but because $c$ is inside a logarithm,
the uncertainty in $\alpha$ is magnified in $c$, so that it lies in range
$0.11 < c < 1.33$. 

We can also obtain a value for $c$ from the intercept, which is $-0.052 \pm
0.1$, another wildly imprecise quantity. This method of calculating $c$ gives
0.38, with limits $0.20 < c < 0.75$. The intercept appears, therefore, to
provide a more precise determination of $c$, though, as we shall see later,
that is not always the case.

There is another technique one can try in extracting information from the
graphs. Instead of allowing only $c$ to be determined from the graphs, we may
attempt to determine the constant $1/\pi$ from the gradient and the intercept,
since each graph has two degrees of freedom. Therefore, we replace $1/\pi$
with a variable, say, $\Omega$, and the equation becomes
\[
    \ln w^2 = -\alpha \ln J + \alpha \ln \left( \frac{\Omega}{2} \ln cL
    \right),
\]
and we may determine $\Omega$ using the data. Here, we have $\Omega \simeq
0.333$, with the uncertainty $0.236 < \Omega < 0.453$. Note that, the value
predicted from theory is $1/\pi = 0.3183$, which lies well within the
experimental error. This is quite encouraging as it agrees with the result
derived theoretically (in spite of its 33\% uncertainty!)

It is possible, for the sake of comparison with \S\ref{cs:min:spr} and
\S\ref{cs:min:varspr}, to analyze the data for $w^2$ as though its scaling law
had no logarithmic correction. It is not really appropriate, because of the
relatively pronounced (--0.919) deviation from --1, but we shall quickly do it
for subsequent comparison. Assuming that $w^2$ is proportional to $D/J$ with
constant of coefficient $\lambda$,
\[
    w^2 = \frac{\lambda D}{J},
\]
we can take the log of both sides
\[
    \ln w^2 = \ln{\lambda D} - \ln J
\]
and calculate $\lambda$ from the intercept, knowing that $D = 1/2$. The
intercept is $-0.052 \pm 0.1$, giving us $\lambda = 1.90$, within the limits
$1.72 < \lambda < 2.10$, or $1.9 \pm 0.2$.

We have another graph from which we may obtain information: the graph of
$Jw^2$ against $\ln w$. For this graph, it is prudent to take the one for all
38 points, since no approximations have been made in the scaling law we are
trying to verify, and the more points we have on the graph, the more confident
we may be of its statistics. It is simply
\begin{equation}
\label{eq:Jw2}
    Jw^2 = \frac{\Omega}{2}(\ln cL - \ln w)
\end{equation}
which comes from a trivial re-arrangement of Eq.~(\ref{eq:w2}). We have once
again allowed $1/\pi$ be determined from the graph. The gradient is $-0.174 \pm
0.013$, from which we may infer $\Omega \simeq 0.348 \pm 0.026$. This is close
to the predicted $1/\pi = 0.318$, although it actually lies outside the range
of experimental error (the discrepancy is 9\%). We can, with reasonable
confidence, now set $\Omega$ to its theoretical value, $1/\pi$. From the
intercept, $1.006 \pm 0.089$, we may infer $c = 0.556$ with the limits being
$0.32 < c < 0.97$. These limits are slightly different from those found from
the first graph (Fig.~\ref{fig:min1000-2Mw2a}), but not by much. Both methods
give the same order of magnitude estimate $c \sim 0.5$.

\subsection{The $Spr$ scaling law}
\label{cs:min:spr} We shall make use of Figs.~\ref{fig:min1000-2MSpr} and
\ref{fig:min1000-2MSpr2}, the latter consisting of the last 20 points of the
former. It is interesting to notice that the correlation coefficients for both
graphs are very high, and indeed, that of the second graph is so close to 1
that it was rounded up to 1 by the spreadsheet that produced the graph. Once
again, the straight lines with gradients so close to --1 confirm the
approximate $Spr \sim \sqrt(D/J)$ law. It is not known (see \cite{kogan})
whether there is a logarithmic correction to $Spr$ or not. If there is, one
might expect it to be of the form
\[
    Spr^2 = \frac{\sigma D}{J} \ln \left( \frac{sL}{Spr} \right),
\]
in analogy with Eq.~(\ref{eq:w2}). We have left the constant in front of $D/J$
as a parameter ($\sigma$) to be determined.

To look for the logarithmic correction, we can perform a similar analysis as
we did in \S\ref{cs:min:w2}, making the approximation $\ln Spr \ll \ln cL$, to
obtain
\[
    \ln Spr^2 = -\alpha \ln J + \alpha \ln \left( \frac{\sigma}{2} \ln sL \right),
\]
where
\[
    \frac{1}{\alpha} = 1 + \frac{1}{2\ln sL}.
\]
To ensure the validity of the approximation, we shall use the graph of the
last 20 points only (Fig.~\ref{fig:min1000-2MSpr2}). There is another reason
for putting more faith in the statistics yielded by the second graph: if one
looks carefully along the points of the first graph, one sees that the points
towards the lower right of the graph lie along a line that does not quite
coincide with the line of best fit constructed from all of the points. This is
because of some deviation from linearity towards the top left of the graph. It
would be wise, therefore, to take only the results towards the bottom right,
as in Fig.~\ref{fig:min1000-2MSpr2}. The gradient is $-0.9904 \pm 0.0024$,
which would imply $s \simeq 2.5 \times 10^{19}$, $7.6 \times 10^{14} < s < 8.8
\times 10^{26}$. The uncertainty covers 12 orders of magnitude! The intercept
is $0.592 \pm 0.017$, giving us $\sigma \simeq 0.07$, with limits $0.054 <
\sigma < 0.087$. We shall try to understand the meaning of the values
obtained. Consider a rearrangement of the scaling law
\[
    Spr^2 = \frac{\sigma D}{J} \left[ \ln s + \ln \left( \frac{L}{Spr} \right) \right],
\]
The largest possible value for $\ln (L/Spr)$ occurs when $Spr$ is at its
smallest, around 5. Thus, the maximum value for the second logarithm is 5.3
(minimum is about 1.6). The first logarithm, however, is $\ln s$, which is 44
($34 < \ln s < 62$). Therefore, the right hand side is dominated completely by
the large constant $\ln s$, and the logarithm involving $L$ has almost no
effect on the scaling law. This is a clear sign that there is no logarithmic
correction, since the variation of the second logarithm has negligible effect
on the variation of the \emph{sum} of the two logarithms, on which the scaling
law depends. This can be approximated by ignoring the variation of the second
logarithm with $Spr$, and taking 3 as the approximate average value for it,
leaving $\ln s \simeq 47$, with limits $37 < \ln s < 65$.
\[
    Spr^2 = \frac{\sigma D}{J} \ln s,
\]
which is,
\[
    Spr^2 = \frac{\mu D}{J},
\]
where the coefficient $\mu = \sigma \ln s \simeq 3.3$.

Now, if we assume that there is no logarithmic correction, we can try to
extract from the data the constant of proportionality for the scaling law
\[
    Spr^2 = \frac{\mu D}{J}.
\]
Taking logs, we have
\[
    \ln Spr^2 = \ln(\mu D) - \ln J.
\]
The gradient agrees well with the predicted value of --1. We know that the
diffusion coefficient $D$ is 1/2, so from the intercept, $\mu \simeq 3.615$,
with limits $3.55 < \mu < 3.68$, or $3.615 \pm 0.065$. Thus, the two methods,
one assuming a logarithmic correction, and the other not, produce results which
agree with one another (the 3.3 obtained earlier has an associated uncertainty
which is rather difficult to work out, so we will not do that here).

\subsection{The $\var(Spr)$ scaling law}
\label{cs:min:varspr} We look at Figs.~\ref{fig:min1000-2MvarSpr} and
\ref{fig:min1000-2MvarSpr2}. Both have extremely high correlations (almost
perfect correlation for Fig.~\ref{fig:min1000-2MvarSpr2}), so we can have
confidence in the results. The results are unambiguously linear, confirming
the scaling relationship $\var(Spr) \sim \sqrt(D/J)$. The only thing remaining
is the determination of the existence of logarithmic correction.

The gradient of the graph is $-1.001 \pm 0.003$, which puts the exact value of
--1 within the range of experimental error. Even if there is some deviation
from --1, it is minute. This is strong evidence \emph{against} the existence of
logarithmic correction for $\var(Spr)$. There is no point in trying to apply
the same analysis as before to find out the constant inside the logarithm,
because here, the gradient can lie on either side of --1, causing the value of
that constant to be anywhere between the exponential of a large negative
number and the exponential of a large positive number, i.e.\ many orders of
magnitude. The fact that the constant can assume these extreme values implies
that the logarithm of the constant would dominate the scaling law, when it is
either very small, in which case the logarithm of the constant would be a
large negative number, or very large, in which case the log would be a large
positive number. It is fair to conclude that $Spr$ appears from the data
\emph{not} to have any logarithmic correction.

We may proceed to find the constant of proportionality, $\nu$, defined by
\[
    \var(Spr) = \frac{\nu D}{J}.
\]
Taking logs,
\[
    \ln \var(Spr) = \ln(\nu D) - \ln J.
\]
The gradient has already been verified to be --1. The intercept is $-0.744 \pm
0.023$. Therefore, $\nu \simeq 0.95$, subject to $0.93 < \nu < 0.97$, or $0.95
\pm 0.02$.

\subsection{Calculation of $c$ for different values of L}
>From Eq.~(\ref{eq:alpha}), we found that plotting graphs of $\ln w^2$ against
$\ln J$ allowed the logarithmic parameter $c$ to be determined. After further
discussion, it was decided to investigate this parameter and try to see
whether and how it varies with $L$. To this end, simulations were run for
different values of $L$, within the correct regime for $J$, such that $a \ll J
\ll L$. The observance of this regime was now especially important, owing to
the $\ln(1-x) \approx -x$ approximation we made in our derivation of
Eq.~(\ref{eq:alpha}).

The results obtained were rather poor. Correlations were low and the value of
$c$ did not appear to show any kind of relationship to the value of $L$. We
shall not present these unhelpful results here. Instead, we shall describe a
new method of simulation, which will be used to obtain better results.

\subsection{MINIMAL1: a new method}
If we take a detailed look at the variation of something like $Spr$ with time,
it will be seen that the MINIMAL method of simulation, which we have been
using from the outset, does \emph{not} allow $Spr$ to go to zero. This is
because of the sequence of steps undertaken at each time quantum $\tau$. Only
after both species of traders have diffused and annihilated do we take
measurements of the system using \vb"getstats()". In practice, annihilations
in the marketplace are not instantaneous, and for some of the time at least,
overlapping buyers and sellers \emph{can} exist. This suggests an alternative
method of simulation, MINIMAL1, based on the following 8 steps:
\begin{enumerate}
  \item Diffuse buyers
  \item Call \vb"getstats()" routine---overlapping traders possible
  \item Annihilate overlapping traders
  \item Call \vb"getstats()" routine---no overlapping traders
  \item Diffuse sellers
  \item Call \vb"getstats()" routine---overlapping traders possible
  \item Annihilate overlapping traders
  \item Call \vb"getstats()" routine---no overlapping traders
\end{enumerate}
This method represents a more detailed observation of the market as it includes
intermediate overlapped states which were ignored in MINIMAL. MINIMAL1 allows
the bid-offer spread to vanish just before an annihilation, and can be said to
produce a more accurate profile of the evolution of the market. However, equal
weight is given to the statistics obtained at each of the even steps shown
above. Whether this is a good picture of real-life trading is another matter
altogether, which we shall not discuss here. 

\subsection{MINIMAL1 in action}
\begin{figure}
\centering \resizebox{\figwidth}{!}{\includegraphics{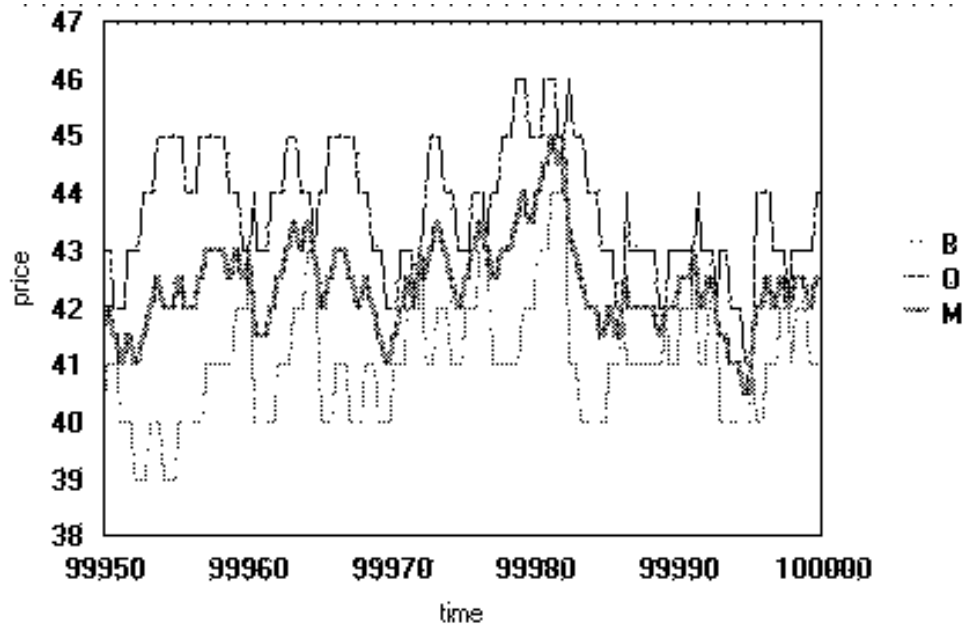}}
\caption{Graph of bid, offer and midmarket with time, for $L
= 100$} \label{fig:min1picbom}
\end{figure}

\begin{figure}
\centering \resizebox{\figwidth}{!}{\includegraphics{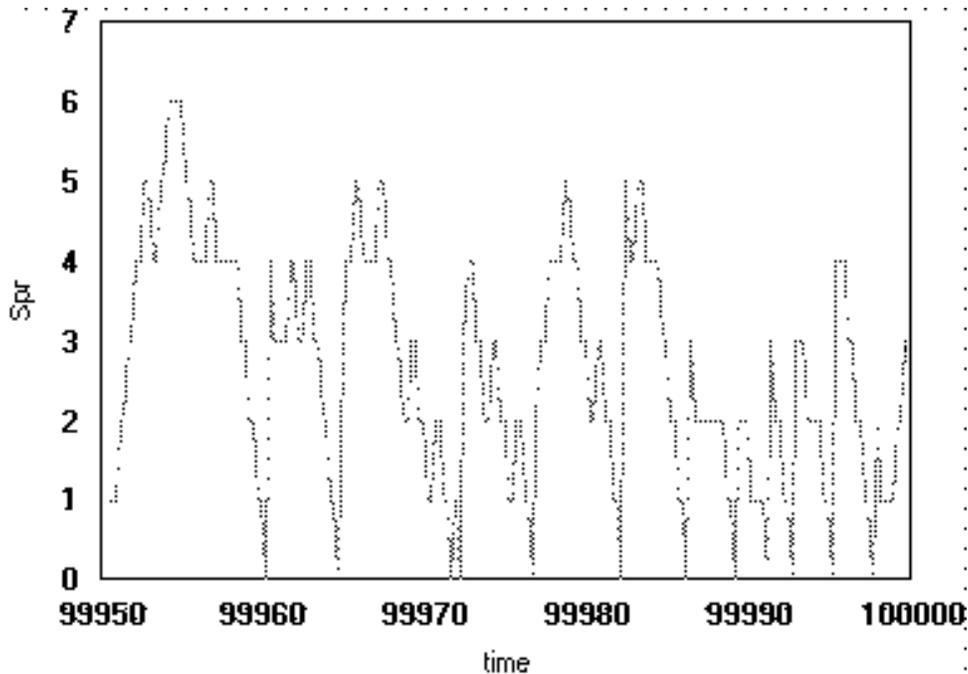}}
\caption{Graph of $Spr$ with time, for $L = 100$}
\label{fig:min1picspr}
\end{figure}

The test of this model was to see whether it would produce better results than
MINIMAL. We still used the method of averaging over \vb"DISP_INT" time steps
to obtain our statistics, as two million time step results were too many to
deal with easily. However, it was possible, by a further modification to the
program, to produce code that instead of averaging over \vb"DISP_INT", wrote
the state of the system at points 2, 4, 6 and 8 in the above sequence of
steps, so that each \vb"getstats()" call was followed by the recording of the
data in a file. This allowed one to produce a graph of how various quantities
such as $B$, $O$ and $Spr$ varied with each and every time step. Besides being
a satisfying thing to look at, such a graph gave a visual demonstration of the
simulation at work, and an assurance that the simulation was working as it
ought (Figs.~\ref{fig:min1picbom} and \ref{fig:min1picspr}).
The graphs show that bid and offer can meet at the same point at certain
times, just before the annihilation of traders at the reaction front.
Fig.~\ref{fig:min1picspr} demonstrates this especially well. $Spr$ gradually
gets closer to zero by the diffusion of the traders at the reaction front,
until eventually it reaches zero and annihilation occurs, after which the best
bid and best offer immediately `snap back' to the position of the next best
bid/offer. This is precisely the way the model is expected to work. Satisfied
that our new simulation is functional, we press on and analyze results
obtained thereby.

\subsection{MINIMAL1 results}
We shall perform the same analysis for MINIMAL1 results as we did for MINIMAL.
To ensure the approximate validity of the approximation $\ln w \ll \ln cL$, we
shall take only results for which $\ln w < 3$, because if we take $c \sim 0.5$,
then for $L = 1000$, $\ln cL \simeq 6$. Values of $\ln w \simeq 3$ cannot
\emph{really} be considered small compared with $\ln cL$, but if we restrict
our range any further, then we would be making use of a small number of
results (10 points or so) in which case reliability of results would suffer.
Thus, we will use points for which $\ln w < 3$, bearing in mind that the points
towards the top left of the graph are less accurate.

\begin{figure}
\centering \resizebox{\figwidth}{!}{\includegraphics{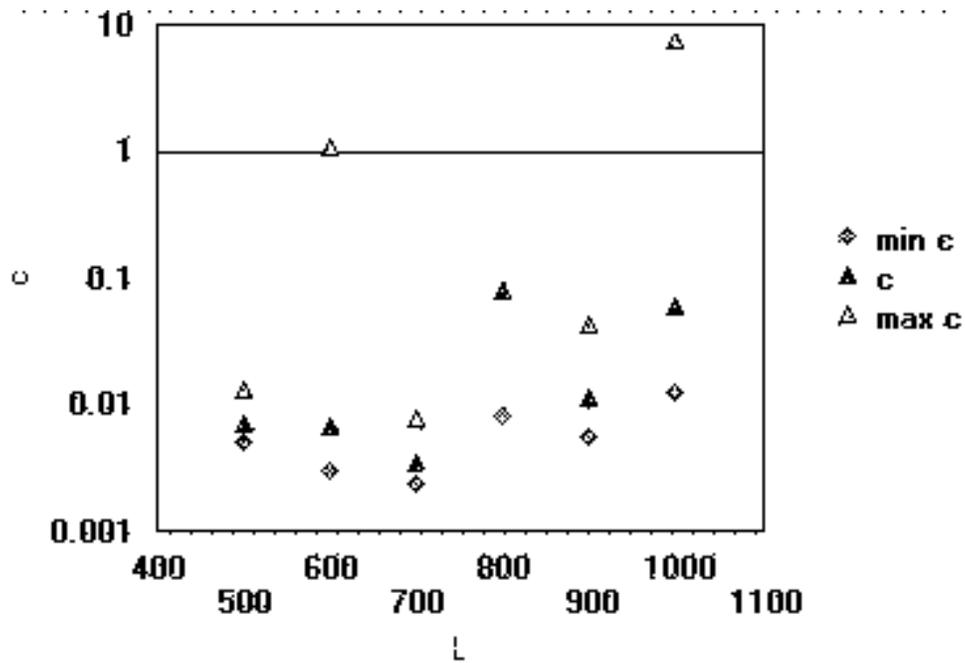}}
\caption{Graph of $c$ (derived from plots of $\ln w^2$ vs.\ $\ln J$) against
$L$} \label{fig:min1w2c}
\end{figure}

\begin{figure}
\centering \resizebox{\figwidth}{!}{\includegraphics{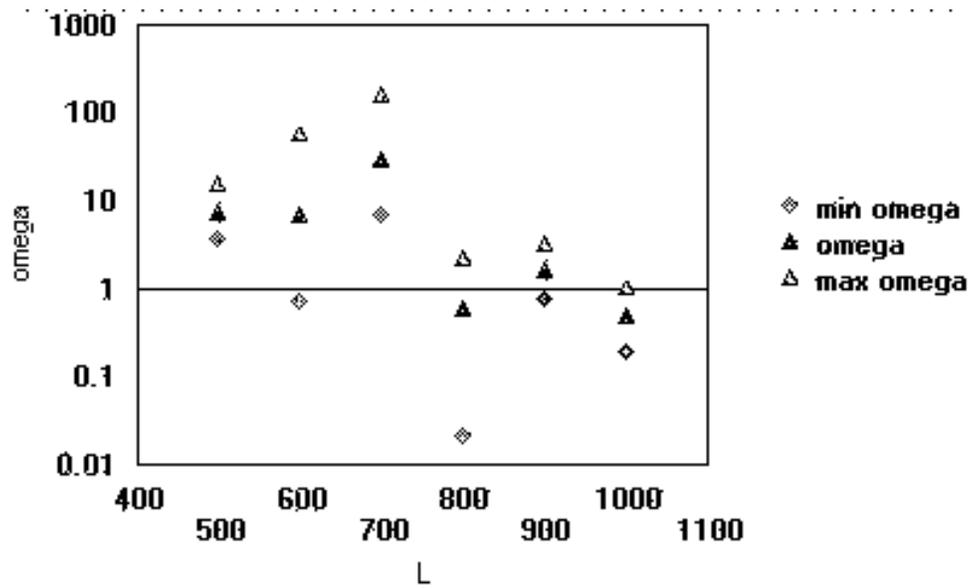}}
\caption{Graph of $\pi\Omega$ (derived from plots of $\ln w^2$ vs.\ $\ln J$)
against $L$} \label{fig:min1w2omega}
\end{figure}

It was decided that the last 12 points would be used to plot a graph from
which the gradient and intercept would be obtained. Two factors influenced this
decision---the desire to have as many points as possible to increase
reliability of the results, and the need to restrict ourselves to values of
$\ln w < 3$---and a compromise was reached. This state of affairs was far from
perfect, as the statistics generated from a mere 12 points were really not
very convincing, and there were points towards the top left of the graphs for
which $\ln w$ was greater than 3 or so. Nevertheless, we present the results
here for comparison with later results (Figs.~\ref{fig:min1w2c} and
\ref{fig:min1w2omega}).

The graph for $c$ appears to suggest $c \sim O(0.01)$ or thereabouts, which is
much smaller than what we had before, from the MINIMAL results. The reason for
this might be that 12 points do not reliably fix the gradient on a graph.
Note, also, that if we take $c \sim 0.01$, we have $\ln cL \approx 2.3$, which
requires $\ln w^2 \ll 4.6$. This was grossly violated in our graphs, as we
allowed $\ln w^2$ to go up to 6 or slightly beyond! Thus, these results are
invalid and useful information cannot be drawn from them. The statistics from
these graphs actually point to their own invalidity. It would be impossible to
restrict our results to $\ln w^2 \ll 4.6$ as our results are all $\ln w^2 > 2$
which already violates the regime. This forces us to take another approach.

\begin{figure}
\centering \resizebox{\figwidth}{!}{\includegraphics{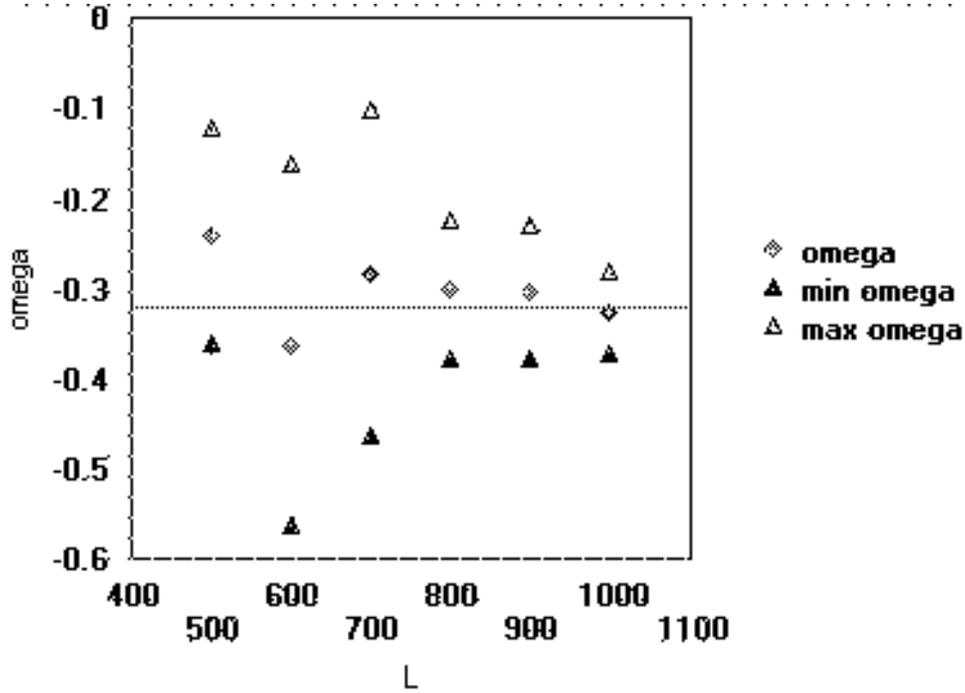}}
\caption{Graph of $\Omega$ against $L$, derived from plots of $Jw^2$ against
$\ln w$ (the line at $\Omega = 0.32$ represents the theoretical value)}
\label{fig:min1Jw2}
\end{figure}

To overcome the restriction of the range of allowed $\ln w$, we can plot $J
w^2$ against $\ln w$ for each value of $L$. This I have done, and the results
are presented below:
\begin{center}
\begin{tabular}{cccc}
L & Gradient & Intercept & $R^2$ \\
\hline 1000 & $-0.16 \pm 0.02$ & $1.01 \pm 0.13$ & 0.559\\
900 & $-0.15 \pm 0.04$ & $0.98 \pm 0.13$ & 0.348\\
800 & $-0.15 \pm 0.04$ & $1.01 \pm 0.12$ & 0.383\\
700 & $-0.14 \pm 0.10$ & $1.08 \pm 0.28$ & 0.0799\\
600 & $-0.18 \pm 0.10$ & $1.16 \pm 0.26$ & 0.112\\
500 & $-0.12 \pm 0.06$ & $0.92 \pm 0.14$ & 0.144
\end{tabular}
\end{center}
>From Eq.~(\ref{eq:Jw2}), $\Omega$ should be $1/\pi \approx 0.32$. We plot the
gradient as a function of $L$ in Fig.~\ref{fig:min1Jw2}.
There are three sets of points: omega, min omega and max omega. The min and
max series represent the minimum and maximum values omega can take, within the
bounds of experimental error. Thus, they can be thought of as the ends of
(imaginary) error bars. The horizontal line in the middle of the graph at
$\Omega = -0.32$ represents the theoretical value. One can see that all the
results agree with this predicted value, albeit with rather poor precision.
Coupled with the earlier results from MINIMAL, this is mildly encouraging.

\begin{figure}
\centering \resizebox{\figwidth}{!}{\includegraphics{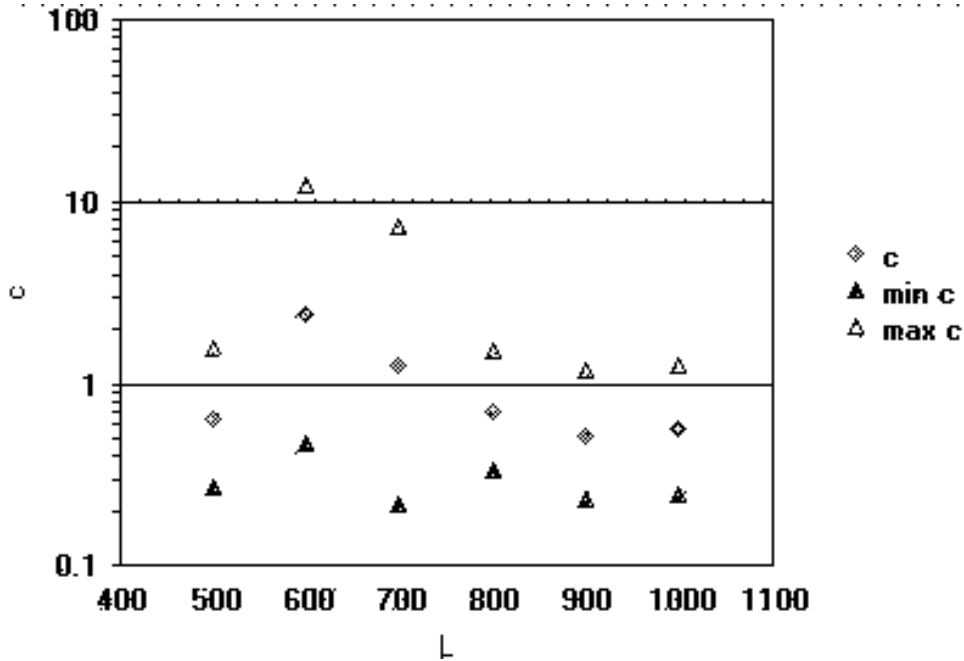}}
\caption{Graph of $c$ against $L$, derived from plots of $Jw^2$ against $\ln
w$} \label{fig:min1Jw2c}
\end{figure}

>From the intercepts, we can work out $c$. We may, with some confidence, take
$\Omega$ to be $1/\pi$. The values of $c$ obtained thus were plotted against
$L$, in Fig.~\ref{fig:min1Jw2c}.
Because of the exponentiation of the intercept involved in obtaining $c$,
errors in the intercept were greatly magnified, leading to relatively poor
precision in our determination of $c$. However, it agrees with the earlier
result of $c \sim 0.5$, though this set of results seems to point, rather
vaguely, to $c \sim O(1)$.

\subsection{MINIMAL2: Cornell's approach}
S.~Cornell, in his paper \cite{corn:sims}, performed extensive simulations of
diffusion-annihilation reactions, and described in detail his method of
tackling the problem. His method, which allows particles of \emph{both}
species to diffuse over the same time steps, can lead to crossed-over buyers
and sellers, which need then to be removed, along with the usual overlapping
traders. One possible advantage of such a method, compared with MINIMAL and
MINIMAL1, is that the asymmetry between buyers and sellers, in which one type
of trader always diffuses before the other type, is removed, and both are
allowed to diffuse simultaneously. 
However, an
inevitable consequence of this, when used with discrete time, is the crossing
over of traders. At first, one might be tempted to think that such a scenario
cannot occur in the market because traders generally try to sell as high as
possible, and buyers buy as low as possible. However, because exchange of
information via the trading screen is not instantaneous (the screen is updated
every so often), and humans can sometimes make mistakes, it is possible to
have the best bid at a higher price than the best offer. In this case, usually
the cross-over would be spotted by a broker, who would then inform the two
traders of this situation so that they can do a deal immediately. The
crossing-over characteristic of Cornell's model, which is perfectly legitimate
in chemical reactions where annihilation is not immediate but follows an
exponential decay law, is actually not as far-fetched as it might seem, when
applied to financial markets.

Strictly speaking, this is not the model described in \cite{kogan}, but a
quick modification of the program was performed to see what such a method
would yield. After several  runs of
this method  the results did not look much different
from those obtained using MINIMAL1. Further investigation is required before
any conclusions may be drawn.

\subsection{The scaling law for $NUM$}
\label{cs:min:num} 
\begin{figure}
\centering \resizebox{\figwidth}{!}{\includegraphics{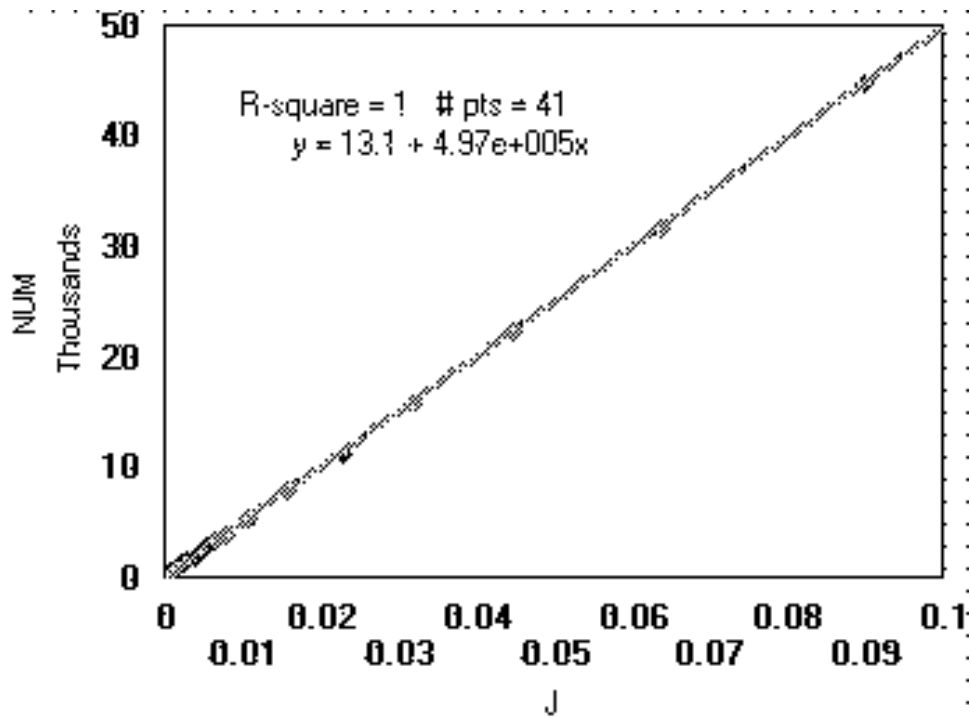}}
\caption{Graph of $NUM$ against $J$ for $L = 1000$} \label{fig:NUMvsJ}
\end{figure}

\begin{figure}
\centering \resizebox{\figwidth}{!}{\includegraphics{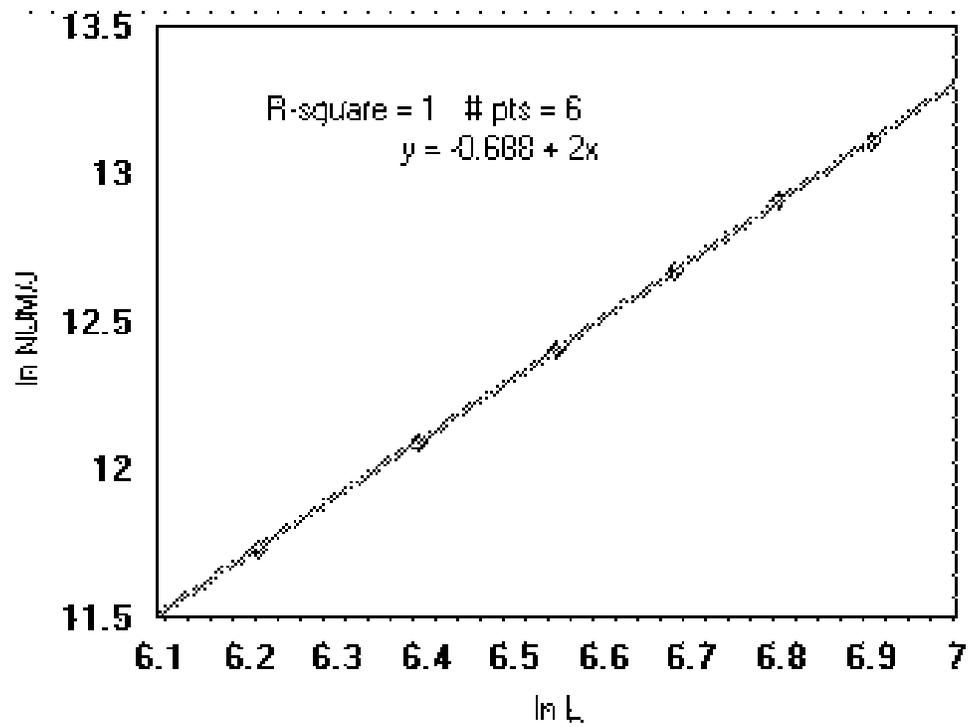}}
\caption{Graph of $\ln (NUM/J)$ against $\ln L$} \label{fig:NUMvsL}
\end{figure}

One interesting scaling law not mentioned in \cite{kogan}
concerns the equilibrium number of traders in the market, or $NUM$, 
where it is understood to mean the value at
equilibrium. This is clearly shown in the two graphs, Figs.~\ref{fig:NUMvsJ}
and \ref{fig:NUMvsL}.
Fig.~\ref{fig:NUMvsJ} shows that $NUM \propto J$ and the gradient is the
coefficient of proportionality. From plotting many such graphs of $NUM$
against $J$ for different $L$, one can collect the gradients ($NUM/J$) together
and thus plot $\ln(NUM/J)$ against $\ln L$ (Fig.~\ref{fig:NUMvsL}). The graph
is a straight line, with gradient $1.999 \pm 0.007$ and intercept $-0.688 \pm
0.004$. We may infer that $NUM/J \propto L^2$ with a constant of
proportionality of 1/2 (from $\exp(-0.688)$). Note the remarkably good
correlations on the graphs and the negligible experimental errors in the
determination of the gradients. We may summarize, with great confidence, that
\begin{equation}
\label{eq:empnum}
    NUM = \frac{JL^2}{2}
\end{equation}
obtained empirically.

Dimensional analysis tells us that $NUM$ is dimensionless. The right hand side
of Eq.~(\ref{eq:empnum}), however, is \emph{not} dimensionless. We can make it
dimensionless by introducing $D$ in the denominator, as $D/J$ and $L^2$ have
the same dimensions. The diffusion coefficient that we have been using is $D =
1/2$, so the true scaling law is
\begin{equation}
\label{eq:numlaw}
    NUM = \frac{JL^2}{4D}.
\end{equation}
There is a firm theoretical justification for this scaling law. Consider a
trader entering the system at the edge of price space. He will remain in price
space until he hits the reaction front and annihilates with a trader of the
opposite species. The reaction front is approximately in the middle of price
space, a distance $L/2$ from the newly-inserted trader. From Einstein's work
on Brownian motion (see \cite{gardiner}), the mean square distance $\lambda^2$
diffused by a particle in time $t$ is $\sim Dt$. The time taken $T$ for the new
trader to diffuse to the reaction front is given by
\[
    \left( \frac{L}{2} \right)^2 \sim DT.
\]
Now, we multiply both sides by $J$, the rate of insertion of traders, and
divide by $D$:
\[
    \frac{JL^2}{4D} \sim JT.
\]
$JT$ is simply the number of traders inserted between the time of the entrance
of our trader into the market, and the time of his disappearance through
annihilation. This is precisely the equilibrium number of traders in the
market. One can see this in the following way: the total number of traders
keeps increasing with every new trader inserted, up till a time $T$ after the
insertion of the first trader, when annihilation starts to occur, and balances
the flux of new traders. This is only approximate, of course, to within a
factor of order 1. Thus, we have shown that $NUM \sim JL^2/4D$.

\subsection{Scaling law for $\tau_S$}
\label{cs:min:Ts} Now, we turn our attention to the scaling law obeyed by the
instantaneous dealing time, $\tau_S$, defined as the time between two consecutive
annihilations. The program was modified so that it could record values of
$\tau_S$ and calculate their average and variance. There is a vital difference
between this quantity and those we have been investigating thus far: the time
between sales is \emph{not} a function of the state of the system. All the
other quantities, such as bid-offer spread and midmarket, could be read off
the trading screen, and they were quantities that were characteristic of the
state of the market. The instantaneous dealing time, however, is different, in that at
any time one does not know when the next sale is going to be, and thus what
the current value of $\tau_S$ is. One can only record $\tau_S$ as a historical
quantity. Furthermore, the number of readings one can take of this quantity is
not fixed, but varies, depending on $J$ as well as the number of time steps
the simulation is run for.

The program was run for the range of values of $J$ between 0.00004 and 0.007,
which has been shown to produce $w$ that satisfies $10 < w < 100$, ensuring
that the midmarket fluctuation, and other important lengths, are at least a
factor of 10 or so away from the size of the system, or the size of the
smallest unit of length. For the smallest value of $J$, there were only 55
values of $\tau_S$ obtained, whilst that number was 7140 for the largest value
of $J$ used. Therefore, the average and variance of $\tau_S$ for the larger
values of $J$ were more reliable than those for smaller $J$.

Originally, a graph of $\ln \tau_S$ against $\ln J$ was plotted for all the
values, but while most of the points lay along a straight line, points
corresponding to the 10 or so smallest values of $J$ were rather scattered and
did not lie so well on the line. This was because of the relatively small
number of data points over which the averages and variances were obtained.
These points were not as reliable as the others and so it was decided to
exclude them in the graph plotting, and include only those for which $J \geq
0.0002$, all of which had been averaged over at least 210 points. A new graph
was plotted. We obtained
\begin{equation} \label{eq:lnTs}
    \ln \tau_S = (0.08 \pm 0.03) - (0.984 \pm 0.005) \ln J
\end{equation}
leading to the $J$ exponent being $-0.984 \pm 0.005$ and the constant in front
of $J$ being $1.085 \pm 0.035$. The exponent is close to the expected value of
-1, but unfortunately it is not \emph{quite} within experimental error. The
constant in front is also close to the expected value of 1, but just outside
experimental error. We plot a graph of the fluctuation of time to midmarket
sale, and find the following:
\begin{equation} \label{eq:lnvarTs}
    \ln \var(\tau_S) = (-0.34 \pm 0.1) - (1.97 \pm 0.02) \ln J.
\end{equation}
The exponent is expected to be 2, since the fluctuation, defined as the
variance of $\tau_S$, goes as the square of time, which is inversely
proportional to $J$, from dimensional analysis. We obtain $0.71 \pm 0.08$ for
the constant in front, for which we have no ``expected'' value, since
intuition does not tell us how scattered the times should be. Dimensional
analysis is certainly correct in telling us the exponent of $J$, but it cannot
give us any clue about what the constant should be, apart from a possible
dimensionless parameter.

It was noted that, because each of the simulations were run for a specified
number of timesteps, namely 2 million, the total number of trades occurring
within the time of each simulation was proportional to $J$. Thus, for smaller
$J$, there were very few values of $\tau_S$ for small $J$ (as few as 55 for $J
= 0.00004$), and more for larger $J$ (7140 for $J = 0.007$). The number of
readings from which each average and variance is obtained should not affect
the accuracy of the results, though it might lower its precision, i.e.\ more
scatter about the ``true'' mean and variance. In order to overcome the problem
of having a variable number of $\tau_S$ data points, the program was further
modified so that the simulation would continue until a \emph{fixed} number of
trades had taken place. The limit was set at 1000, and the simulations
terminated when 1000 trades had taken place and all their accompanying
$\tau_S$ values were recorded. The total running time was somewhere in the
region of 18 hours.

\begin{figure}
\centering \resizebox{\figwidth}{!}{\includegraphics{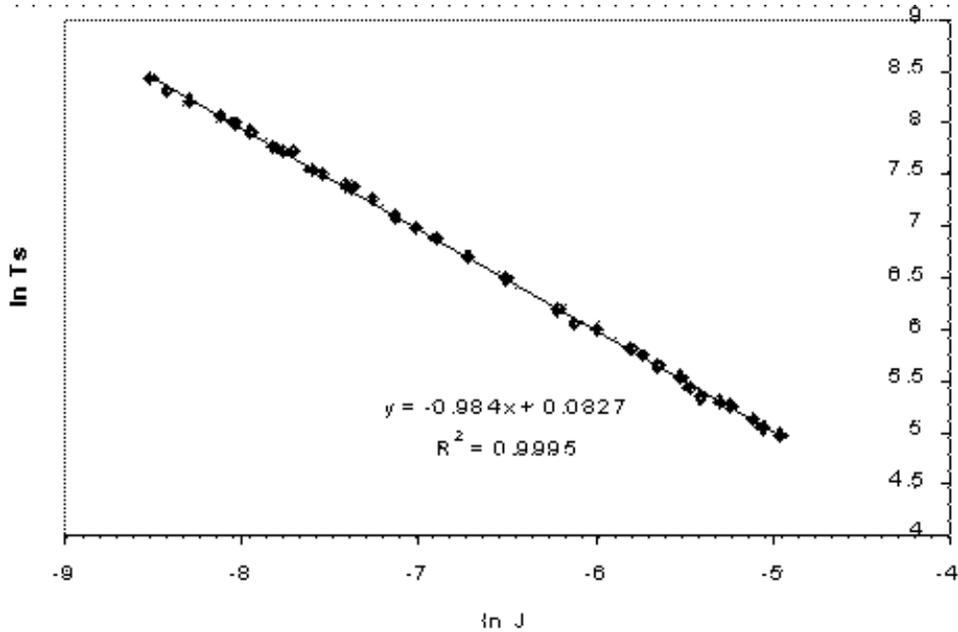}}
\caption{Graph of $\ln \tau_S$ against $\ln J$ for $L = 1000$}
\label{fig:TsvsJ}
\end{figure}

\begin{figure}
\centering \resizebox{\figwidth}{!}{\includegraphics{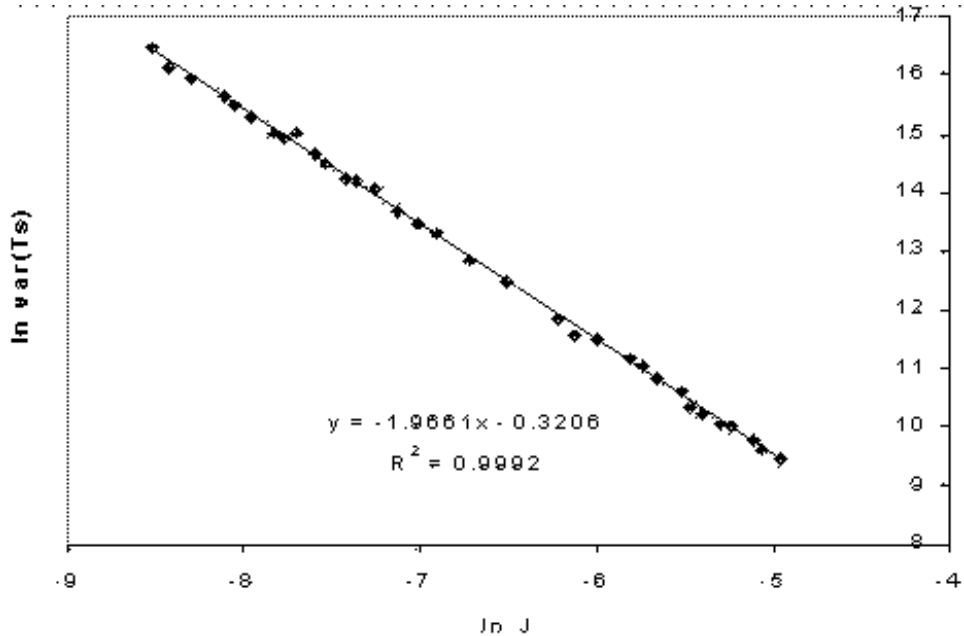}}
\caption{Graph of $\ln \var(\tau_S)$ against $\ln J$ for $L = 1000$}
\label{fig:varTsvsJ}
\end{figure}

The results obtained were plotted on two graphs, one of $\ln \tau_S$ against
$\ln J$ (Fig.~\ref{fig:TsvsJ}), and the other of $\ln \var(\tau_S)$ against
$\ln J$ (Fig.~\ref{fig:varTsvsJ}).
When plotting the graphs, it was found that the data for the smaller values of
$J$, though lying approximately along a straight line joining them and the
other points, were rather more scattered about that line than the other points
were. They appeared less reliable and were therefore excluded from the two
graphs presented. Only values for which $J \geq 0.0002$ were plotted. From the
gradients and intercepts, the following information was gleaned:
\begin{eqnarray}
    \ln \tau_S & = & (0.08 \pm 0.03) - (0.984 \pm 0.004) \ln J \label{eq:lnTs-1000} \\
    \ln \var(\tau_S) & = & (-0.32 \pm 0.07) - (1.97 \pm 0.01) \ln J \label{eq:lnvarTs-1000}
\end{eqnarray}
In the case of $\tau_S$, the constant in front is $1.086 \pm 0.03$, whilst the
corresponding constant is $0.73 \pm 0.05$ for $\var(\tau_S)$. One can see a
remarkable similarity between these results in Eqs.~(\ref{eq:lnTs-1000}) and
(\ref{eq:lnvarTs-1000}), and the earlier ones in Eqs.~(\ref{eq:lnTs}) and
(\ref{eq:lnvarTs}). In fact, the only real difference is the reduction in the
experimental uncertainty in the parameters. Thus, the second, more detailed,
set of results not only confirms the validity of the first, but adds to it by
narrowing the margins of uncertainty.

It is fair, in the light of such compelling evidence, to believe that $\tau_S$
goes as $1/J$, as both the exponent and the constant were found to be
\emph{very nearly} -1 and 1, respectively. Such a conclusion is supported also
by intuition, which tells us that $\tau_S = 1/J$. The constant in front has
also been confirmed. For $\var(\tau_S)$, the exponent can be taken to be -2,
while the constant in front is $0.73 \pm 0.05$. It seems very likely that
neither quantity has a logarithmic correction.

\subsection{Scaling law for $\tau_{\rm reduced}$}
\label{cs:min:tau_reduced} There is another way of defining the instantaneous dealing time. 
It is identical to that given in \S\ref{cs:min:Ts} for single
annihilations, but differs for multi-annihilation. We can call this $\tau_{\rm
reduced}$, a sort of ``reduced'' instantaneous dealing time. If the last
annihilation contained $n$ simultaneous trades, and the next one contained $m$
simultaneous trades, then
\begin{equation}\label{eq:tau_reduced}
  \tau_{\rm reduced} = \frac{\tau}{n+m-1}.
\end{equation}
$n$ and $m$ may be considered the respective degeneracies of the trades. It
would be interesting to see what statistics this new quantity obeys.

\begin{figure}
\centering \resizebox{\figwidth}{!}{\includegraphics{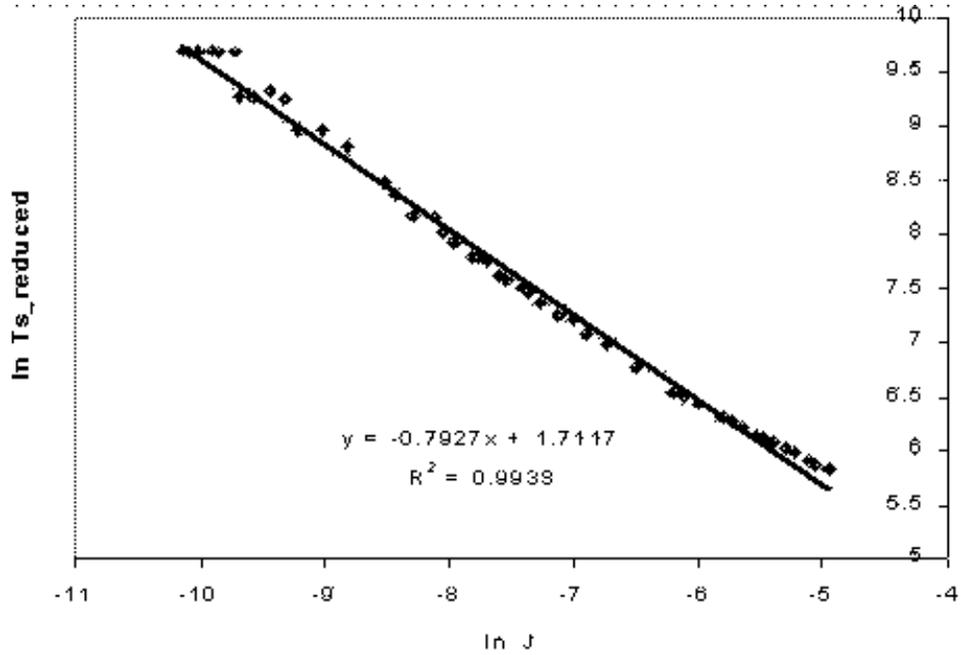}}
\caption{Graph of $\ln \tau_{\rm reduced}$ against $\ln J$ for $L = 1000$}
\label{fig:Ts_rvsJ}
\end{figure}

We repeated the simulations for this reduced $\tau$, doing them for exactly
the same range of $J$ as we did for normal $\tau$. 1000 data points were
obtained for each $J$, making each point as reliable as any other. A graph of
$\ln \tau_{\rm reduced}$ against $\ln J$ was plotted (Fig.~\ref{fig:Ts_rvsJ}).
This graph is interesting in that the points together form a precise
\emph{curve} rather than a line. There is very little scatter, except in those
points with the smallest $J$'s. The rest of the points trace out a curve that
tends to bend upwards as $J$ is increased. If we only take the points with the
smallest $J$ in isolation, we obtain a gradient of --0.84. If we take the
points with the highest $J$ in isolation, we obtain --0.58 instead. Thus, there
is some variation of the exponent with $J$. If by brute force we calculate a
line of best fit, its statistics would be
\[
    \ln \tau_{\rm reduced} = (-0.793 \pm 0.009) \ln J + (1.71 \pm 0.07)
\]
which gives an exponent of --0.793 and a ``constant'' of proportionality of
$5.5 \pm 0.4$.

\begin{figure}
\centering \resizebox{\figwidth}{!}{\includegraphics{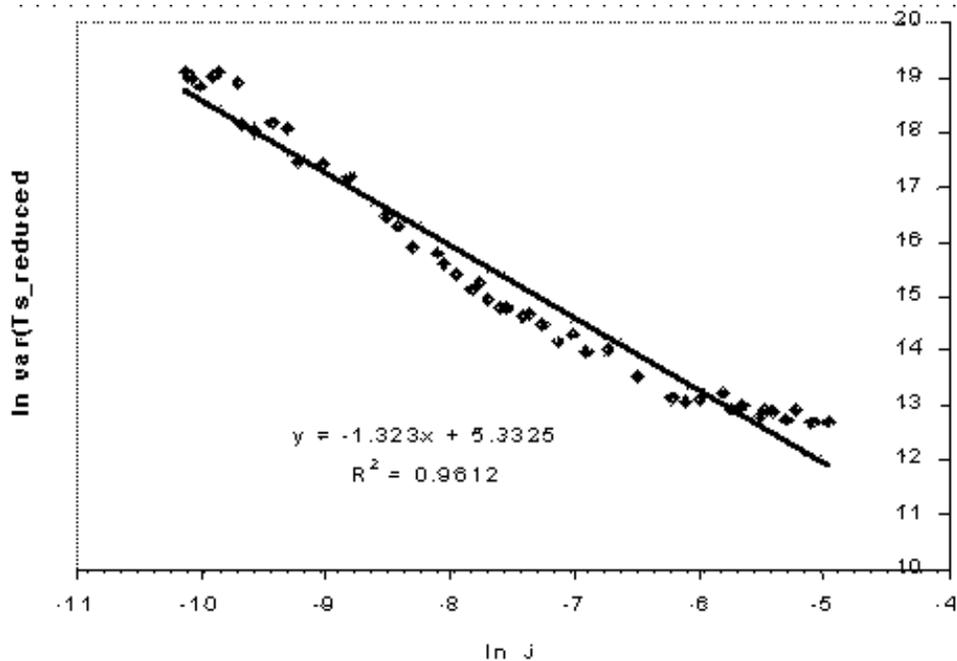}}
\caption{Graph of $\ln \var(\tau_{\rm reduced})$ against $\ln J$ for $L =
1000$} \label{fig:varTs_rvsJ}
\end{figure}

Similarly, with the fluctuation (i.e.\ variance) in $\tau_{\rm reduced}$, the
graph is not a straight line but a curve that curves upwards as $J$ is
increased. This is shown in Fig.~\ref{fig:varTs_rvsJ}.
Its gradient varies from --0.47 at its absolute minimum to --1.73 at its
maximum. The mean gradient and intercept are
\[
    \ln \var(\tau_{\rm reduced}) = (-1.32 \pm 0.04) \ln J + (5.3 \pm 0.3)
\]
giving a very imprecise estimate of the constant as $210 \pm 70$. The
explanation of why the points do not fit a straight line well has not been
found, and further investigations are needed before definite conclusions can
be drawn. We will not pursue this question here.  

\subsection{The minimal model with asymmetric fluxes}
\label{cs:min:asym} It was now decided to generalize the situation to one in
which asymmetric fluxes of buyers and sellers are permitted. Intuitively, one
would expect that, in the steady-state, the midmarket moves linearly with time
towards one end or the other, away from the edge with the higher flux. This is
a particularly productive phenomenon to investigate, since it will tie in very
well with our later work on the two-liquid model, where different fluxes of
limit-order and market-order traders will be introduced, in order to reproduce
the well-known widening of the bid-offer spread in the prelude to a crash. The
analytical result, Eq.~(\ref{eq:asym}), for the speed of the moving midmarket,
is suggested in \cite{kogan} . This is the result we shall test.

\begin{figure}
\centering \resizebox{\figwidth}{!}{\includegraphics{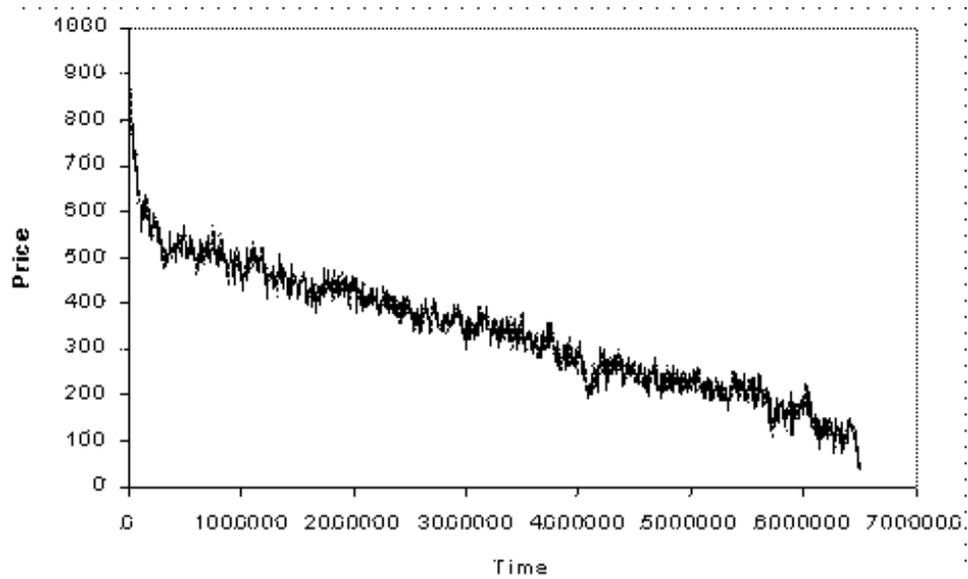}}
\caption{Midmarket variation with time for $J = 0.001$ and $\Delta J = 0.0001$}
\label{fig:asymbom1}
\end{figure}

\begin{figure}
\centering \resizebox{\figwidth}{!}{\includegraphics{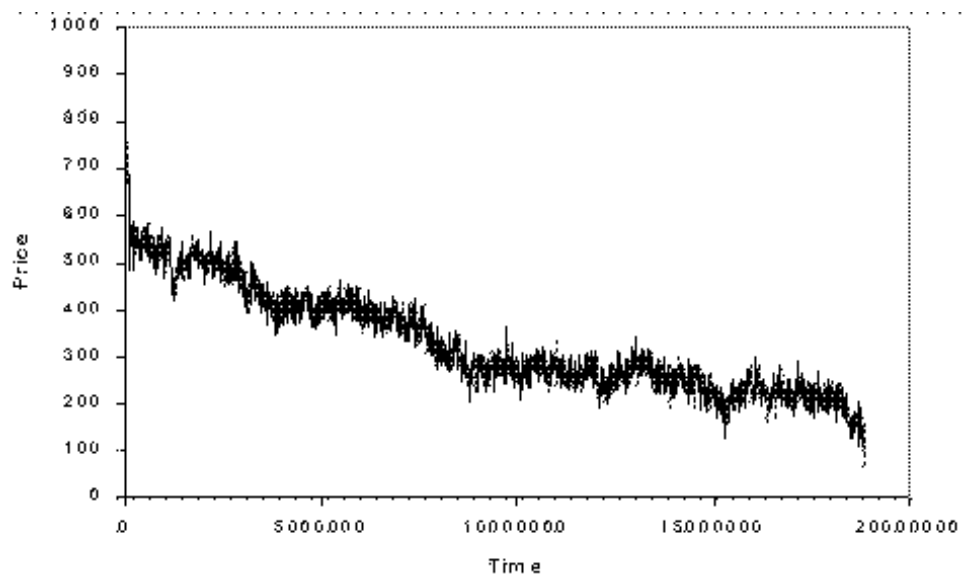}}
\caption{Midmarket variation with time for $J = 0.001$ and $\Delta J =
0.00001$} \label{fig:asymbom2}
\end{figure}

Initially, the program was run for very long times, in order to get a feel for
what happened to the midmarket when the fluxes were asymmetric. It was
observed that the midmarket did move, though it did not always do so at a
constant speed. To illustrate this, two graphs of midmarket variation are
shown: Fig.~\ref{fig:asymbom1} and \ref{fig:asymbom2}.
Fig.~\ref{fig:asymbom1} appears to be quite straight and linear, but
Fig.~\ref{fig:asymbom2} does not. The latter exhibits some irregularity. In
fact, there were many other examples of such graphs, where the midmarket
movement was initially quite linear, but after some time, deviated from
linearity. Most tended to bend downwards, indicating a higher speed of
movement. As the scaling law  Eq.~(\ref{eq:asym})   applies only to small
times, $T < L^2/D$, we should restrict our attention to times less than 2
million, and for these, the graphs were approximately linear.

\begin{figure}
\centering \resizebox{\figwidth}{!}{\includegraphics{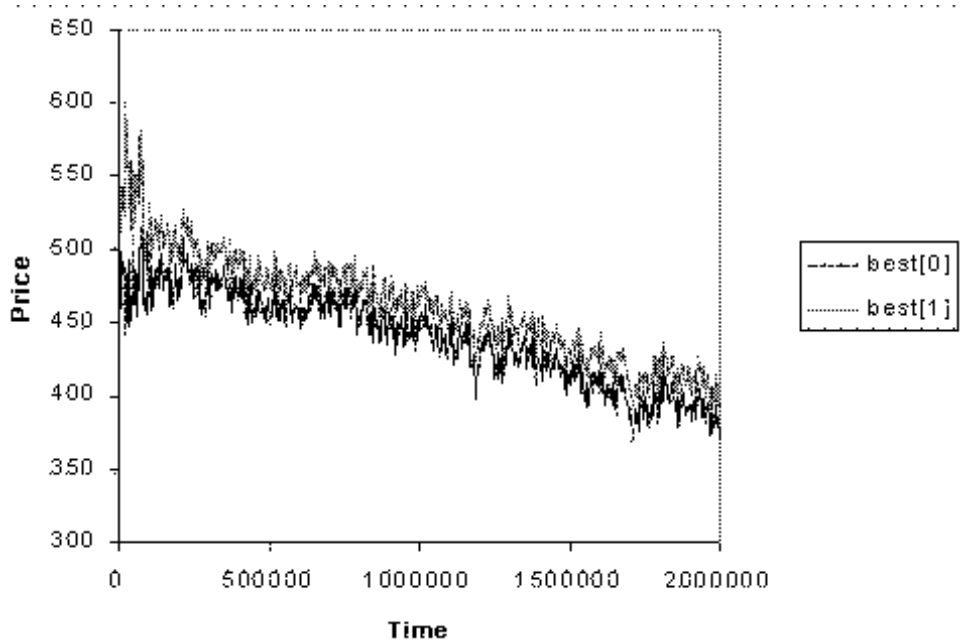}}
\caption{Variation of best bid and best offer with time for $J = 0.005$ and
$\Delta J = 0.0003$} \label{fig:asymbo}
\end{figure}

\begin{figure}
\centering \resizebox{\figwidth}{!}{\includegraphics{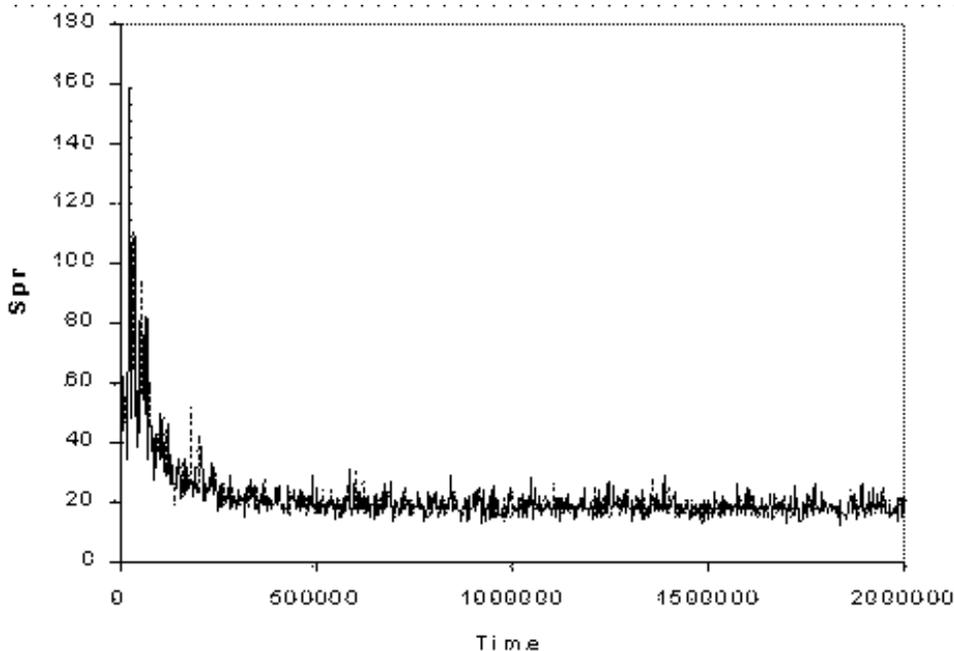}}
\caption{Variation of bid-offer spread with time for $J = 0.005$ and $\Delta J
= 0.0003$} \label{fig:asymspr}
\end{figure}

It is significant that the introduction of asymmetric fluxes of buyers and
sellers has not caused any change in the bid-offer spread. A graph was plotted
of the variation of the best bid and best offer with time
(Fig.~\ref{fig:asymbo}), from which it may be seen that the spread has
\emph{not} increased with time. In addition, a graph of spread itself was
plotted as a function of time (Fig.~\ref{fig:asymspr}). Spread appears to stay
constant with time (after steady-state was reached).

\subsubsection{$\Delta J$ dependence}
\label{cs:min:asym:dJ} Our first task was to investigate the dependence of the
speed of midmarket movement on $\Delta J$. This was done by fixing all the
other variables ($J$ and $L$) and allowing $\Delta J$ alone to vary. A graph
would then be plotted of $\ln |dx/dt|$ against $\ln \Delta J$, from which the
exponent of the $\Delta J$ dependence could be deduced. A wide range of values
for $\Delta J$ were tried, but it was found that not all such values produced
good results. For $\Delta J / J$ that was greater than 0.4 or so, it was seen
that the midmarket moved so quickly that it hit rock bottom (price = 0) in
very little time, sometimes less 1 million time steps, which would have been
insufficient time for the system to come to any sort of steady-state. In the
past, it was observed that 1 million time steps or so were required before the
system could come to equilibrium; now, although equilibrium can never be
reached in this asymmetric situation, a sort of steady state can reasonably be
achieved in that time. At the other end of the scale, it is possible to choose
a $\Delta J$ to $J$ ratio that is too small. The midmarket drifts so slowly
that the gradient of the graph is negligibly small. Although not a problem in
itself, when one plots the logarithm of this quantity, it is a large negative
number. A small uncertainty in the gradient is magnified into a large
uncertainty in the logarithm. In practice, what this means is that the points
for which the ratio is smaller than 0.02 or so are quite scattered and do not
provide much useful information. Therefore, we shall limit our investigations
to the range $0.02 < \Delta J / J < 0.4$.

\begin{figure}
\centering \resizebox{\figwidth}{!}{\includegraphics{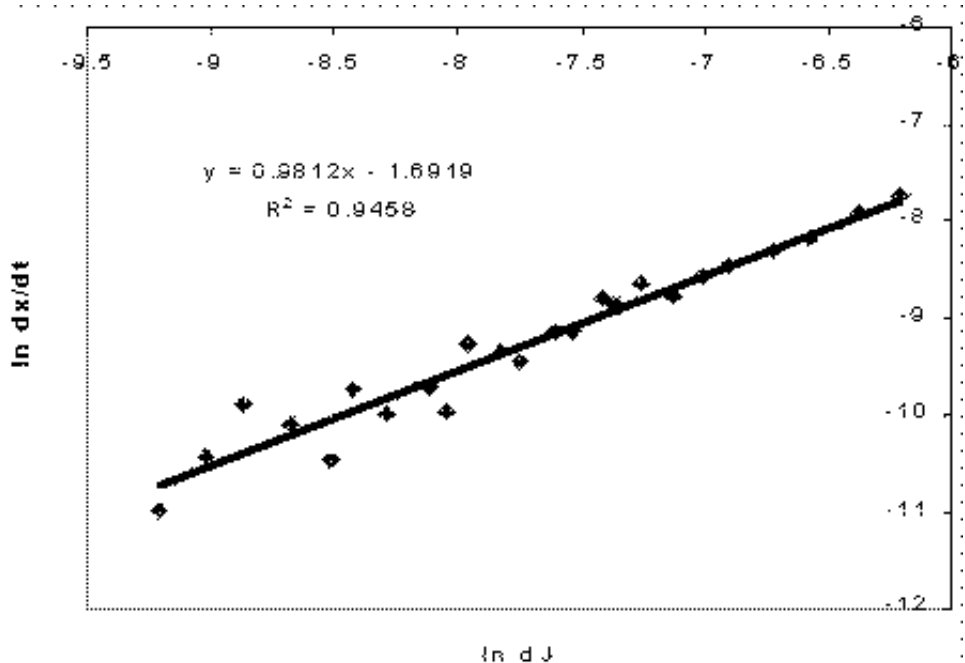}}
\caption{Graph of $\ln |dx/dt|$ against $\ln \Delta J$ for $J = 0.005$}
\label{fig:asymdJ1}
\end{figure}

Our choice of $J$ is still subject to the restrictions described in the
earlier sections, that the lengths produced by such a $J$ (e.g.\ $w$, $Spr$)
must be much larger than $a$, the lattice spacing, and much smaller than $L$.
Thus, we require $0.007 < J < 0.00004$. In order to allow $\Delta J$ to range
over as large an interval as possible, it is wise to choose a large $J$, as
$\Delta J$ is limited by it. It was decided to use $J = 0.005$, and $L =
1000$, as usual. $\Delta J$ was allowed to range from 0.0001 to 0.002. The
graph is shown in Fig.~\ref{fig:asymdJ1}.
As one can see, there is noticeably more scatter in the points towards the
lower end of the graph (small $\Delta J$). The statistics from the graph
indicate that
\[
    \ln |dx/dt| = (0.98 \pm 0.05) \ln \Delta J - (1.7 \pm 0.4).
\]
Thus, the observed exponent of $\Delta J$ is in accordance with the
theoretical value of 1. The constant of proportionality appears to be 0.18,
with the minimum being 0.12 and the maximum 0.27.

\begin{figure}
  \centering
  \resizebox{\figwidth}{!}{\includegraphics{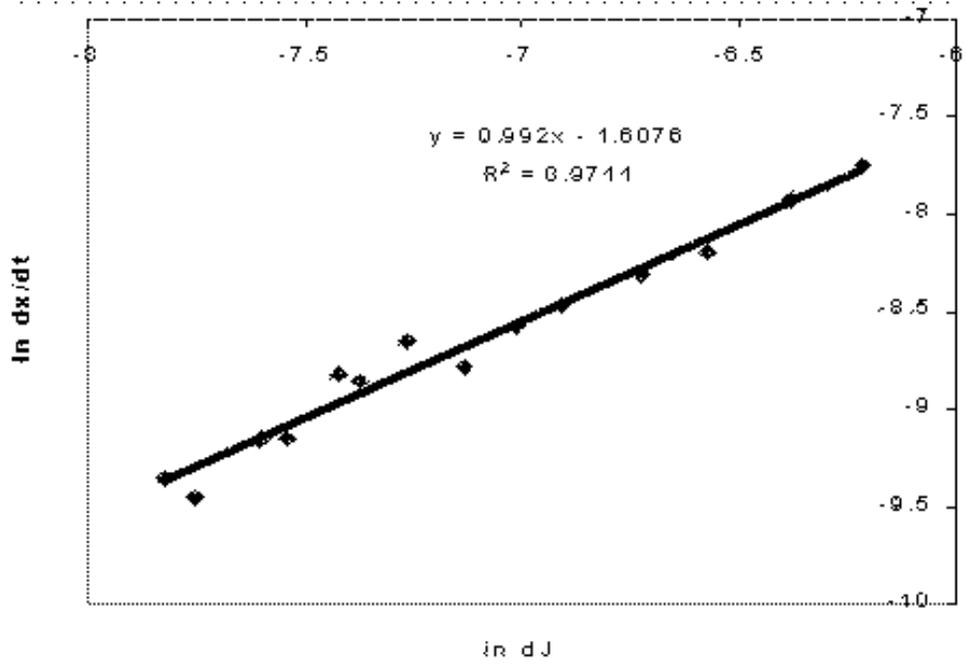}}
  \caption{Graph of $\ln |dx/dt|$ against $\ln \Delta J$ for $J = 0.005$, without the first 10 points}
  \label{fig:asymdJ2}
\end{figure}

\begin{figure}
  \centering
  \resizebox{\figwidth}{!}{\includegraphics{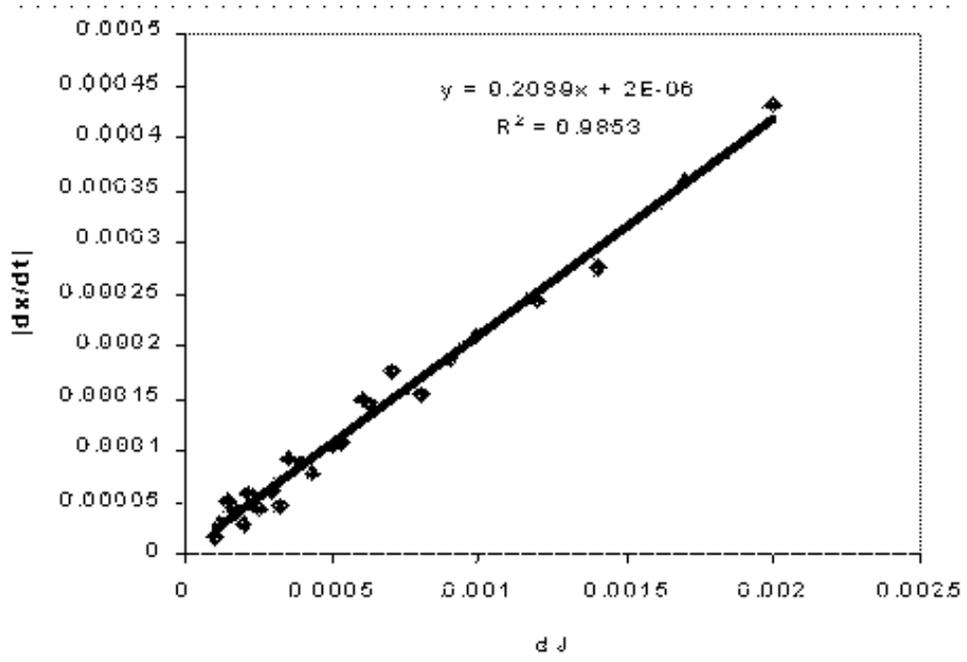}}
  \caption{Graph of $|dx/dt|$ against $\Delta J$ for $J = 0.005$}
  \label{fig:asymdJ3}
\end{figure}

There is no reason why we should not try to get more accurate and precise
statistics from our results, which are perfectly valid. Looking at the graph,
one can see that apart from the first 10 points or so, the data do fit a line
very well, and that the fit is spoilt by the scatter in the lower 10 points.
Therefore, we plot another graph, Fig.~\ref{fig:asymdJ2}, of the same two
quantities, this time without the first 10 points.
The statistics are:
\[
    \ln |dx/dt| = (0.99 \pm 0.05) \ln \Delta J - (1.61 \pm 0.35).
\]
The exponent is now even closer to the expected value of 1. The constant in
front is 0.20, with the limits being 0.14 and 0.29. In the light of such
compelling evidence, we can conclude that there is no logarithmic dependence
of any sort on $\Delta J$ and that $|dx/dt|$ is proportional to $\Delta J$.
Moreover, we can now plot a graph of $|dx/dt|$ against $\Delta J$ directly,
from which we hope to obtain a more precise estimate of the constant of
proportionality. This was done in Fig.~\ref{fig:asymdJ3}.
Because we are no longer plotting the logarithm of small values of $|dx/dt|$,
it is possible to re-include the first 10 points in our graph. The graph gives
us an intercept of $2.4 \times 10^{-6}$ with an uncertainty of $4.4 \times
10^{-6}$, which makes the intercept effectively zero. The gradient is $0.209
\pm 0.005$. Thus, plotting a direct graph can give us a \emph{much} more
precise estimate of the constant of proportionality.

\subsubsection{$J$ dependence}
\label{cs:min:asym:J} Our next task is to determine the $J$ dependence of the
formula for the speed of a moving midmarket. An inverse dependence is
expected. The limits on the ratio $\Delta J$ to $J$ dictate that the maximum
value of $J$ used be $J_{max} = 50 \Delta J$ and the minimum be $J_{min} = 2.5
\Delta J$. We require that $J_{max} < 0.007$ in order to keep all lengths at
least a factor of 10 away from $L$ or $a$, implying $\Delta J < 0.00014$. Thus,
in the interests of maximizing the range of $J$ available to us, whilst using
a ``round'' number, it was decided to set $\Delta J = 0.0001$, with $J_{min} =
0.00025$ and $J_{max} = 0.005$.

\begin{figure}
  \centering
  \resizebox{\figwidth}{!}{\includegraphics{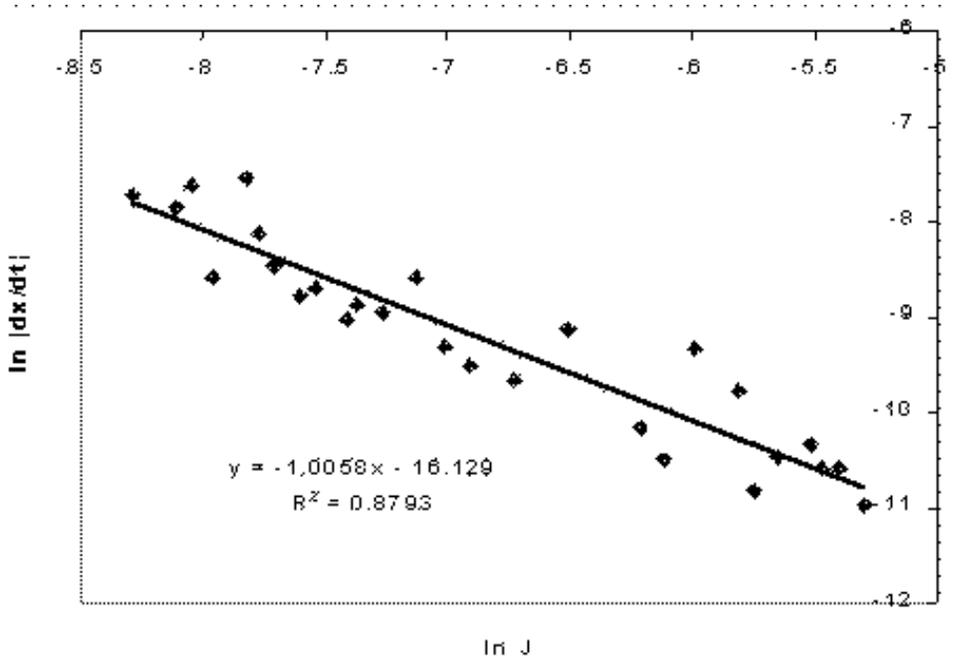}}
  \caption{Graph of $\ln |dx/dt|$ against $\ln J$ for $\Delta J = 0.0001$}
  \label{fig:asymJ1}
\end{figure}

\begin{figure}
  \centering
  \resizebox{\figwidth}{!}{\includegraphics{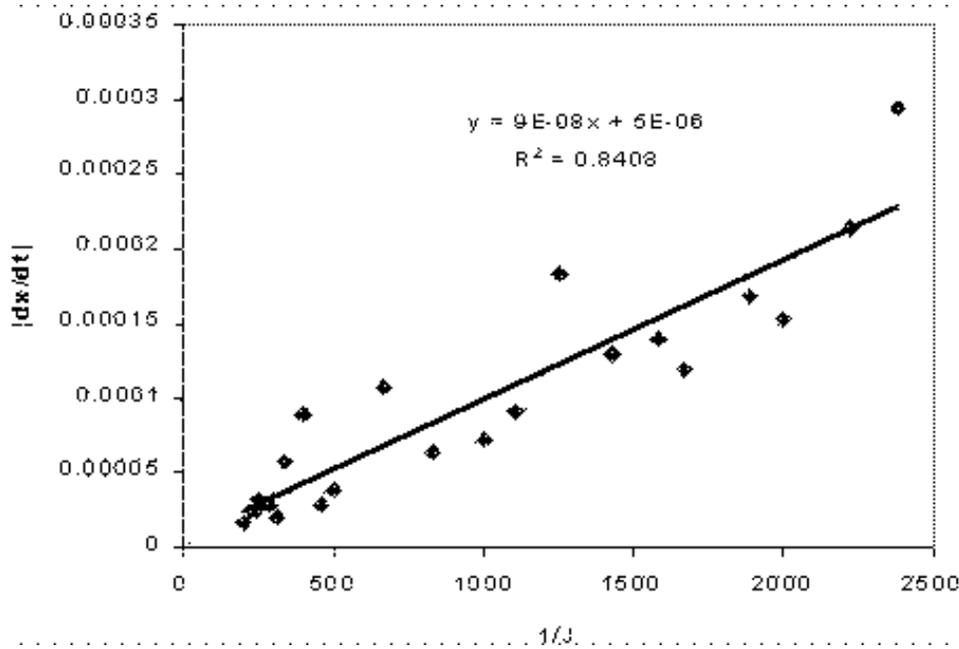}}
  \caption{Graph of $|dx/dt|$ against $1/J$ for $\Delta J = 0.0001$ (without first 5 points)}
  \label{fig:asymJ2}
\end{figure}

The results were plotted on a graph of $\ln |dx/dt|$ against $\ln J$
(Fig.~\ref{fig:asymJ1}).
This gave us the following:
\[
    \ln |dx/dt| = (-1.01 \pm 0.07) \ln J - (16.1 \pm 0.5)
\]
from which a constant of proportionality of $9.9 \times 10^{-8}$ could be
obtained (limits $5.9 \times 10^{-8}$ and $16.5 \times 10^{-8}$). Once again,
the exponent confirms the analytic result. In order to refine our estimate of
the constant of proportionality (which will be of use later), we do a direct
plot of $|dx/dt|$ against $1/J$. Here, it was noticed that the 5 points
corresponding to the smallest $J$ values we used were very scattered on that
plot, so it was decided to exclude them (Fig.~\ref{fig:asymJ2}). It does indeed
yield a more precise estimate: $(9.3 \pm 0.9) \times 10^{-8}$. The intercept,
found to be $(0.5 \pm 1.1) \times 10^{-5}$, is effectively zero.
Therefore, we may conclude with confidence that the speed is inversely
proportional to $J$.

\subsubsection{$L$ dependence}
\label{cs:min:asym:L} Finally, we test the $L$ dependence of the formula. This
is, again, expected to be an inverse relationship. We tried many different
values of $L$, from 100 up to 2000. Initially, $J = 0.001$ and $\Delta J =
0.00002$ were used, but it was observed that the results were unusable for
small $L$, such as 100 or 200, because there was far too much fluctuation from
the mean midmarket. This is only to be expected from our earlier work on the
fluctuation of the midmarket, $w^2$, which is known to depend on $1/J$.
Therefore, it was decided to increase $J$ whilst keeping the ratio $\Delta J /
J$ the same in order to reduce the fluctuation from the mean midmarket.

\begin{figure}
  \centering
  \resizebox{\figwidth}{!}{\includegraphics{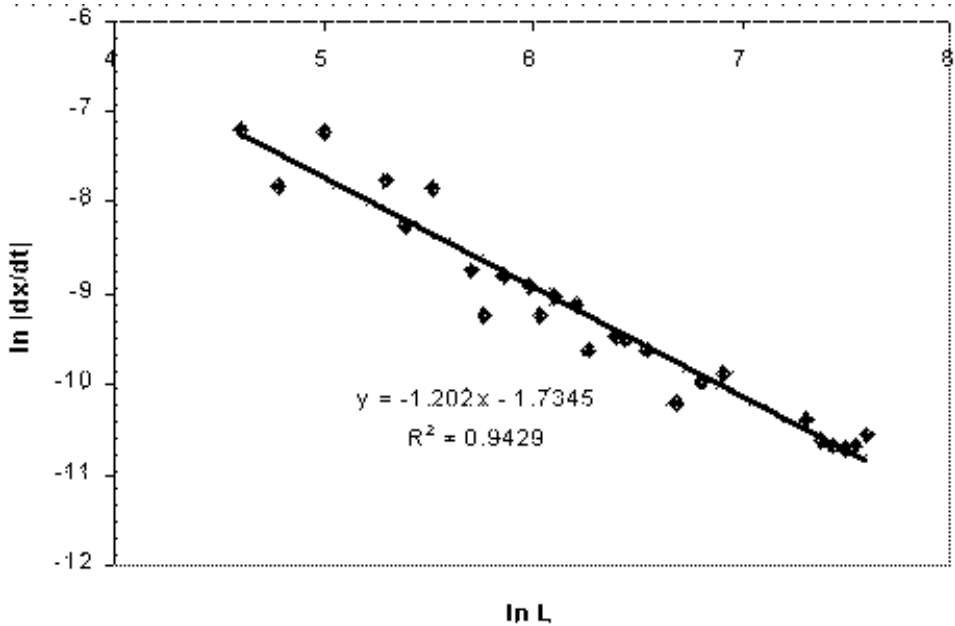}}
  \caption{Graph of $\ln |dx/dt|$ against $\ln L$ for $\Delta J = 0.005$}
  \label{fig:asymL1}
\end{figure}

\begin{figure}
  \centering
  \resizebox{\figwidth}{!}{\includegraphics{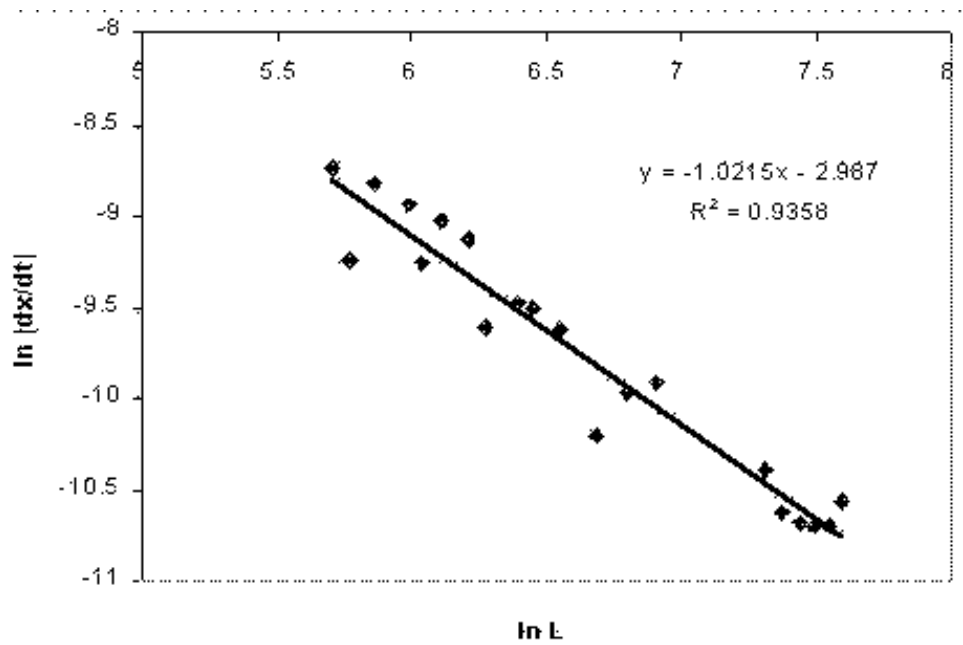}}
  \caption{Graph of $\ln |dx/dt|$ against $\ln L$ for $\Delta J = 0.005$ (without the first 6 points)}
  \label{fig:asymL2}
\end{figure}

\begin{figure}
  \centering
  \resizebox{\figwidth}{!}{\includegraphics{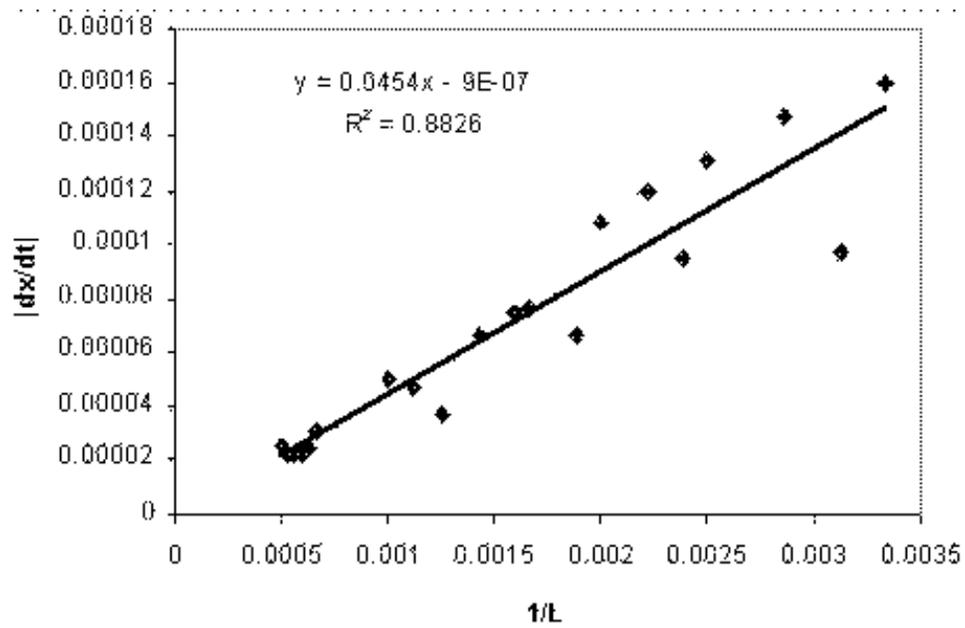}}
  \caption{Graph of $|dx/dt|$ against $1/L$ for $\Delta J = 0.005$ (without the last 6 points)}
  \label{fig:asymL3}
\end{figure}

$J = 0.005$ and $\Delta J = 0.0002$ were used, which allowed the ratio $\Delta
J$ to $J$ to stay within the allowed limits (see \S\ref{cs:min:asym:dJ}) for
$100 \leq L \leq 2000$. A log-log graph was plotted (Fig.~\ref{fig:asymL1}).
The line is fairly straight, but with some scatter, especially for small $L$.
The reason is that for these small $L$, the midmarket moved towards zero so
quickly that not much time elapsed before it crashed into the lower boundary.
There were not many points from which one could derive a gradient, and in any
case, there was much fluctuation in the instantaneous speed of the moving
midmarket, and sometimes the fluctuations dominated, leaving the drift barely
observable. All these factors made it very hard to determine $|dx/dt|$. The
gradient of the log-log plot is -1.2, which is not too bad, though we can
certainly do better by excluding from our graph the first 6 points with the
smallest $L$, which are the most scattered. This amounted to ignoring all the
results for $L < 300$. This new graph is shown in Fig.~\ref{fig:asymL2}.
The improved statistics are:
\[
    \ln |dx/dt| = (-1.02 \pm 0.06) \ln L - (3.0 \pm 0.4).
\]
The exponent is now much closer to the theoretical value of -1. The constant
in front is $0.05 \pm 0.02$. A more precise estimate may be obtained by a
simple plot of $|dx/dt|$ against $1/L$, as we discovered earlier. Once again,
the 6 points corresponding to $L < 300$ (now appearing last) will be excluded
from the plot, which is shown in Fig.~\ref{fig:asymL3}.
The intercept is $(-0.9 \pm 7) \times 10^{-6}$ which is basically equivalent
to zero. The slope is $0.045 \pm 0.004$, thus giving us an estimate of the
constant of proportionality that is 5 times more precise than that provided by
the log-log plot. We are therefore satisfied that the $L$ dependence really is
an inverse one.

\subsubsection{Constant of proportionality}
Having established the functional dependence of the speed of the moving
midmarket, $|dx/dt|$, on $J$, $\Delta J$, and $L$, we have demonstrated the
truth of the formula suggested by Kogan and Eliezer, to within a constant of
order 1. We shall now attempt to determine what that constant is. It is not
necessary to calculate numerically the $D$ dependence (which we have never
done for any quantity) because dimensional analysis alone tells us that if the
speed, with units of dollars/sec, is proportional to $\Delta J / J L$, there
\emph{must} also be a factor of $D$ there, in order to balance the units.

For each of the dependences we investigated, a partial constant of
proportionality, which included in it the other two variables, was obtained.
>From each of these we may infer an estimate of the constant in front of all
three variables, i.e.\ the constant $k$, defined
\[
    x_0(t) = - \frac{k D \Delta J}{J L} t.
\]
>From the first set of results, concerning the $\Delta J$ dependence, we deduce
that $k = 2.09 \pm 0.05$, whilst for the second and third data sets, we obtain
$1.86 \pm 0.18$ and $2.25 \pm 0.2$, respectively. We summarize this with a
table:

\begin{center}
\begin{tabular}{cccc}
  Set & $k$ & Min & Max \\
\hline
  1st ($\Delta J$)  & $2.09 \pm 0.05$ & 2.04 & 2.14 \\
  2nd ($J$)         & $1.86 \pm 0.18$ & 1.68 & 2.04 \\
  3rd ($L$)         & $2.25 \pm 0.20$ & 2.05 & 2.45
\end{tabular}
\end{center}

Although not tremendously precise, the data do tell us that the constant is
close to 2, which was the factor suggested in Eq.~(3.37) of \cite{kogan}. Our
most precise set of results was the 1st set, without a doubt, because
\emph{all} the points lay on a straight line, for both the direct and the
log-log plots, and it is reasonable that we should pay more attention to its
conclusions than to those of the subsequent sets, for which certain wildly
inaccurate points had to be excluded. We can therefore conclude that our
numerical simulations favour a constant of about 2.1, with an error of 0.1 or
so.

This is supported by the theory. The formula in \cite{kogan} was obtained from
Eq.~(3.36) in \cite{kogan} by setting $\zeta(x,t)$, the density difference
between buyers and sellers, to zero for the midmarket, and then solving the
resulting quadratic in $x$ with the stochastic part (the Fourier series with
noise-dependent coefficients) neglected. Expanding the square root in the
quadratic formula gives Eq.~\ref{eq:asym}. Of course, for the two-liquid
model, when the flux of market-order traders exceeds that of limit-order
traders, the system decouples into two reaction fronts, between limit and
market order traders, with asymmetric fluxes. Here, the bid-offer spread
widens with a speed equal to \emph{twice} that calculated here for the minimal
model, because there are two fronts. We will investigate the two-liquid model
in the next section.

\nsection{Computer simulations of the two-liquid model} \label{cs:2liq} Having
spent much of our time and effort investigating the minimal model, it is now
appropriate to turn our attention to the two-liquid model. It is more
complicated than the minimal model to investigate, because there are four
kinds of traders: limit-order buyer, limit-order seller, market-order buyer
and market-order seller, and for this reason it is less analytically
tractable. Many of Cardy et.\ al.'s methods which worked well with the minimal
model cannot be applied here. Thus, there is good reason to resort to
numerical simulations to solve the two-liquid model.

The intention of the two-liquid model is to provide a more realistic
description of a financial market in the prelude to a crash. In particular,
what motivated the inclusion of the ``second liquid'' was the desire to
reproduce the well-known phenomenon, observed by market practitioners, of the
widening of the bid-offer spread. It has been demonstrated
(Fig.~\ref{fig:asymbo} and \ref{fig:asymspr}) that the asymmetric minimal
model, though able to produce a steadily moving trade price, does not exhibit
bid-offer spread widening. Market-order traders, believed to be the missing
element, were added to reproduce this phenomenon. The rules of trading can be
represented by the following equations, if we take $B$ and $S$ as the
limit-order traders, and $B'$ and $S'$ as the market-order ones:
\begin{eqnarray*}
    B + S \rightarrow 0 \\
    B + S' \rightarrow 0 \\
    B' + S \rightarrow 0
\end{eqnarray*}
Note that $B' + S' \rightarrow 0$ does not occur, since market-order traders
do not put up their prices on the trading screen and so they cannot see one
another. In numerical simulation, the problem arises of which ones to
annihilate when one has more than two types of traders at a reaction front.
For example, if there are $B$ and $B'$ at the same point in price space as
$S$, do we allow the limit-order traders to trade first, or let the trades
take place randomly? In the interests of computational efficiency, it was
decided to allow limit-order traders to trade first, and then let the
remainding LO buyers trade with MO sellers. The three types of annihilations
occur in the order shown above. Although this might not be representative of a
real market, it should suffice for a first attempt at simulating one.

There are so many things in this model worthy of investigation that our
research could do no more than scratch the surface, since our time was limited.
Nevertheless, we did manage to get some interesting results, which are
described in the following sections. We shall not dwell on the intricacies of
the computer program used to perform these simulations, since the principles
used here are no different from those used for the minimal model, and those we
have already discussed in ample detail in \S\ref{cs:min:sim}.


\subsection{Variation of spread with fraction of market-order traders}
\label{cs:2liq:spr} It was decided to measure the bid-offer spread as a
function of the fraction of market-order traders. Because we are using a model
in which only the flux of traders is controllable, and the total number of
traders of either type fluctuates, we can only measure the fraction of
market-order traders with respect to the relative fluxes of the two types of
traders. In other words, the fraction of market-order traders (MO) is here
defined as the MO flux divided by the \emph{total} flux of both types of
traders. Using this definition, it is possible to change the fraction of MO
traders by changing the MO flux, while keeping the total flux the same. It
will be seen that this definition of MO fraction is supported by theory (which
does \emph{not} support a definition based on trader number).

\begin{figure}
\centering \resizebox{\figwidth}{!}{\includegraphics{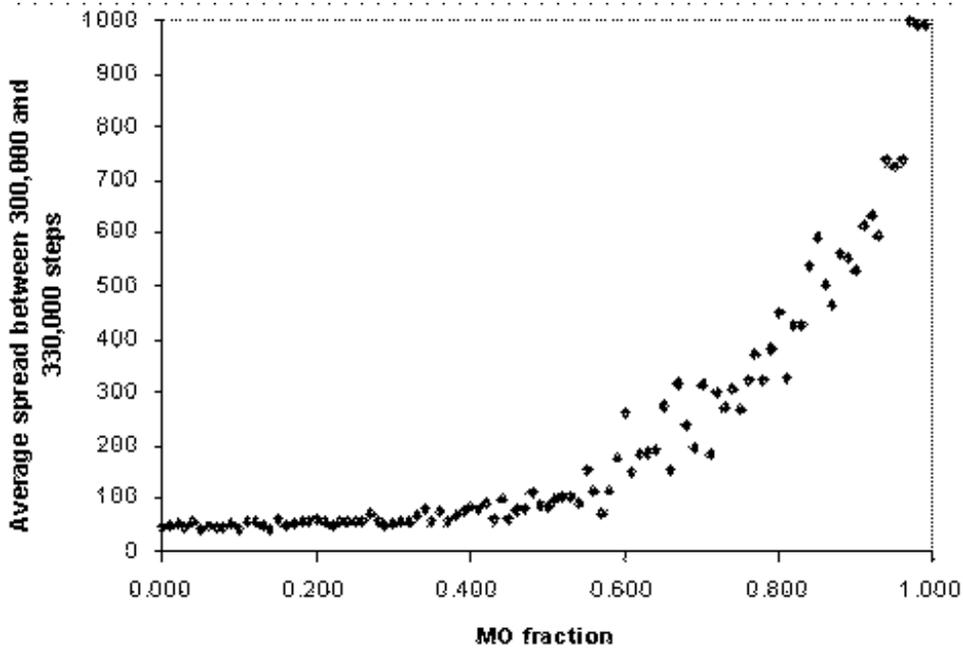}}
\caption{Variation of bid-offer spread with fraction of market-order traders}
\label{fig:2liqspr}
\end{figure}

As time progresses, we know that in cases where MO flux exceeds LO
(limit-order) flux, the bid-offer spread widens. The problem of what value to
take as the spread for a particular flux configuration thereby arises. It was
decided to define the spread here as the average spread between $t = 300,000$
and $t = 330,000$. This interval was chosen because it was \emph{believed} to
be sufficient time for the system to come to a sort of steady-state, yet not
too long so that the two separating reaction fronts crash into the boundaries
of price space. The mean flux was chosen to be 0.0005, a relatively high flux,
because midmarket fluctuations, $w^2$, which we wanted to minimize in order to
have more accurate results, are known to go as $1/J$. Thus, a 10\% MO flux
would correspond to $J_{MO} = 0.0001$ and $J_{LO} = 0.0009$. Simulations were
performed for many different fractions and the results were plotted on a graph
(Fig.~\ref{fig:2liqspr}).

The critical point seems to be approximately 0.5, above which the spread begins
to grow very rapidly \footnote{Preliminary results that bifurcation
point in the Two-Liquid Model occurs at $f = 1/2$ was obtained by
Adrian McGowan and one of us (I.I.K.) in the spring 1999.}.
 There is some fluctuation in the results, but they do
fit a curve quite well, though they would fit a straight line probably just as
well---we cannot tell at this stage. The value of the spread is undefined for
100\% market-order traders.

It is interesting to consider what would happen to the spread as the MO
fraction tends to 1. We may apply Eq.~\ref{eq:asym} to the two-liquid model if
we assume that it consists of two decoupled minimal systems, of MO against LO
traders, with reaction fronts which move apart (this only works if $J_{MO} >
J_{LO}$). We must also assume that the two fronts act independently of one
another. This is only an approximation and does not apply at the beginning of
the simulation, when the system is still trying to reach a steady-state.
Eq.~\ref{eq:asym} tells us that for a system with asymmetric fluxes $\bar J +
\Delta J$ and $\bar J - \Delta J$, the speed of movement of the midmarket
(which is effectively the same as the rate of movement of the reaction
\emph{front}, since the spread stays constant) is $2D\Delta J / \bar J L$. In
our case, $J_{MO} = \bar J + \Delta J$ and $J_{LO} = \bar J - \Delta J$.
Making the necessary substitutions, and remembering that the fraction $f$ of MO
traders is defined $f = J_{MO} / (J_{MO} + J_{LO})$, we obtain
\[
    \frac{d(Spr)}{dt} = \frac{4D(2f-1)}{L}
\]
which tells us how the spread increases with time. Integrating, we have
\begin{equation}
\label{eq:2liqspr}
    Spr(t) = \frac{4D(2f-1)t}{L} + S_0
\end{equation}
where $S_0$ is the ``natural'' spread which exists in the absence of the
drift, at $f = 0.5$. The formula predicts, therefore, that the graph of spread
against fraction $f$ would be linear for $f > 0.5$, with a gradient of
$8Dt/L$. In our case, this would have been 1260, if we use $t = 315,000$. The
spread at $f = 0.5$, $S_0$, is about 84, leading to a prediction of 714 for
the y-intercept.

\begin{figure}
\centering \resizebox{\figwidth}{!}{\includegraphics{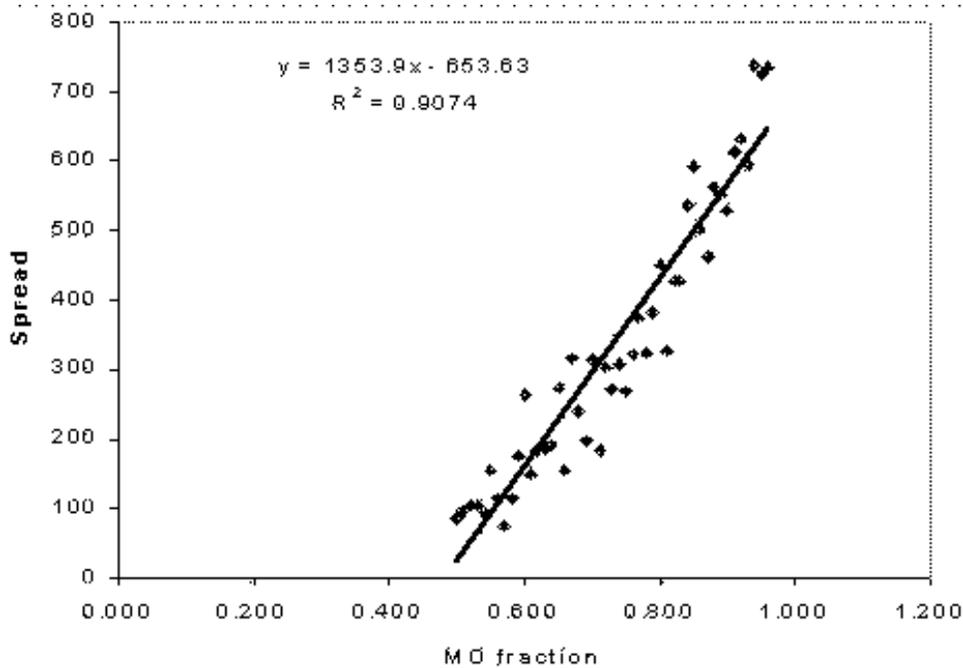}}
\caption{Variation of bid-offer spread with fraction of market-order traders,
for $f \geq 0.5$ and excluding 3 outliers} \label{fig:2liqspra}
\end{figure}

We have already observed that the portion of the graph (Fig.~\ref{fig:2liqspr})
for $f > 0.5$ is more like a curve than a straight line. However, if we use a
straight line as a first approximation, and fit it to the points $f > 0.5$, we
can see how the theoretical prediction matches up with the data. This was done
in Fig.~\ref{fig:2liqspra} which was a plot of all the points for which $f
\geq 0.5$, excluding the last three points which were obvious outliers.
The graph gives a gradient of $1354 \pm 64$ and an intercept of $654 \pm 48$.
Comparing the gradient with the theoretical value of 1260, one can see that
the agreement is quite good. The y-intercept, too, is in surprisingly good
agreement with the calculated value. Having said that, it is clear that
Eq.~(\ref{eq:2liqspr}) gives little more than an estimate of what the spread
is. In any case, the equation only works in the region $f \geq 0.5$.

\begin{figure}
\centering \resizebox{\figwidth}{!}{\includegraphics{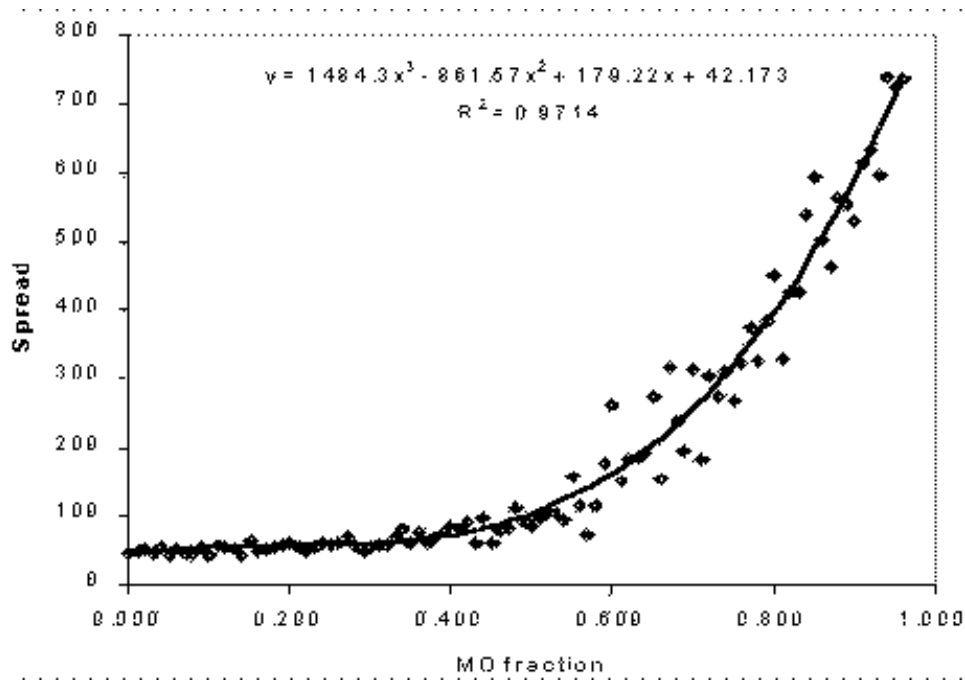}}
\caption{Cubic fit to variation of spread with MO fraction, excluding 3
outliers} \label{fig:2liqsprfit}
\end{figure}

A market practitioner would want a better fit to the data than a crude
straight line. To this end, we have fitted the curve (less the 3 outliers) to a
cubic, as shown in Fig.~\ref{fig:2liqsprfit}. The fit is remarkably good.
It cannot be determined with any certainty at this stage whether the portion
of $f > 0.5$ is a curve, like a cubic, or a straight line.

One wonders whether, in spite of the excellent cubic fit performed above, a
single description of the variation of spread, or indeed any quantity, can be
valid for all $0 < f < 1$. A cause for concern lies in the very sharp
transition from a stable system $f < 0.5$ to an unstable one $f > 0.5$, with
$f = 0.5$ being the critical point. We expect 0.5 to be the critical point,
because an infinitesimal increase of $f$ beyond this point would produce an
imbalance of MO and LO fluxes, which would, in our model, inevitably lead to a
steady widening of the spread. Yet is this the case? We see a sharp rise in
spread at around 0.5, but does a phase transition really occur at exactly 0.5?
The spread plotted was only the average around $t = 315,000$; perhaps we have
been observing the spread too soon after the beginning of the simulation, and
the system has not had time to reach a steady-state? These are all valid
questions, which we now try to answer by performing more extensive simulations.

\subsection{More definite results for spread}
\label{cs:liq:sprdef} Admittedly, $t = 300,000$ to $t = 315,000$ is rather soon
to start observing market behaviour; we know that for symmetric fluxes, the
market requires approximately 1 million time steps to come into equilibrium.
We were forced to do so in order to maximize the range of $f$'s accessible to
us. It was necessary to compare spreads measured at the \emph{same point in
time} for each simulation. Had we chosen a later time to observe, e.g.\ 1
million steps, we would not have been able to observe the variation of spread
with $f$ for $f \geq 0.75$ or so, for which the spread at $t = 10^6$ would
have exceeded 1000\footnote{One might think that the problem can be solved by
simply increasing $L$. Not only does this increase the size of price space,
but it also slows down the widening process (speed of widening $\sim 1/L$) and
overall, the time taken for the spread to reach $L$ goes as $L^2$. However,
the time taken for the system to reach equilibrium as a whole goes as $L^2$, 
cancelling exactly with the increase in time gained from increasing $L$.}. 
If we look at $f = 0.0$, and compare the spread 
obtained by averaging from $t = 300,000$ to $t = 330,000$, and that obtained by
averaging from $t = 10^6$ to $t = 2 \times 10^6$, we see that they are 45.5
and 41.7, respectively. An acceptable difference, if we bear in mind the
inherent fluctuations. However, for $f = 0.50$, the spreads are 84.0 and
127.5, respectively. They differ by a factor of 1.5. Therefore, it is probable
that the system had \emph{not} reached a steady-state by the 300,000th time
step.

\begin{figure}
\centering \resizebox{\figwidth}{!}{\includegraphics{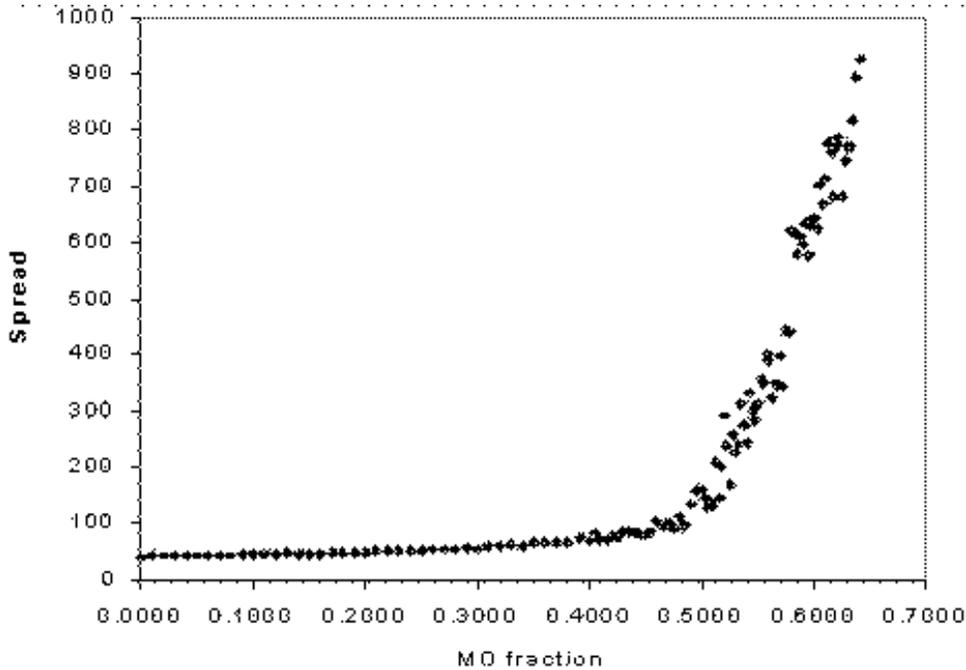}}
\caption{Variation of spread with MO fraction, with spread averaged over 1
million steps} \label{fig:2liqspr1M}
\end{figure}

\begin{figure}
\centering \resizebox{\figwidth}{!}{\includegraphics{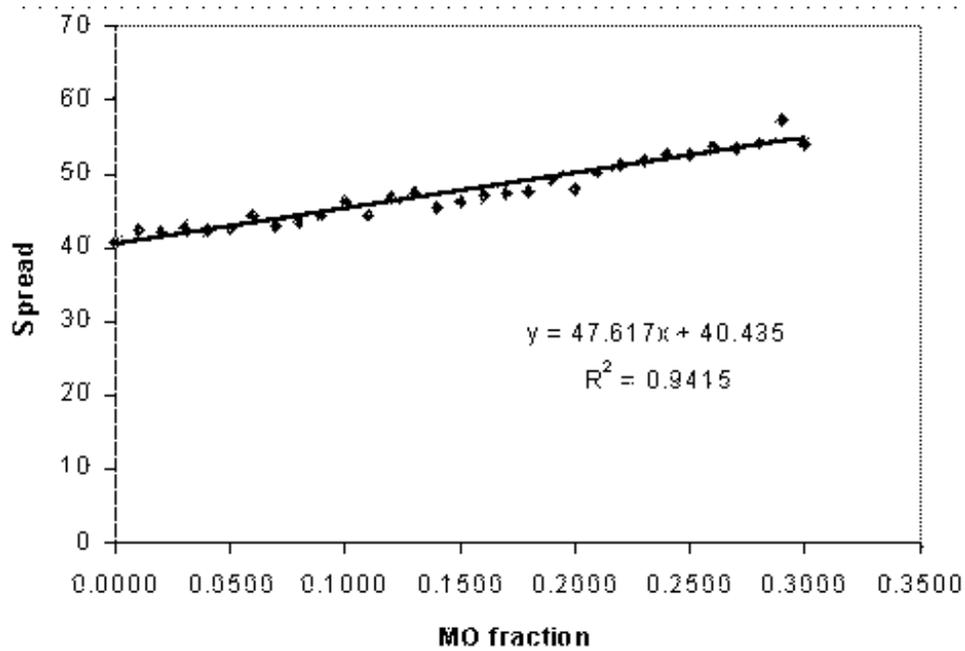}}
\caption{Variation of spread with $f$ for $f \leq 0.3$, with spread averaged
over 1 million steps} \label{fig:2liqspr1Ma}
\end{figure}

\begin{figure}
\centering \resizebox{\figwidth}{!}{\includegraphics{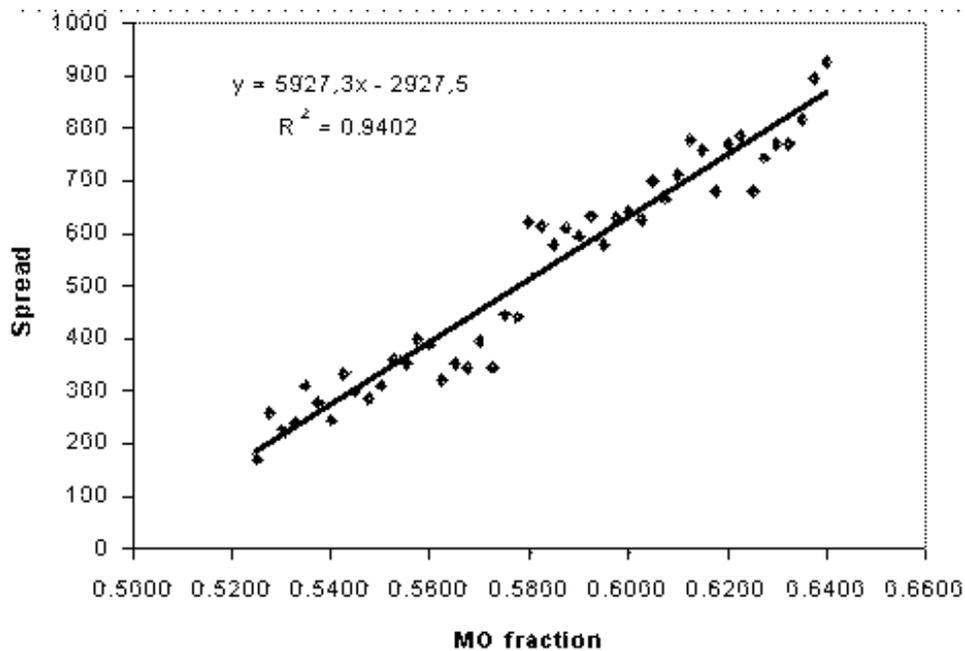}}
\caption{Variation of spread with $f$ for $f \geq 0.525$, with spread averaged
over 1 million steps} \label{fig:2liqspr1Mb}
\end{figure}

In order to probe this further, the spread calculations performed earlier were
repeated, this time taking the spread as the average spread between 1 million
and 2 million time steps (Fig.~\ref{fig:2liqspr1M}). This is the most
comprehensive set of results for two-liquid spread so far.
Looking at the graph, one is immediately struck by how well the points $f <
0.5$ fit a straight line. The reason for this is that the average over 1
million time steps is less prone to fluctuation than that over a mere 30,000
steps. We are limited in what region of $f$ we can probe, since the spread
shoots upwards very quickly once $f$ exceeds 0.5, and reaches 1000 around 0.64
or so. One can also see that the graph divides neatly into two straight
sections: $f < 0.5$ and $f > 0.5$. This is encouraging, as it accords with
what we predicted using the formula for asymmetric fluxes applied to the
two-liquid model, Eq.~(\ref{eq:2liqspr}). Once again, we plot the two
sections, $f < 0.5$ and $f > 0.5$ separately to determine the gradient and
intercept (Fig.~\ref{fig:2liqspr1Ma} and \ref{fig:2liqspr1Mb}).

We consider first the section corresponding to $f \leq 0.3$, which is the
straightest part of the $f < 0.5$ half. The full plot in
Fig.~\ref{fig:2liqspr1M} shows an almost perfect straight line. The expanded
version in Fig.~\ref{fig:2liqspr1Ma} supports this, with the fitted line
achieving a correlation coefficient in excess of 0.94. The gradient and
intercept are, respectively, $47.6 \pm 2.2$ and $40.4 \pm 0.4$. Initially,
attempts were made to include points up to $f = 0.4$ but it was found that the
points did not lie so well on a straight line. At the moment, we do not have a
theory that can explain this straight portion. Note that the spread at 0.5
cannot be worked out by extrapolation from this formula, since the spread
starts to deviate from this line at $f \sim 0.4$.

Turning to the $f \geq 0.525$ portion, we note first that, contrary to previous
results, the points quite clearly fit a line. The gradient and intercept are
$5927 \pm 223$ and $-2928 \pm 130$, respectively. Theory predicts that, for $t
\simeq 1,500,000$, the gradient $d(Spr)/df$ would be $8Dt/L = 6000$. The
intercept is $S_0 - 4Dt/L$, the first term of which has to experimentally
determined. We find that the equation describing $f \geq 0.5$ is
\[
    Spr(f) = (5927 \pm 223)f - (2928 \pm 130).
\]
The gradient agrees with theory, to well within experimental error. From
Fig.~\ref{fig:2liqspr1M}, we find, by visual inspection, that the spread at the
point of symmetry, $f = 0.5$, is $140 \pm 20$. Incidentally, \emph{neither}
straight line section, when extrapolated to $f = 0.5$, gives the correct
result, since 0.5 is in the curved section joining the two straight sections,
and is larger than either straight line formula would predict. Now that we
know $S_0$, we also know what theory predicts for the intercept of the $f \geq
0.525$ graph: it is $-2860 \pm 20$ (the uncertainty comes from our experimental
determination of $S_0$). The actual intercept agrees well with our prediction.
Thus, we are pleased to see good agreement with theory.

In the light of these much more detailed results, especially
Fig.~\ref{fig:2liqspr1M}, it appears that the spread starts to rise rapidly
\emph{just before} $f = 0.5$, because we have already observed that the spread
at that point of symmetry deviates from either linear regime. We recall that,
in our method of simulation, whenever there was an abundance of traders that
needed to be annihilated at a point in price space, we decided to always allow
the LO traders to annihilate first. Such a technique would, in general, cause
the best LO bid/offer to be further apart than it would be if we had not made
that imposition and instead had allowed the LO and MO traders to trade with
equal probability. This is because the spread is determined by LO traders
alone, and if LO traders are more likely than MO traders to get annihilated,
it is reasonable that the best bids and offers would `snap back' more often,
resulting in an increased average spread. Immediately, we can see that this
preference for LO-LO trades only affects those trades where both LO buyers and
LO sellers are present, i.e.\ when the spread is zero. Therefore, for $f >
0.5$, where the spread widens at a constant rate, and apart from at the
beginning, the LO traders are completely separated, and we would expect the
LO-LO preference to make no difference, since the LO traders on either side
are no longer in contact. For $f$ just below 0.5, we expect no steady drift of
the best bid or offer due to asymmetric fluxes, and so the situation is less
clear. During all those times when the best LO bid and offer are not in
contact, the priority given to LO-LO trades makes no difference. When they are
coincident, however, \emph{more} of the LO traders are going to be annihilated
than they would otherwise. As a result, in such an annihilation where traders
of all types are present, the number of LO traders will be more greatly
diminished than that of MO traders. To consider what happens to the LO best
bid, envisage a situation where, at the midmarket, the number of LO buyers is
less than the total number of LO/MO sellers, which in turn is less than the
total number of LO/MO buyers. Following our scheme, LO-LO trades occur first,
followed by LO-MO trades. All LO buyers are eliminated. The only way for the
best bid \emph{not} to change would be for the number of LO buyers to exceed
the total number of sellers. Had we used a random scheme of mutual
annihilation, however, since buyers outnumber sellers, there would be no reason
why some LO buyers should not be left. Similarly, if we perform this analysis
for sellers, we would find a similar prejudice against the best offer staying
where it is. In both cases, the number of LO traders has to outnumber the
total number of the opposite type of trader in order for the best bid/offer to
remain unchanged. For a random annihilation scheme, the only condition is that
one type of trader outnumber the other, and then there is the
\emph{possibility} of either the best bid or best offer to remain unaltered.
Therefore, it can be seen that increased spread is an artefact of the our
simulation.

By itself, the above analysis does not explain why the critical point might
occur below 0.5. To do that would require one to know quantitatively how the
LO-LO preference affects the spread. Alternatively, to obviate this problem,
it might be useful to change the simulation algorithm so that trades are
indeed random, and that no preference is given to any type of trade. Such a
modification would greatly increase processing time though it would probably
resolve our uncertainty.

It is interesting to note the numerical prefactor that is defined as the ratio
of the spread at the point of symmetry ($f = 0.5$) to that for the minimal
model ($f = 0.0$). From the results we already have, we can immediately give a
value for it. The former is $140 \pm 20$ and the latter $40.4 \pm 0.4$, both
coming from the graph of spread in the region $f < 0.5$. The ratio is
therefore $3.47 \pm 0.50$.

\nsection{Conclusion} \label{conclusion} The  numerical analysis
performed in this paper
 confirmed that $w^2$, $Spr^2$ and $\var(Spr)$ are all proportional to $D/J$,
as predicted by dimensional analysis. The results also confirmed the existence
of logarithmic correction for $w^2$, but not for $Spr$ or $\var(Spr)$. This is
not surprising, because the midmarket is ergodic and, given enough time, it
will cover the entire width of price space. Increasing $L$ increases the range
in price space over which the midmarket can roam; thus, the midmarket variance
diverges as $L \rightarrow \infty$. The constant in front of the $D/J$ in the
$w^2$ scaling law was confirmed to be $1/\pi$. The logarithmic parameter for
$w^2$ is $c \sim 0.56$, $0.32 < c < 0.97$, for $L = 1000$. It appears that $c
\sim O(1)$ for $500 < L < 1000$. $Spr$ and $\var(Spr)$ do not seem to have
logarithmic corrections. This is because $Spr$ (and $\var(Spr)$, which is
derived from $Spr$) is a fundamentally different quantity from the midmarket
variance. Intuitively, changing $L$ should not change $Spr$, since it is a
property of the reaction front. The constants in front of $D/J$ for $Spr$ and
$\var(Spr)$ are $3.615 \pm 0.065$ and $0.95 \pm 0.02$, respectively, whilst
the approximate `constant' in front of $D/J$ for $w^2$ is $1.9 \pm 0.2$. We
can therefore observe a hierarchy in the three major lengths: $\var(Spr) < w^2
< Spr$ according to $0.95 < 1.9 < 3.6$. It was discovered that
$NUM$ scaled as $JL^2/4D$, and a theoretical justification was given. This law
was the most exact of all the scaling laws that the data supported; there was
almost perfect correlation for all graphs of $NUM$ against $J$ and $L$. Time
to midmarket sale ($\tau_S$) was investigated, and found to be equal to $1/J$,
as expected. Its fluctuation also scaled as $1/J$, but with $0.73 \pm 0.05$ as
the constant of proportionality. Neither quantity showed signs of having
logarithmic corrections. A further quantity, $\tau_{reduced}$, a sort of
scaled time, was investigated. Its scaling law seemed a little less
straightforward, as the points on the log-log plot quite clearly followed a
curve rather than a line. Further investigation is needed to determine its
exact nature. The minimal model was extended by allowing asymmetric fluxes of
buyers and sellers. Simulations verified the $\Delta J$, $J$ and $L$
dependences, and the constant of proportionality, of the formula
(Eq.~(\ref{eq:asym}) in this paper) predicted by Kogan et.\ al.\ in
\cite{kogan} for the speed of a moving midmarket in the presence of asymmetric
fluxes.

Investigations into the two-liquid model produced widening of the bid-offer
spread in the prelude to a crash, when the flux of market order traders
exceeded the flux of limit order traders. Application of Eq.~(\ref{eq:asym})
to the spread yielded approximate, though uncertain, agreement. A cubic fit to
the spread as a function of MO fraction was performed. Further, more detailed,
results confirmed that the graph of $Spr$ against $f$ was made up of two
straight sections, in agreement with theory. The predicted gradient and
intercept of the second straight section were corroborated by experiment. The
formula for asymmetric fluxes was found to describe the bifurcated regime
well, but sheds no light on the equilibrium regime, where $f < 0.5$. The
critical point was found to be approximately $f = 0.5$, or perhaps just before.
The ratio of spread for $f = 0.0$ and $f = 0.5$ was calculated to be $3.47 \pm
0.50$.

The  analysis presented in this paper 
 has covered five major scaling laws described in \cite{kogan}.
However, there are many more to be tested, and so there is much scope for
further research. Other scaling laws that might be investigated include density
near the best bid/offer as a function of deal rate, and higher correlation
functions describing equilibrium. We have only scratched the surface of the
two-liquid model; it would be interesting to see how the other scaling laws
are modified in the prelude to a crash. Finally, \cite{kogan} describes a third
model, the bias model, which attempts to incorporate herd, or crowd, effects.
It does so by modifying the diffusion operators to contain a drift, which
depends on whether the last trade has moved upwards or downwards. This
modification models momentum trading, where the traders have a tendency to go
with the crowd. Such a model is designed to develop instabilities, and the
drift parameter may be considered to be in a meta-stable basin, buffeted back
and forth by random market movements (diffusion), until the drift is so large
that it is knocked out of its basin, into the surrounding unstable region.
Once there, it gathers momentum by positive feedback, leading to a crash. In future
work we shall investigate the bias model.

\nsection{Acknowledgements}
 The work of one  of us (D.L.C.C.)  was partly supported by the  summer
project grant  from  Physics Department, Oxford University.
D.L.C.C. would like to thank Ming Lee  and Stella Chan for their
support and understanding. D.E. would like
 to thank Matt Moen, James Lee, Monique Elwell. I.I.K. would like to
thank John Cardy, Juan Garrahan, Zoltan R\'{a}cz, David Sherrington  
and Oleg Zaboronski and acknowledge 
 stimulating atmosphere of ESF programme SPHINX.
\appendix
\nsection{Appendix}
\subsection{The master equation and the diffusion coefficient}
\label{app:master} From the master equation for the trader diffusion process,
we may derive the diffusion equation and thus an expression for the diffusion
coefficient $D$.

Let $P(x|t)$ be the probability of having a particle at point $x$. Let $p$ be
the probability for a particle to hop \emph{away} from a given point, in each
time quantum $\tau$. Each hop is to a point a distance $a$ away from the
original point of the particle. Let the probability of hopping to the left and
to the right be equal, i.e.\ $p/2$. Therefore, the probability of moving to
the left is $p/2$, moving to the right $p/2$, and staying put $(1-p)$. We
write the master equation by considering the probability for the particle to be
at $x$ at a time $t + \tau$:
\begin{eqnarray*}
    P(x|t+\tau) & = & \frac{p}{2} P(x+a|t) + \frac{p}{2} P(x-a|t) + (1-p)P(x|t) \\
    P(x|t+\tau) - P(x|t) & = & \frac{p}{2} [P(x+a|t) + P(x-a|t) - 2 P(x|t)]
\end{eqnarray*}
We expand either side in a Taylor series about $t$ and $x$. The odd
derivatives cancel out on the right hand side, leaving just the even
derivatives, which add. For small $a$ and $\tau$, we may neglect terms of
second order or higher in $\tau$ on the LHS and fourth order or higher in $a$
on the RHS. The partial derivatives are all evaluated at $x$ and $t$.
\begin{eqnarray*}
    \tau \frac{\partial P(x|t)}{\partial t} + \ldots & = & \frac{p}{2}\left[
    \frac{\partial^2 P(x|t)}{\partial x^2} \left( \frac{a^2}{2} \right) +
    \ldots + \frac{\partial^2 P(x|t)}{\partial x^2} \left(\frac{a^2}{2}\right) + \ldots
    \right]\\
    \frac{\partial P(x|t)}{\partial t} & = & \left(\frac{a^2 p}{2\tau}\right)
    \frac{\partial^2 P(x|t)}{\partial x^2}.
\end{eqnarray*}
This is the diffusion equation for $P(x|t)$, with diffusion coefficient $D =
a^2 p / (2\tau) $. In our simulations, $p = 1$ so $D = a^2 / (2\tau)$, or $D =
1/2$, for $a = \tau = 1$.

\subsection{Proof of $\expect{X} = \expect{\overline{X}}$ and $\var(X) =
\expect{\overline{X^2}} - \expect{\overline{X}}^2$} \label{app:meanvar} If we
have a set of $N$ results $\set{X_i}$ and we choose to record only the
averages over $n$ steps, $\overline{X}_j$, of the $j^{\mathrm{th}}$ set of $n$
values, such that
\begin{eqnarray*}
\overline{X}_j = \frac{1}{n} \sum_{i = n(j-1)+1}^{nj} X_i, & 1 \leq j \leq N/n,
\end{eqnarray*}
then we may obtain the average ($\langle \ldots \rangle$) over the entire set
of data by
\[
\expect{X} = \frac{1}{N} \sum_i X_i = \frac{1}{N} \sum_{j=1}^{N/n}
n\overline{X}_j.
\]
This we may rewrite as
\[
\expect{X} = \frac{1}{N/n}\sum_{j=1}^{N/n} \overline{X}_j =
\expect{\overline{X}}
\]
which is simply the mean of the set of recorded averages ($n$ is a constant).
Similarly, we may obtain the variance of an entire set of data from knowing
the sum of the squares, and the mean, of sets of $n$ values:
\begin{eqnarray*}
\var(X) & = & \frac{\sum_i X_i^2}{N} - \left( \frac{\sum_i X_i}{N} \right)^2 \\
        & = & \frac{\sum_j n\overline{X^2}_j}{N} - \left( \frac{\sum_j
        n\overline{X}_j}{N} \right)^2 \\
        & = & \frac{\sum_j \overline{X^2}_j}{N/n} - \left( \frac{\sum_j
        \overline{X}_j}{N/n} \right) ^2 \\
        & = & \expect{\overline{X^2}} - \expect{\overline{X}}^2.
\end{eqnarray*}
Thus, the variance of an entire set of data is equal to the mean of the
averaged squares minus the square of the mean of the recorded averages. Armed
with these relations, we may reconstruct the mean and variance of the whole
set of data from the averaged recorded data.



\normalsize
\bibliography{paper}
\bibliographystyle{amsplain}

\end{document}